\documentclass[11pt,letterpaper]{article}

\usepackage{jheppub_mod}
\usepackage{slashed}
\usepackage{graphicx}
\usepackage{caption}
\usepackage{subcaption}
\usepackage{bbm}
\usepackage{psfrag}
\usepackage{mathrsfs}

\makeatletter
\newsavebox\myboxA
\newsavebox\myboxB
\newlength\mylenA

\newcommand*\xoverline[2][0.75]{%
    \sbox{\myboxA}{$\m@th#2$}%
    \setbox\myboxB\null
    \ht\myboxB=\ht\myboxA%
    \dp\myboxB=\dp\myboxA%
    \wd\myboxB=#1\wd\myboxA
    \sbox\myboxB{$\m@th\overline{\copy\myboxB}$}
    \setlength\mylenA{\the\wd\myboxA}
    \addtolength\mylenA{-\the\wd\myboxB}%
    \ifdim\wd\myboxB<\wd\myboxA%
       \rlap{\hskip 0.5\mylenA\usebox\myboxB}{\usebox\myboxA}%
    \else
        \hskip -0.5\mylenA\rlap{\usebox\myboxA}{\hskip 0.5\mylenA\usebox\myboxB}%
    \fi}
\makeatother

\newcommand{\cC}{\mathcal{C}}
\newcommand{\cD}{\mathcal{D}}

\newcommand{\cH}{\mathcal{H}}

\newcommand{\cK}{\mathcal{K}}

\newcommand{\cN}{\mathcal{N}}
\newcommand{\cO}{\mathcal{O}}

\newcommand{\cR}{\mathcal{R}}

\newcommand{\cV}{\mathcal{V}}

\newcommand{\cY}{\mathcal{Y}}

\newcommand{\uB}{\mathrm{B}}

\newcommand{\uW}{\mathrm{W}}

\newcommand{\ud}{\mathrm{d}}
\newcommand{\ue}{{e}}

\newcommand{\CC}{\mathbb{C}}

\newcommand{\RR}{\mathbb{R}}
\newcommand{\ZZ}{\mathbb{Z}}

\newcommand{\Othree}{\mathrm{CY}}

\newcommand{\WWW}{\mathrm{WWW}}
\newcommand{\WW}{\mathrm{WW}}
\newcommand{\iid}{\mathrm{i.i.d.}}

\newcommand{\ga}{\gamma}
\newcommand{\ka}{\kappa}

\newcommand{\sig}{\sigma}

\newcommand{\vt}{\vartheta}

\newcommand{\CY}{\mathrm{CY}}

\newcommand{\ex}{\cY_{25}}
\newcommand{\Ri}{\mathscr{Ri}}

\newcommand{\U}[1]{\mathrm{U}\!\left(#1\right)}

\newcommand{\la}[1]{\mathtt{#1}}

\title{Heavy Tails in Calabi-Yau Moduli Spaces}

\author{Cody Long, Liam McAllister, and Paul McGuirk}

\affiliation{Department of Physics, Cornell University, Ithaca, New
  York, 14853, USA}

\emailAdd{cel89@cornell.edu}
\emailAdd{mcallister@cornell.edu}
\emailAdd{mcguirk@cornell.edu}

\abstract{We study the statistics of the metric on K\"ahler moduli
  space in compactifications of string theory on Calabi-Yau
  hypersurfaces in toric varieties.  We find striking hierarchies in
  the eigenvalues of the metric and of the Riemann curvature
  contribution to the Hessian matrix: both spectra display heavy
  tails.  The curvature contribution to the Hessian is non-positive,
  suggesting a reduced probability of metastability compared to cases
  in which the derivatives of the K\"ahler potential are uncorrelated.
  To facilitate our analysis, we have developed a novel triangulation
  algorithm that allows efficient study of hypersurfaces with
  $h^{1,1}$ as large as 25, which is difficult using algorithms
  internal to packages such as Sage.  Our results serve as input for
  statistical studies of the vacuum structure in flux
  compactifications, and of the distribution of axion decay constants
  in string theory.}

\begin{document}

\maketitle

\newpage

\section{\label{sec:intro}Introduction}

The vacuum structure of compactifications of string theory to four
dimensions is extremely complex, and direct enumeration of all vacua
appears impossible.  Moreover, it is plausible that the number of
vacua of string theory that are consistent with all present
observations vastly exceeds the number of measurements of fundamental
parameters that could be performed in the future.  In this situation
it is impractical, and for many purposes misguided, to
pursue individual vacua of string theory as candidate models of the
observed universe.  A more fruitful course is to determine the
characteristic properties of classes of vacua, and make statistical
predictions \cite{Douglas:2003um}.

An idealized enumerative approach to vacuum statistics would begin
with discrete data, including the compactification topology and a
specification of quantized fluxes, and compute the full
four-dimensional low-energy effective action as a function of these
integer inputs.  One could then study the statistics of physical
observables of interest in the ensemble of effective actions resulting
from all possible choices of discrete data.  However, despite
extensive efforts to understand the statistics of string vacua (for
reviews with references,
see~\cite{Grana:2005jc,Douglas:2006es,Denef:2008wq}), the present
reality falls short of this ideal.  In particular, quantum corrections
to the effective action are poorly understood, and many of the most
incisive results concern the distribution of supersymmetric Minkowski
vacua of a subsector of the theory, in the classical approximation.
For example, the elegant result of \cite{Ashok:2003gk} in type~IIB
flux vacua describes an index approximating the distribution of
configurations for which the classical F-flatness conditions for the
complex structure moduli hold.  The K\"ahler moduli, which can
parameterize important instabilities, are not included.

A central idea advanced in
\cite{Douglas:2003um,Ashok:2003gk,Denef:2004ze,Denef:2004cf} is that a
reasonable approximation to the true distribution of string vacua can
be obtained by determining the statistics of vacua in an associated
ensemble of ${\cal N}=1$ supergravity theories that are random in an
appropriate sense.  The expectation is that the Wilson coefficients of
operators in the associated ensemble, when viewed as stochastic
variables, will closely follow the distribution of the true Wilson
coefficients that would be obtained if one could produce and study a
large ensemble of genuine string vacua.  Ideally, this expectation can
be checked and refined by comparison to small ensembles of string
vacua in computable special cases.

To specify an appropriate ensemble of random supergravity
theories, one must provide an ensemble of superpotentials $W$, each of
which is a holomorphic section of a nontrivial holomorphic line bundle
over the moduli space ${\cal M}$ parameterized by the scalar fields
$\phi^{a}$ of the theory, as well as a corresponding ensemble of
K\"ahler potentials, which in local coordinates
are real analytic functions of the $\phi^a$
and their conjugates $\bar{\phi}^{\bar{a}}$.
In the simple special case of supersymmetric Minkowski solutions, the vacua
are determined by holomorphic equations of the form
\begin{equation}
  \frac{\partial W}{\partial \phi^{a}} =0\,,
\end{equation} with no dependence on the K\"ahler potential.
The task of computing the distribution of supersymmetric Minkowski
vacua therefore requires only the specification of statistical
properties of superpotentials, and some significant mathematical
results are applicable to this problem: there exists a precise
notion of a random holomorphic section of a random holomorphic line
bundle in this context \cite{Douglas:2004zu}.

Unfortunately, knowledge of the superpotential alone is insufficient
for studying supersymmetric vacua with nonvanishing cosmological
constant, or non-supersymmetric vacua of any type: one must also
specify the K\"ahler potential.  Rather little is known about what one
could reasonably call a {\it{random K\"ahler potential}}, or a
{\it{random K\"ahler metric}}, especially for ${\rm{dim}}_{\mathbb
  C}{\cal{M}}>1$, which is the case of primary interest (though
see~\cite{Ferrari:2011we,*Ferrari:2011is,*Ferrari:2011sf} for recent
progress).  Because possible K\"ahler potentials are poorly
characterized, it appears difficult to give a precise definition of a
{\it{random supergravity theory}}, and thereby to create a
well-defined ensemble of effective theories whose statistics could
accurately model the statistics of non-supersymmetric solutions of
string theory.

In this work we begin to address this question, by exploring the
characteristics of K\"ahler metrics in a well-motivated ensemble of
effective theories. Specifically, we study the K\"ahler
moduli spaces of compactifications of string theory on Calabi-Yau
hypersurfaces in toric varieties, considering in detail O3/O7
orientifolds of type~IIB.  A primary goal is to understand the extent
to which the resulting ensemble of K\"ahler metrics can be viewed as
an ensemble of random matrices without ``special'' properties.  Said
differently, our null hypothesis --- assumed implicitly or explicitly
in earlier works such as
\cite{Douglas:2003um,Ashok:2003gk,Denef:2004ze,
  Denef:2004cf,Marsh:2011aa} --- is that the various covariant
derivatives of the K\"ahler potential at a fixed point in field space
are tensors whose components are statistically independent and
identically distributed (i.i.d.).   For brevity, we will say that the K\"ahler potential is an \textit{i.i.d.~function}.\footnote{We use the term ``i.i.d.~function''
  to avoid confusion with other notions of random functions, which can
  entail correlations among derivatives.
  See~\cite{Bachlechner:2014rqa} for an analysis of Gaussian random
  functions in a related context.}  This is a rather strong
assumption, which few would argue describes the actual set of K\"ahler
potentials arising in Calabi-Yau compactifications, but it has been
used in the past for want of a ready alternative.  By exploring an
ensemble of explicit Calabi-Yau compactifications gathered from the
Kreuzer-Skarke database~\cite{KS_database} we will be able to reject
the null hypothesis and expose some of the underlying structure of the
metrics on Calabi-Yau moduli spaces.\footnote{K\"ahler potential data
  of an ensemble of Calabi-Yau manifolds gathered from the
  Kreuzer-Skarke database were also used in~\cite{Gray:2012jy} to
  explore the prevalence of the Large Volume Scenario vacua
  of~\cite{Balasubramanian:2005zx}.}  This is a step toward specifying
the ensemble of possible non-holomorphic data along the lines that
possible superpotentials were characterized in \cite{Douglas:2004zu}.

More specifically, our null hypothesis is that the curvature
contribution to the Hessian matrix,
\begin{equation}
  \label{eq:HR}
  \bigl(\cH_{R}\bigr)_{a\bar{b}} :=  -
  e^{\cK} R_{a\bar{b}c\bar{d}}\bar{F}^{c}F^{\bar{d}}\,,
\end{equation}
can be modeled, as proposed in \cite{Marsh:2011aa}, as the sum of a
Wigner matrix and a Wishart matrix,
\begin{equation}
  \label{eq:WW_curvature_Hessian_NH}
  \bigl(\cH_{R}\bigr)_{a\bar{b}} \approx
  \bigl(\cH^{\mathrm{WW}}_{R}\bigr)_{a\bar{b}} :=
  \underbrace{\cK_{ac}^{\phantom{ac}e}\cK_{\bar{b}\bar{d}e}\bar{F}^{c}F^{\bar{d}}}_{\mathrm{Wishart}}
  -\underbrace{\cK_{a\bar{b}c\bar{d}}\bar{F}^{c}F^{\bar{d}}}_{\mathrm{Wigner}}\,.
\end{equation} where $\cK_{\bar{b}\bar{d}e}$ and $\cK_{a\bar{b}c\bar{d}}$ denote derivatives of the K\"ahler potential $\cK$, and $F_{a}$  is the F-term.

We find compelling evidence against
(\ref{eq:WW_curvature_Hessian_NH}).  For Calabi-Yau hypersurfaces in
toric varieties, the eigenvalue distributions of $\cH_{R}$ constructed
from explicit K\"ahler potential data have {\it{much longer tails}}
than predicted by (\ref{eq:WW_curvature_Hessian_NH}).  The spectra we
obtain are very leptokurtic, with a sharp peak and heavy tails.  We
obtain a related result for the K\"ahler metrics themselves: the
eigenvalue spectra of the metrics on K\"ahler moduli spaces exhibit
large hierarchies and heavy tails, at least in the case of O3/O7
orientifolds that we study directly.

We then argue that for Calabi-Yau hypersurfaces in toric varieties,
the structure of the K\"ahler metric on ${\cal M}$ leads to
distinctive changes in a basic statistical observable, the probability
of metastability, in comparison to the probability resulting from the
null hypothesis for the K\"ahler metric.  Specifically, significant
correlations originating in the geometry of the moduli space ensure
that the covariant derivatives of $\cK$ are {\it{not}} all
independent, and we argue that, in the case of O3/O7 projections of
type~IIB, these correlations turn out to make the probability of
metastability lower in the actual effective theory than it is in an
effective theory where the K\"ahler potential is an i.i.d.~function.
As a result, even though our analysis establishes that one of the
working assumptions of \cite{Marsh:2011aa}, namely
(\ref{eq:WW_curvature_Hessian_NH}), must be modified substantially to
reflect actual Calabi-Yau compactifications, the qualitative
conclusion of \cite{Marsh:2011aa}, which is that instabilities are
extremely prevalent at generic points in the moduli space, appears to be 
\textit{strengthened} by our findings, at least in this corner of the landscape.

The presence of lengthy tails in the distribution of eigenvalues of
the metric also has significant consequences for model-building with
axions in string theory, because axion decay constants $f_a$ are given
by (the square roots of) eigenvalues of the K\"ahler metric.  While
individual axion decay constants have been computed in special cases
\cite{Banks:2003sx,Svrcek:2006yi}, trends in models with multiple
axions have not previously been characterized.  Our finding that the
largest eigenvalues are hierarchically larger than the median
eigenvalues implies qualitative changes in scenarios for axion
inflation in string theory.

This paper is organized as follows.  In~\S\ref{sec:RMT} we review the
application of random matrix theory to the supergravity Hessian, and
then discuss how to include deterministic data on K\"ahler metrics in
this approach.  In~\S\ref{sec:modulisp} we review the K\"ahler moduli
space of O3/O7 orientifolds of Calabi-Yau manifolds and prove an
important negativity result for the curvature Hessian
$\cH_{R}$~\eqref{eq:HR}.  In~\S\ref{sec:method} we describe our method
for generating an ensemble of moduli space metrics and curvature
Hessians.  In~\S\ref{sec:results}, we present our primary results,
demonstrating the existence of hierarchies and heavy tails
in the spectra of the metric and $\cH_{R}$. We also elucidate the
source of the heavy tails in $\cH_{R}$ and discuss potential
implications for metastability. In~\S\ref{sec:alt} we discuss alternatives to the
i.i.d.~model for the curvature Hessian, including a qualitatively
successful approach in which the metric on moduli space is modeled as
a Bergman metric.  In~\S\ref{sec:axions}, we consider the consequences
of the eigenvalue distribution of the K\"ahler metric for the spectra
of axion decay constants.  Our conclusions appear in~\S\ref{sec:conc}.
Appendix~\ref{app:toric} contains an abbreviated review of toric
varieties and their Calabi-Yau hypersurfaces, which are the focus of
our study.  Finally, in Appendix~\ref{app:algorithm} we present an efficient algorithm to triangulate the
polytopes encoding the toric variety data.

\section{\label{sec:RMT}Random K\"ahler Metrics and Random Supergravities}

In this section, we
review the Hessian matrix in four-dimensional supergravity
(\S\ref{sec:sugraH}) and recall how it can be modeled using the tools
of random matrix theory,
following~\cite{Denef:2004cf,Marsh:2011aa}~(\S\ref{sec:RMT_sugra}).

\subsection{The supergravity Hessian}  \label{sec:sugraH}

A four-dimensional $\cN=1$ supergravity theory with $N$ chiral
supermultiplets containing complex scalar fields $\phi^{a}$ is
specified by a real K\"ahler potential
$\cK\bigl(\phi^{a},\bar{\phi}^{\bar{b}}\bigr)$ and a holomorphic
superpotential $W(\phi^{a})$. The bosonic action is
\begin{equation}
  \label{eq:4d_action}
  S=\int\ud^{4}x\,\sqrt{-g}\biggl\{\frac{1}{2}R
  -\cK_{a\bar{b}}\,
  \partial_{\mu}\phi^{a}\partial^{\mu}\bar{\phi}^{\bar{b}}
  - V\biggr\}\,,
\end{equation}
in which we have set the reduced Planck mass to unity,
$R$ is the Ricci scalar associated with the four-dimensional spacetime metric
$g_{\mu\nu}$, and the F-term potential is
\begin{equation}
  \label{eq:F_term_pot}
  V=e^{\cK}\biggl\{F_{a}\bar{F}^{a}-3\bigl\lvert W\bigr\rvert^{2}
  \biggr\}\,,
\end{equation}
where $F_{a}$ is the K\"ahler-covariant derivative of the superpotential,
\begin{equation}
  F_{a}=\partial_{a}W+\cK_{a}W\,.
\end{equation}
Indices are raised and lowered with the K\"ahler metric
$\cK_{a\bar{b}}$.  The action~\eqref{eq:4d_action} is invariant under
the K\"ahler transformations $\cK\to \cK+f+\bar{f}$, $W\to e^{-f}W$,
where $f$ is a holomorphic function of the scalar fields. The Hessian
matrix is constructed from second derivatives of the potential
\begin{subequations}
  \label{eq:Hessian}
\begin{equation}
  \cH=\begin{pmatrix}
    V_{a\bar{b}} & V_{ab} \\ V_{\bar{a}\bar{b}} & V_{\bar{a}b}
  \end{pmatrix}\,,
\end{equation} and takes the form
\begin{align}
  V_{a\bar{b}}&=
  e^{\cK}\bigl\{
  Z_{a}^{\phantom{a}\bar{c}}
  \bar{Z}_{\bar{c}\bar{b}}
  -F_{a}\bar{F}_{\bar{b}}
  -R_{a\bar{b}c\bar{d}}\bar{F}^{c}F^{\bar{d}}+
  \cK_{a\overline{b}}\big( F_{c}\bar{F}^{c}
  -2\bigl\lvert W\bigr\rvert^{2} \big)\bigr\}\,,
  \label{eq:mixed_Hessian}\\
  V_{ab}&=
  e^{\cK}\bigl\{U_{abc}\bar{F}^{c}-Z_{ab}\bar{W}\bigr\}\,,
\end{align}
\end{subequations}
in which
\begin{equation}
  \label{eq:iid_Z_U}
  Z_{ab}=\cD_{a}F_{b}\,,\quad U_{abc}=\cD_{a}\cD_{b}F_{c}\,,
\end{equation}
where $\cD_{a}$ is the K\"ahler-covariant and geometrically-covariant
derivative.  Since the field space is K\"ahler, the Riemann tensor can
be expressed in terms of ordinary derivatives of the K\"ahler
potential as
\begin{equation}
  \label{eq:Riemann}
  R_{a\bar{b}c\bar{d}}=\cK_{a\bar{b}c\bar{d}}-
  \cK_{ac}^{\phantom{ac}e}\cK_{\bar{b}\bar{d}e}\,.
\end{equation}

\subsection{\label{sec:RMT_sugra}Random supergravities}

Creating a globally well-defined potential energy function depending
on $N$ fields is exponentially expensive at large $N$.  However, to
understand the behavior at a collection of critical points, including
the particularly interesting case of metastable minima, only certain
local data about the potential are necessary: it suffices to specify
the first four derivatives of the K\"ahler potential $\cK$, and the
first three derivatives of the superpotential $W$, at each point of
interest.  By performing Taylor expansions of the ${\cal N}=1$ data at
a set of well-separated points, one can construct an ensemble of
Hessian matrices, and characterize the probability of metastability
(or of any other outcome of interest) in this ensemble.  In this
approach, {\it{no
    information}} about correlations from point to point is retained.
Understanding correlations in field space is left as a problem for
future work.\footnote{See~\cite{Bachlechner:2014rqa} for recent
  progress on incorporating global constraints.}

The essential assumption made in random supergravity approaches to the
Hessian~\eqref{eq:Hessian} is that the K\"ahler potential and the
superpotential are
real and holomorphic i.i.d.~functions,
respectively, on the moduli space~\cite{Denef:2004cf,Marsh:2011aa}.
That is, expressing the K\"ahler potential via a Taylor series  around a fixed origin,
\begin{equation}
  \cK=\cK_{0}+\cK_{a}\phi^{a}+\cK_{\bar{a}}\bar{\phi}^{\bar{a}}+
  \cK_{a\bar{b}}\phi^{a}\bar{\phi}^{\bar{b}}+\cdots\,,
\end{equation}
the elements of each tensor $\cK_{ab\cdots\bar{c}\bar{d}\cdots}$ are
taken to be independent and identically distributed variables: each element is drawn from some statistical distribution
$\Omega$, and although the same distribution is used for every
element, each element is drawn independently from the others (up to
the symmetries of the tensor).  We use the notation
$x\in\Omega\left(\mu,\sig\right)$ to denote that the real number $x$
is drawn from a distribution with mean $\mu$ and standard deviation
$\sig$.  When $z$ is complex, $z\in\Omega\left(\mu,\sig\right)$ means
that $z=x\,e^{i\theta}$, where $x$ is a real number drawn from
$\Omega$ and the phase $\theta$ is drawn from a uniform
distribution.  We will often work with normal distributions, though
universality in random matrix theory ensures that  most results will be
independent of the details of the distribution $\Omega$, for
sufficiently large $N$.

At a particular critical point of the potential, we can perform a
K\"ahler transformation and coordinate redefinition to set
\begin{equation}
  \label{eq:local_kahler_frame}
  \cK=0\,,\quad \cK_{a\bar{b}}=\delta_{a\bar{b}}\,,\quad
  F_{a}=\delta_{a}^{~1}\,F\, e^{i\vt_{F}}\,,
\end{equation}
in which $F$ is positive and $\vt_{F}$ is an unimportant phase.  The
random supergravity approach is then to model the higher derivatives
as i.i.d.~variables,
\begin{equation}
  \cK_{a\bar{b}c}\in \Omega\bigl(0,\sig\bigr)\,,\quad
  \cK_{a\bar{b}c\bar{d}}\in \Omega\bigl(0,\sig\bigr)\,,\quad
  \sig=N^{-1/2}\,.     \label{Kiid}
\end{equation}
The choice $\sig=N^{-1/2}$ is purely for later convenience.

The covariant derivatives of the superpotential are likewise taken to
be i.i.d.~variables\footnote{The requirement of being at a critical point necessarily spoils the i.i.d.~assumption, as
the potential is subject to the $2N$ conditions $\partial_{a}V=0$ at
that point.  However, since these are $\cO\left(N\right)$ conditions
on a  matrix with $\cO\left(N^{2}\right)$ entries, at large $N$ imposing the
critical point equation does not significantly affect positivity
properties~\cite{Marsh:2011aa}.} with
\begin{equation}
  Z_{ab}\in m_{\mathrm{susy}}\,\Omega\bigl(0,\sig\bigr)\,,\quad  \label{Wiid}
  U_{abc}\in m_{\mathrm{susy}}\,\Omega\bigl(0,\sig\bigr)\,.\quad
\end{equation}
in which $m_{\mathrm{susy}}$ is a characteristic scale for
supersymmetric mass terms.  At generic critical points, we expect that
$m_{\mathrm{susy}} \sim F$ (in Planck units).

\begin{figure}
\begin{center}
  \begin{subfigure}{0.45\textwidth}
    \psfrag{a}[Bc]{$\la{-2}$}
    \psfrag{b}[Bc]{$\la{-1}$}
    \psfrag{c}[Bc]{$\la{1}$}
    \psfrag{d}[Bc]{$\la{2}$}
    {\includegraphics{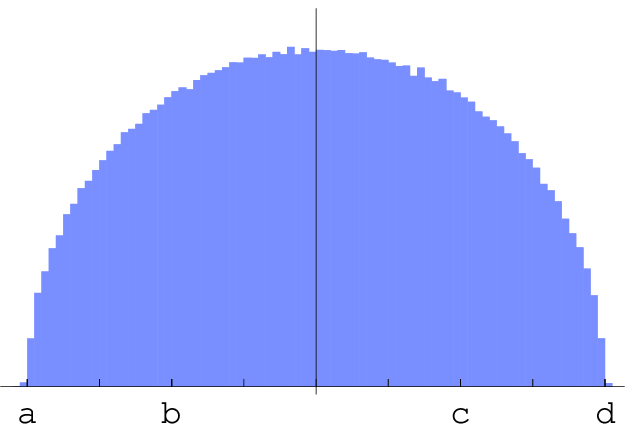}}
    \caption{Wigner ensemble~\eqref{eq:Wigner}}
  \end{subfigure}
  \quad
  \begin{subfigure}{0.45\textwidth}
    \psfrag{a}{$\la{1}$}
    \psfrag{b}{$\la{2}$}
    \psfrag{c}{$\la{3}$}
    \psfrag{d}{$\la{4}$}
    {\includegraphics{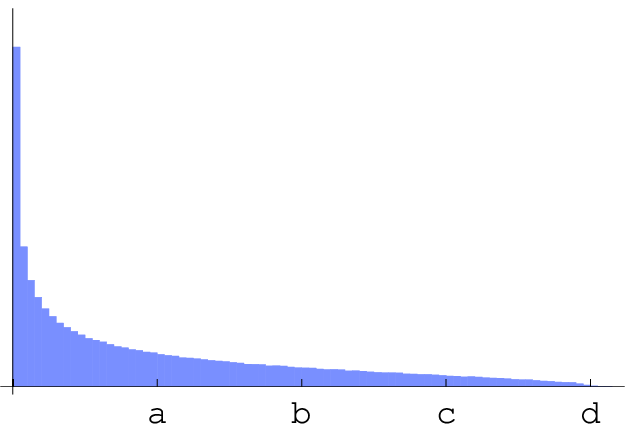}}
    \caption{Wishart ensemble~\eqref{eq:Wishart}
      with $q=n$}
  \end{subfigure}
  \caption{\label{fig:classical_spectra}Eigenvalue distributions for
    two classical matrix ensembles.  The data displayed result from collections of 1000
    matrices with $200\times 200$ i.i.d.~entries with $\sig=1/\sqrt{200}$.}
\end{center}
\end{figure}

Two classical ensembles of matrices are particularly relevant for our
analysis and for the models of~\cite{Denef:2004cf,Marsh:2011aa}.  The
first is the complex Wigner ensemble~\cite{Wigner,*Wigner2,*Wigner3,*Wigner4},
which consists of $n\times n$ Hermitian matrices of the form
\begin{equation}
  \label{eq:Wigner}
  M=A+A^{\dagger}\,,
\end{equation}
in which each element of $A$ is an i.i.d.~complex variable.  The other
relevant ensemble is the complex Wishart ensemble~\cite{Wishart}, consisting of $n\times n$ Hermitian matrices of the form
\begin{equation}
  \label{eq:Wishart}
  M=AA^{\dagger}\,,
\end{equation}
in which $A$ is an $n\times q$ matrix ($q\ge n$) with i.i.d.~complex
entries.  The eigenvalue spectra for these ensembles take on the
characteristic shapes shown in Figure~\ref{fig:classical_spectra}.

\begin{figure}
\begin{center}
  \begin{subfigure}{0.45\textwidth}
  \psfrag{a}{$\la{2}$}
  \psfrag{b}{$\la{4}$}
  \psfrag{c}{$\la{6}$}
  {\includegraphics{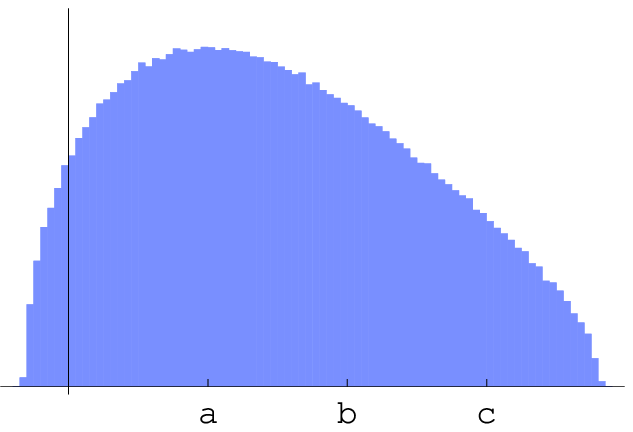}}
  \caption{\label{fig:wasteland}$\cH^{\WWW}$}
  \end{subfigure}
  \quad
  \begin{subfigure}{0.45\textwidth}
  \psfrag{a}{$\la{-1}$}
  \psfrag{b}{$\la{1}$}
  \psfrag{c}{$\la{2}$}
  \psfrag{d}{$\la{3}$}
  \psfrag{e}{$\la{4}$}
  {\includegraphics{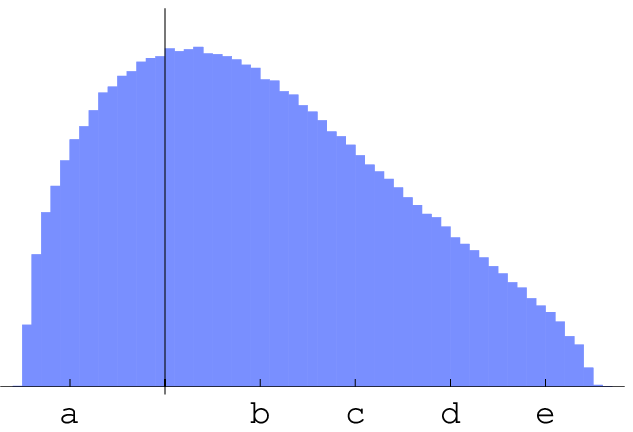}}
  \caption{\label{fig:WW_spectra}$\cH_{R}^{\WW}$}
  \end{subfigure}
  \caption{\label{fig:RMT_models} Eigenvalue distributions
    for random matrix models of the i.i.d.~Hessian $\cH^{\iid}\approx
    \cH^{\WWW}$~\eqref{eq:RMT_Hess} (taken with $3\left\lvert
    W\right\rvert^{2}/F^{2}=.01$) and of the Riemann contribution
    $\cH_{R}^{\iid}\approx
    \cH^{\WW}_{R}$~\eqref{eq:WW_curvature_Hessian_NH}.  In both cases,
    the spectra are in units of $\left\lvert
    F\right\rvert^{2}$. Figure~\ref{fig:wasteland} is adapted from
    Figure 3 of~\cite{Marsh:2011aa}.}
\end{center}
\end{figure}

The structure of the Hessian~\eqref{eq:Hessian} and the special
structure of the Riemann tensor~\eqref{eq:Riemann} suggest that, under
the i.i.d.~assumptions (\ref{Kiid}) and (\ref{Wiid}), the various
contributions to the Hessian can be approximated by these classical
distributions and a constant shift matrix~\cite{Marsh:2011aa}
\begin{align}
  \cH=&
  \underbrace{
    \begin{pmatrix}
      0 & U_{ab1}\bar{F}^{1}-Z_{ab}\bar{W} \\
      \bar{U}_{\bar{a}\bar{b}\bar{1}}F^{\bar{1}}-\bar{Z}_{\bar{a}\bar{b}}\bar{W} & 0
    \end{pmatrix}
    +F^{2}
    \begin{pmatrix}
      -\cK_{a\bar{b}1\bar{1}} & 0\\
      0 & \cK_{b\bar{a}1\bar{1}}\end{pmatrix}}_{\approx\mathrm{Wigner}} \notag\\
  &+\underbrace{
    \begin{pmatrix}
      Z_{a}^{\phantom{a}\bar{c}}\bar{Z}_{\bar{b}\bar{c}} & 0 \\
      0 & \bar{Z}_{\bar{a}}^{\phantom{a}c}Z_{bc}
    \end{pmatrix}}_{\approx\mathrm{Wishart}}
  +\underbrace{
    F^{2}\begin{pmatrix}
      \cK_{a1}^{\phantom{a1}e}\cK_{\bar{b}\bar{1}e} & 0\\
      0 & \cK_{\bar{a}\bar{1}}^{\phantom{a1}\bar{e}}\cK_{b1\bar{e}}
    \end{pmatrix}}_{\approx\mathrm{Wishart}} \label{eq:RMT_Hess}\\
  &+\underbrace{\bigl(F^{2}-2\bigl\lvert W\bigr\rvert^{2}\bigr)\mathbbm{1}
  -F^{2}\delta_{a}^{\phantom{a}1}\delta_{\bar{b}}^{\phantom{b}\bar{1}}
  -F^{2}\delta_{\bar{a}}^{\phantom{a}\bar{1}}
  \delta_{b}^{\phantom{b}1}}_{\approx\mathrm{shift}}\,,\notag
\end{align}
where we are locally working in a basis such
that~\eqref{eq:local_kahler_frame} holds at the critical point under
consideration.  In~\cite{Marsh:2011aa}, it was  established that the  typical spectrum of eigenvalues of an
ensemble of i.i.d.~supergravity
Hessians is well-modeled by the sum
\begin{equation}
  \cH^{\mathrm{i.i.d.}}\approx\cH^{\uW\uW\uW}=\mathrm{Wigner}
  +\mathrm{Wishart}+\mathrm{Wishart}\,.
\end{equation}
The spectrum of the  ensemble $\cH^{\WWW}$ is shown in
Figure~\ref{fig:wasteland}.  In~\cite{Marsh:2011aa}, it was
found that the probability for a generic critical point to be
metastable scales as $e^{-cN^{p}}$, where $p\approx 1.5$ for
$\cH^{\mathrm{i.i.d.}}$ and $p\approx 1.9$ for $\cH^{\WWW}$.

This random supergravity approach must be modified when one or more of
the i.i.d.~assumptions (\ref{Kiid}) and (\ref{Wiid})
is invalid.  Some form of modification is clearly necessary to give a
detailed model of string compactifications, as the resulting
supergravity data are expected to exhibit significant correlations.
Although a complete incorporation of all such correlations is
impossible given current knowledge, we can make considerable progress
by incorporating information about the classical K\"ahler potential.

In the following sections, we revisit the supergravity
Hessian~\eqref{eq:Hessian} using ensembles of K\"ahler potentials for
the classical metrics on K\"ahler moduli space in explicit Calabi-Yau
compactifications.  Since our focus is on how correlations in the
K\"ahler geometry impact the Hessian, we retain an i.i.d.~approach to
the F-terms and focus on the Riemann contribution to the Hessian,
$\cH_{R}$, which can be analyzed without further modeling of the
superpotential and its higher derivatives $Z_{ab}$ and $U_{abc}$.

\section{\label{sec:modulisp}The K\"ahler Moduli Space of O3/O7
  Orientifolds}

In this section, we review the K\"ahler moduli space of
compactifications of type IIB string theory on O3/O7 orientifolds of
Calabi-Yau manifolds.  These compactifications are well-explored, and allow for a
separation of scales between the masses of the stabilized complex
structure moduli and K\"ahler moduli as
in~\cite{Kachru:2003aw,Balasubramanian:2005zx}.  We will assume that
such a separation has been achieved and so limit our treatment to the
K\"ahler moduli sector.  In the following sections, we will make use
of the corresponding classical K\"ahler potential to improve upon the
i.i.d.~treatments discussed in \S\ref{sec:RMT}.

After reviewing some of the basic properties of the moduli space, we
provide expressions for the metric and Riemann curvature that will be
useful for our scan of moduli spaces, and then prove a simple but
important negativity property of the curvature contribution to the
Hessian~\eqref{eq:Hessian}.

\subsection{The K\"ahler cone}

A compactification of type~IIB  string theory on a Calabi-Yau threefold yields an
$\cN=2$ theory in four dimensions.\footnote{See, for
  example,~\cite{Grimm:2004uq,Blumenhagen:2006ci,Conlon:2006gv} for
  reviews of the effective field theories arising in these constructions and in their
  orientifolds.} The deformations of the K\"ahler structure, together
with scalars provided by the $p$-form potentials $C_{4}$, $C_{2}$, and
$B_{2}$, form a quaternionic K\"ahler moduli space of real dimension
$4h^{1,1}$, where $h^{p,q}$ are the Hodge numbers of the Calabi-Yau.
Reduction to an $\cN=1$ theory is accomplished by the introduction of
orientifold planes, and the structure of the resulting supergravity
theory depends on which orientifold projection is used.  We will
assume an O3/O7 orientifold action, and moreover will take all of the
divisors $D^{a}$ of the Calabi-Yau  to be even under the geometric involution
(i.e.~$h^{1,1}_-=0$), so that all K\"ahler deformations of the
Calabi-Yau are projected in by the orientifold action.  In the
resulting $\mathcal{N}=1$ theory, the 4-cycle volumes $\tau^a$ are
paired with R-R axions $c^{a}=\int_{D^a}C_{4}$ to form the  complex scalar
components $\rho^{a}=\tau^{a}+i\, c^{a}$ of $h^{1,1}$ chiral superfields.  The
classical K\"ahler potential for the K\"ahler moduli space metric, at leading
order in $\alpha^{\prime}$, is given in terms of the volume $\cV$ of
the Calabi-Yau by
\begin{equation}
  \label{eq:kahler_pot}
  \cK=-2\log\cV\,.
\end{equation}
If $\left\{D^{a}\right\}$ is a basis of divisors of the
Calabi-Yau, then the K\"ahler form of the Calabi-Yau can be expanded
in terms of the Poincar\'e duals $[D^{a}]$ of these divisors,
\begin{equation}
  \label{eq:expand_Kahler}
  J=t_{a}\left[D^{a}\right]\,.
\end{equation}
The volume of the Calabi-Yau is
\begin{equation}
  \label{eq:vol}
  \cV=\frac{1}{3!}\int J^{3}=\frac{1}{3!}\ka^{abc}t_{a}t_{b}t_{c}\,,
\end{equation}
in which $\ka^{abc}=D^{a}\cdot D^{b}\cdot D^{c}$ are the triple
intersection numbers of the divisors.  The volume of the divisor
$D^{a}$ is then
\begin{equation}
  \label{eq:tau_def}
  \tau^{a}=\frac{\partial\cV}{\partial t_{a}}=\frac{1}{2}
  \ka^{abc}t_{b}t_{c}\,.
\end{equation}

The Calabi-Yau volume $\cV$ and the K\"ahler potential on moduli space
are easily expressed in terms of the variables $t_{a}$, which
parameterize the volumes of 2-cycles. However, the 2-cycle volumes
do not provide good K\"{a}hler coordinates in O3/O7 projections, in
that we cannot write the metric as
$\partial_{w^{\phantom{\bar{b}}\!\!\! a}}\bar{\partial}_{\bar{w}^{\bar{b}}}\cK$
where $w^{a}$ is a complexification of $t_{a}$.
Because the $\tau^{a}$, rather than the $t_{a}$, provide good
coordinates on moduli space, we find it convenient to use lowered
indices for 2-cycle volumes, even though this goes against
convention.

The K\"ahler moduli space is specified by the requirement that the
volumes of all holomorphic curves and divisors, and  of the
Calabi-Yau itself, are positive.  Since the volumes of such
objects are of the form $\int J^{k}$, where $J$ is the K\"ahler
form~\eqref{eq:expand_Kahler}, the moduli space is a cone: if
$\vec{t}=\left(t_{1},\cdots, t_{h^{1,1}}\right)$ is a point in the
moduli space, then so is $\lambda\vec{t}$
for $\lambda>0$.

\begin{figure}
\begin{center}
  \includegraphics[angle=180]{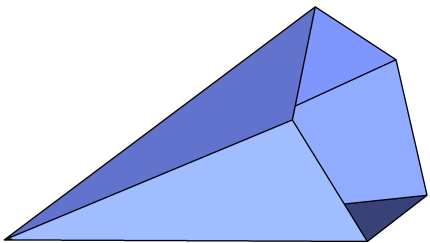}
  \caption{\label{fig:cone}A non-simplicial cone in three dimensions.
    The cone is generated by the five rays originating from the apex.}
\end{center}
\end{figure}

The space of curves also
forms a cone, called the Mori cone, that is generated by a finite set of
curves $C_{i}$.  The condition that all curves have positive volume is
equivalent to the condition
\begin{equation}
  \label{eq:positive_curve}
  \int_{C_{i}}J>0\,,\quad \forall\, C_{i}\,.
\end{equation}
The condition~\eqref{eq:positive_curve} suffices to ensure positivity
of the volumes of all divisors and of the Calabi-Yau, and so the
K\"ahler cone is the set of $t_{a}$ that
satisfy~\eqref{eq:positive_curve}.  Writing the intersections of the
divisors with the curves $ C_{i}$ as
\begin{equation}
  Q_{i}^{a}=D^{a}\cdot C_{i}\,,
\end{equation}
a sufficient condition to be in the interior of the K\"ahler cone is
\begin{equation}
  Q_{i}^{a}t_{a}>0\,, \quad \forall\, i\,.
\end{equation}
In general, there are more than $h^{1,1}$ generators of the Mori cone,
and so $Q$ is not a square matrix (though see the discussion in
Appendix~\ref{app:toric}). In this case the K\"ahler cone
is
said to be {\it{non-simplicial}}, meaning that the number of
generators of the cone is larger than the dimension of the cone (see
Figure~\ref{fig:cone}).  When $Q$ is square, the K\"ahler cone is
simplicial.

\subsection{The metric and curvature on moduli space}

The metric and curvature on moduli space follow from taking
derivatives of the K\"ahler potential with respect to the K\"ahler
coordinates $\rho^{a}=\tau^{a}+i\,c ^{a}$.  However, since the  classical
K\"{a}hler potential is independent of the axions $c^{a}$, we can make
the replacement
\begin{equation}
  \frac{\partial}{\partial \rho^a} \rightarrow
  \frac{1}{2}\frac{\partial}{\partial \tau^a}\,.
\end{equation}
Expressed as~\eqref{eq:vol}, the volume of the Calabi-Yau, and hence
the K\"ahler potential, is only an implicit function of the
coordinates $\tau^{a}$.
We therefore make
extensive use of the chain rule
\begin{equation}
  \label{eq:def_A}
  \frac{\partial}{\partial \tau^a} =
  A_{ba}
  \frac{\partial}{\partial t_b},\qquad
  A_{ba}:=\frac{\partial t_{a}}{\partial\tau^{b}}\,.
\end{equation}
One can show that $A$ has mostly negative eigenvalues, with signature
$\left(1,h^{1,1}-1\right)$. The inverse of $A$ takes the
simple form\footnote{Our notation differs from that
  appearing elsewhere in the literature, where $A$ and its inverse are
  interchanged.}
\begin{equation}
  \label{eq:A_inv}
  \left(A^{-1}\right)^{ab}=
  \frac{\partial \tau^a}{\partial t_b} =
  \frac{\partial^2 \cV}{\partial t_a \partial t_b} =
  \kappa^{abc}t_c\,.
\end{equation}
Note that due to the symmetry properties of $\ka^{abc}$, $A^{-1}$ and
$A$ are symmetric matrices.

Since the coordinates $\tau^{a}$ are defined via~\eqref{eq:tau_def},
$A_{ab}$ is difficult to calculate directly from its
definition~\eqref{eq:def_A}, but can be calculated by
inverting~\eqref{eq:A_inv}.  In what follows, we will perform the
inversion numerically, as the analytic computation is expensive when
$h^{1,1}$ is large.  Higher derivatives of the K\"ahler potential will
involve differentiating $A_{ab}$, and numerical differentiation can be
avoided by using~\eqref{eq:A_inv}:
\begin{equation}
  \frac{\partial A_{bc}}{\partial t_a}  = -A_{bd}
  \frac{\partial \left(A^{-1}\right)^{de}}{\partial t_a}A_{ec}
  =-A_{bd}\kappa^{ade}A_{ec}.
\end{equation}

Using the identities $A_{ab}\tau^{b}=\frac{1}{2}t_{a}$ and
$t_{a}\tau^{a}=3\cV$, the relevant derivatives of the
K\"ahler potential take the form
\begin{subequations}
\label{eq:Kahler_derivs}
\begin{align}
  \cK_{a\bar{b}}=&\frac{t_{a}t_{b}}{8\cV^{2}}-\frac{A_{ab}}{4\cV}\,,
  \label{eq:metric}\\
  \cK^{a\bar{b}}=&4\tau^{a}\tau^{b}-4\cV\bigl(A^{-1}\bigr)^{ab}\,,\\
  \cK_{ac}^{\phantom{ab}e}\cK_{\bar{b}\bar{d}e}
  =&\frac{1}{32\cV^{4}}t_{a}t_{c}t_{b}t_{d}
  -\frac{1}{64\cV^{3}}
  \bigl[
  6A_{\left(ac\right.}t_{b}t_{d\left.\right)}
  +A_{ac}t_{b}t_{d}+A_{bd}t_{a}t_{c}\bigr]\notag\\
  &+\frac{3}{32\cV^{2}}A_{ac}A_{bd}
  -\frac{1}{8\cV^{2}}t_{\left(a\right.}S_{\left.bcd\right)}
  -\frac{1}{16\cV}S_{ace}S_{bdf}\bigl(A^{-1}\bigr)^{ef}\,,
  \label{eq:K3_squared}\\
  \cK_{a\bar{b}c\bar{d}}=&\frac{3}{64\cV^{4}}t_{a}t_{b}t_{c}t_{d}
  -\frac{3}{16\cV^{3}}A_{\left(ab\right.}t_{c}t_{\left.d\right)}
  +\frac{3}{32\cV^{2}}A_{\left(ab\right.}A_{\left.cd\right)}\notag\\
  &-\frac{1}{8\cV^{2}}t_{\left(a\right.}S_{\left.bcd\right)}
  -\frac{3}{16\cV}S_{e\left(ab\right.}S_{\left.cd\right)f}
  \bigl(A^{-1}\bigr)^{ef}
  \label{eq:K4}\,,
\end{align}
\end{subequations}
in which we have defined the totally symmetric tensor
\begin{equation}
  S_{abc}=\ka^{def}A_{ad}A_{be}A_{cf}\,,
\end{equation}
and $M_{\left(a_{1}\cdots a_{n}\right)}$ denotes symmetrization of $M$.
Combining~\eqref{eq:K4} and~\eqref{eq:K3_squared} gives the full
Riemann tensor~\eqref{eq:Riemann}
\begin{align}
  R_{a\bar{b}c\bar{d}}=&
  \frac{1}{64\cV^{4}}t_{a}t_{b}t_{c}t_{d}
  -\frac{1}{64\cV^{3}}\bigl[6A_{\left(ac\right.}t_{b}t_{d\left.\right)}
  -A_{ac}t_{b}t_{d}-A_{bd}t_{a}t_{c}\bigr]\notag\\
  &+\frac{3}{32\cV^{2}}\bigl[A_{\left(ab\right.}A_{\left.cd\right)}
  -A_{ac}A_{bd}\bigr]
  -\frac{1}{8\cV}S_{ea\left(b\right.}S_{\left.d\right)cf}
  \bigl(A^{-1}\bigr)^{ef}
  \label{eq:Riemann2}\,.
\end{align}

\subsection{The Riemann contribution to the Hessian}
\label{sec:rhess}

The Hessian~\eqref{eq:Hessian} receives a contribution from the
Riemann tensor on moduli space contracted with
F-terms~\eqref{eq:mixed_Hessian},
\begin{equation}
\bigl(\cH_{R}\bigr)_{a\bar{b}} :=  - e^{\cK} R_{a\bar{b}c\bar{d}}\bar{F}^{c}F^{\bar{d}}\,,  \label{equ:hrdef}
\end{equation}
This contribution in turn breaks
into two pieces involving the third and fourth derivatives of the
K\"ahler metric~\eqref{eq:Riemann},
\begin{equation}
\cH_{R}  = \cH_{\cK^{\left(3\right)}} + \cH_{\cK^{\left(4\right)}}
\end{equation}
where
\begin{equation}
\label{eq:curvature_Hessian}
  \left(\cH_{\cK^{\left(4\right)}}\right)_{a\bar{b}} :=
  -e^{\cK}\cK_{a\overline{b}c\overline{d}}\bar{F}^{c}F^{\bar{d}}\,, \qquad
  \left(\cH_{\cK^{\left(3\right)}}\right)_{a\bar{b}} :=
  e^{\cK}\cK_{ac}^{\phantom{ac}e}\cK_{\bar{b}\bar{d}e}
  \bar{F}^{c}F^{\bar{d}}\,.
\end{equation}
Although the presentation~\eqref{eq:K3_squared} does not make the
positivity of $\cH_{\cK^{\left(3\right)}}$ manifest,
it does demonstrate that $\cH_{\cK^{\left(3\right)}}$ and
$\cH_{\cK^{\left(4\right)}}$ have very similar structures.
This
similarity is a
manifestation of the fact that the higher derivatives of the K\"ahler
potential cannot be taken to be entirely independent, a fact which
will have important implications for the spectrum of the Hessian.
Indeed, both $\cH_{\cK^{\left(4\right)}}$ and $\cH_{\cK^{\left(3\right)}}$
will be seen to exhibit dramatic tails in their eigenvalue
distributions that are  cancelled to some degree in the full curvature
contribution to the Hessian,
$\cH_{R}=\cH_{\cK^{\left(3\right)}}+\cH_{\cK^{\left(4\right)}}$.
Even so, the  residual tails in $\cH_{R}$ are substantial.

Despite the relative complexity of~\eqref{eq:Riemann2}, the resulting
contribution to the Hessian, $\cH_{R}$, exhibits a striking
property: when $F\neq 0$, $\cH_{R}$ always has at least one strictly
negative eigenvalue.  The proof of this is quite simple, and relies on
the fact that the K\"ahler metric is a homogeneous function of the
K\"ahler coordinates.  Indeed, under $\rho^{a}\to\lambda\rho^{a}$  we have
\begin{equation}
  \cK_{a\bar{b}}\to \lambda^{-2}\cK_{a\bar{b}}\,.
\end{equation}
In order to show that $\cH_{R}$ has a negative eigenvalue, it suffices
to show that
$\left(\cH_{R}\right)_{a\bar{b}}\xi^{a}\bar{\xi}^{\bar{b}}<0$ for some
$\xi^{a}$.  Taking $\xi^{a}=\tau^{a}$, we have
\begin{equation}
  \tau^{a}\tau^{\bar{b}}\left(\cH_{R}\right)_{a\bar{b}}
  =\frac{1}{4}e^{\cK}\tau^{a}\tau^{b}
  \biggl\{\frac{\partial \cK_{c\bar{f}}}{\partial\tau^{a}}
  \frac{\partial \cK_{\bar{d}e}}{\partial\tau^{b}}\cK^{e\bar{f}}
  -\frac{\partial}{\partial\tau^{a}}\frac{\partial}{\partial\tau^{b}}
  \cK_{c\bar{d}}\biggr\}\bar{F}^{c}F^{\bar{d}}\,.
\end{equation}
If $f(x^{1},\cdots x^{n})$ is a homogeneous function of degree $n$,
then it satisfies the well known relation
\begin{equation}
  x^{i}\frac{\partial f}{\partial x^{i}}=nf\,.
\end{equation}
Because $\cK_{a\bar{b}}$ is a homogeneous function of degree $-2$, we have
\begin{equation}
  \tau^{a}\frac{\partial \cK_{c\bar{f}}}{\partial \tau^{a}}
  =-2\cK_{c\bar{f}}\,.
\end{equation}
On the other hand, $\partial \cK_{c\bar{d}}/\partial\tau^{b}$ is a
homogeneous function of degree $-3$, so
\begin{equation}
  \tau^{a}\frac{\partial}{\partial \tau^{a}}
  \biggl[\frac{\partial \cK_{c\bar{d}}}{\partial\tau^{b}}\biggr]
  =-3\frac{\partial \cK_{c\bar{d}}}{\partial\tau^{b}}\,.
\end{equation}
Together,
\begin{equation}
  \tau^{a}\tau^{\bar{b}}\left(\cH_{R}\right)_{a\bar{b}}
  =\frac{1}{4}e^{\cK}
  \biggl\{4\,\cK_{c\bar{f}}
   \cK_{\bar{d}e}\cK^{e\bar{f}}
  -6\,\cK_{c\bar{d}}\biggr\}\bar{F}^{c}F^{\bar{d}}
  =-\frac{1}{2}e^{\cK}\left\lvert F\right\rvert^{2}\,.
\end{equation}
This is non-positive, and so the smallest eigenvalue
$\lambda_{{\rm{min}}}(\cH_{R})$ obeys
\begin{equation}
  \lambda_{{\rm{min}}}(\cH_{R}) \le 0\,,  \label{negativityresult}
\end{equation}
with equality only when $F=0$, in which case $\cH_{R}$ is identically
zero.  It is easy to show that this generalizes: if the metric is a
homogeneous function of negative degree, then $\cH_{R}$ will always
have at least one negative eigenvalue.  Moreover, this result holds
for K\"ahler moduli spaces in other string theories. However, quantum
corrections to the K\"ahler potential will necessarily spoil the
homogeneity of the metric (indeed, the homogeneity of the K\"ahler
metric is closely related to no-scale structure) and so
(\ref{negativityresult}) is guaranteed to hold only at weak coupling
and large volume.

The negativity result (\ref{negativityresult}) is quite different from
the expectation of the random supergravity approach discussed in
\S\ref{sec:RMT_sugra}.  In particular, the i.i.d.~model for the
curvature contribution to the Hessian is well-modeled by the sum of a
Wishart matrix and a Wigner matrix, as in
(\ref{eq:WW_curvature_Hessian_NH}) \cite{Marsh:2011aa}.  The resulting
distribution of eigenvalues is shown in
Figure~\ref{fig:WW_spectra}. Although the vast majority of the members
of an ensemble of such matrices will have at least one negative
eigenvalue, there is a small probability $P\sim \ue^{-c N^{p}}$ that a
randomly chosen matrix will be positive-definite.

We now turn to constructing an ensemble of K\"ahler metrics on
Calabi-Yau hypersurfaces, where we will find even more dramatic
deviations from the model (\ref{eq:WW_curvature_Hessian_NH}).

\section{\label{sec:method}Method}

Although  the  classical metric on the K\"ahler moduli space  of a Calabi-Yau compactification is given
explicitly in terms of 2-cycle volumes and triple intersection
numbers by (\ref{eq:metric}), for $h_{1,1}\gtrsim 5$ it is difficult to  extract meaningful characteristics of the metric analytically. Moreover, for the reasons explained in \S\ref{sec:intro},  it is most useful to determine general statistical properties  that hold  for most or all members of an ensemble of Calabi-Yau manifolds.

Our approach is to construct an ensemble of classical K\"ahler metrics
$\cK_{a\bar{b}}$ that describe the K\"ahler moduli spaces of
Calabi-Yau hypersurfaces in toric varieties.  Moreover, by combining
this explicit metric data with i.i.d.~F-terms, we construct an
ensemble of curvature contributions $\cH_R$ to the supergravity
Hessian matrix.  In this section, we provide a detailed prescription
for generating these ensembles.  In short, we triangulate a number of
reflexive polytopes to obtain the topological data of desingularized
Calabi-Yau hypersurfaces in toric varieties, making use of the
Kreuzer-Skarke database~\cite{KS_database} generated by
\texttt{PALP}~\cite{PALP,PALP2}, and then sample the associated
K\"ahler cones.

\subsection{\label{sec:approach}Generating metrics}

The functional form of the classical K\"ahler potential $\cK$ is fully
specified by the intersection numbers, which in the case of a
Calabi-Yau hypersurface in a toric variety are determined by a
triangulation of the associated polytope (see appendix~\ref{app:toric}
for a brief review of toric geometry).  The computational cost of the
triangulation process grows very quickly with the number of K\"ahler
moduli.  In appendix~\ref{app:algorithm}, we present an efficient
triangulation algorithm that allows us to study hypersurfaces with
$h^{1,1}\lesssim 25$ using modest computational resources.

The relevant outputs of the triangulation algorithm, for a given
Calabi-Yau, are the intersection numbers among divisors, $\ka^{abc}$,
and between the divisors and generators of the Mori cone, $Q_{i}^{a}$.
The K\"ahler cone is generated by a finite number $\nu$ of
$h^{1,1}$-dimensional vectors $\vec{n}_{\alpha}$, with $\alpha=1,\ldots,\nu$.
That is, every point
in the interior of the cone can be written in the form
\begin{equation}
  \label{eq:expand_t}
  \vec{t}=s_{\alpha}\vec{n}_{\alpha}\,
\end{equation}
for positive $s_{\alpha}$.
In general, there may be more
than $h^{1,1}$ generators $\vec{n}_{\alpha}$ (i.e., $\nu >h^{1,1}$)
in which case the cone is
non-simplicial (see Figure~\ref{fig:cone}).

Given the intersection numbers $\ka^{abc}$, the components of the
K\"ahler metric, expressed in a fixed coordinate system, are
determined by the 2-cycle volumes $t_{a}$, i.e.~by the vacuum
expectation values of the K\"ahler moduli.  Thus, to build an ensemble
of K\"ahler metrics arising at points in the moduli space of any one
Calabi-Yau threefold, we must specify a collection of points inside
the K\"ahler cone.  Next, to assemble the larger ensemble of metrics
arising at points in the moduli spaces of a collection of distinct
threefolds, we must specify how we choose a set of polytopes, a set of
associated triangulations, and a collection of points in the K\"ahler
cone determined by each triangulation.

Schematically, the probability of some  measurable outcome
${\cal O}$ (e.g., a given hierarchy in the eigenvalues of the metric) may be written  in the form
\begin{equation}
  P({\cal{O}}) = P({\rm{polytope}}) \times P({\rm{triangulation}})
  \times P(t_{i})
\end{equation}
where $P(t_i)$ denotes the measure chosen for sampling the K\"ahler
cone. The first two factors in (\ref{drake}) encode the relative
weighting given to distinct choices of topological data.  Lacking any
principle to determine this measure, we give equal weight to each
distinct set of intersection numbers obtained in our approach.

Next, one would aspire to choose a sampling measure $P(t_i)$  that  reflects  broad characteristics of
the actual distribution $P_{\rm{vacua}}(t_i)$ of string vacua inside the K\"ahler cone, which can be written on general grounds as
\begin{equation}
  P_{\rm{vacua}}(t_i) = P_{\rm{geometric}}(t_i) \times P_{\rm{selection}}(t_i) \,,  \label{drake}
\end{equation}
where the first factor denotes the geometric measure induced on the
K\"ahler cone by the K\"ahler metric on moduli space, and tends to be
peaked near the boundaries of the cone.\footnote{The volume of
  moduli space as measured by the K\"ahler metric is divergent, even
  on a constant volume slice, due to divergences of the metric near the boundaries of the K\"ahler cone.}  The second factor encodes
additional selection effects: for example, one might expect that  string vacua
are more common near the walls of the K\"ahler cone
because the superpotential for the K\"ahler moduli is exponentially
small when all volumes are large.  (Here we have in mind the general
lore that corrections are important where they are not
computable~\cite{Dine:1985he}.)  Characterizing $P_{\rm{selection}}$
requires a complete understanding of dynamics, which is
inaccessible at present.  As a result, we must be agnostic about whether the
sampling measure $P(t_{i})$ that we choose provides a reasonable approximation to
$P_{\rm{vacua}}(t_i)$.

The choice of $P(t_{i})$ can be encoded in the choice of coefficients
$s_{\alpha}$ appearing in~\eqref{eq:expand_t}.  Because the K\"ahler
cone has infinite volume, it is necessary to impose a long-distance
cutoff on the sampled region.  Since the metric and its derivatives
are homogeneous functions of the 2-cycle volumes, taking $s_{\alpha}$
to be within a finite interval can provide a representative sampling
of the cone.  A natural choice is to draw $s_{\alpha}$ from a uniform
distribution, and this is the choice that we make for much of our
analysis.  However, with this prescription the ratios of 2-cycle
volumes remain within a comparatively small range, and potentially
interesting behavior near the walls of the K\"ahler cone, where the
ratios of 2-cycle volumes are large, is not fully explored.  To
address this deficiency we draw the $s_{\alpha}$ from a logarithmic
distribution for a portion of our dataset.

\subsection{\label{sec:F-terms}Constructing curvature Hessians}

The method presented in \S\ref{sec:approach} suffices to generate an
ensemble of K\"ahler metrics, but in order to specify a scalar
potential on field space we must also give a prescription for
determining $W$ and its derivatives.  As discussed previously, we are
assuming a hierarchical separation of scales between the K\"ahler
moduli and other closed-string moduli, and therefore the effective
number of chiral multiplets is $N=h^{1,1}$.

The curvature contribution $\cH_R$ to the Hessian, (\ref{equ:hrdef}),
depends on the superpotential only through the $F^a$, while the
remainder of the Hessian receives contributions from $\partial_a
\partial_b W$ and $\partial_a \partial_b \partial_c W$ as well.  For
this reason, we focus in this work on the properties of $\cH_R$, which
are more directly geometric, and less affected by our limited
knowledge of the superpotential, than the properties of the full
Hessian $\cH$. Developing an accurate model of the superpotential and
characterizing the resulting Hessian matrix, perhaps along the lines
of~\cite{Rummel:2013yta}, is an important problem for the future.

To construct $\cH_{R}$, we take the F-terms
$F_{a}$ of the K\"ahler moduli $\rho^a$, expressed in the toric basis,
to be i.i.d.~variables (cf.~\eqref{eq:local_kahler_frame}),
\begin{equation}
  F_{a}\in F\,\Omega\left(0,1/\sqrt{h^{1,1}}\right)\,.
\end{equation}
The K\"ahler
metric and Riemann tensor scale homogeneously with the volume, so that
the Hessian  for the canonically-normalized fields scales as
\begin{equation}
  \cH_{R}^{\mathrm{c}}\sim \frac{F^{2}}{\cV^{4/3}}\,.
\end{equation}
We are interested in the shape of the  spectrum of
$\cH_{R}$, rather than the
overall scale,
so throughout we report on the spectrum of the rescaled
curvature Hessian for the canonically-normalized fields,
\begin{equation}
  \label{eq:defHRothree}
  \cH_{R}^{\Othree}=\frac{\cV^{4/3}}{F^{2}}\cH_{R}^{\mathrm{c}}\,.
\end{equation}

\section{\label{sec:results}Results}

Following all of the steps laid out in the previous sections, we can
now present the eigenvalue spectra of the K\"ahler metrics
$\cK_{a\bar{b}}$ and the curvature Hessians ${\cH}^{\CY}_{R}$ for a
collection of 120 Calabi-Yau hypersurfaces in toric varieties.
Although most of this section pertains to O3/O7 projections of
type~IIB, \S\ref{sec:special_geometry} contains a brief analysis of
the curvature Hessians in other corners of the string landscape.

\subsection{Metric eigenvalue spectra}

\begin{figure}
\begin{center}
    \psfrag{a}[Bc]{$\la{0.1}$}
    \psfrag{b}[Bc]{$\la{0.2}$}
    \psfrag{c}[Bc]{$\la{0.3}$}
    \psfrag{d}[Bc]{$\la{0.4}$}
  {\includegraphics{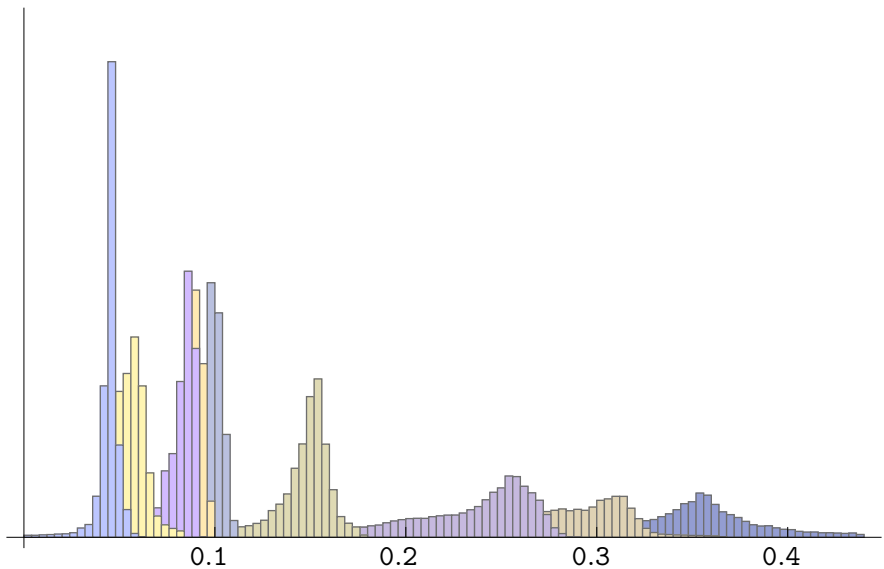}}
  \caption{\label{fig:h11-15-metric-mins}The nine smallest
    eigenvalues, coded by color, of each of $10^{5}$ metrics on the
    K\"ahler moduli space of a threefold with $h^{1,1}=15$.}
\end{center}
\end{figure}

\begin{figure}
\begin{center}
  \begin{subfigure}{0.45\textwidth}
    \psfrag{a}[Bc]{$\la{200}$}
    \psfrag{b}[Bc]{$\la{400}$}
    \psfrag{c}[Bc]{$\la{600}$}
    \psfrag{d}[Bc]{$\la{800}$}
    \includegraphics{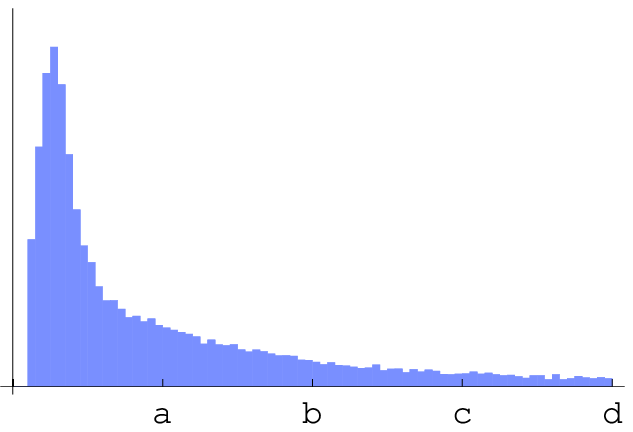}
    \caption{\label{fig:h11-15-metric-maxes}Largest eigenvalues}
  \end{subfigure}
  \quad
  \begin{subfigure}{0.45\textwidth}
    \psfrag{x}[Bc]{$\la{10^{-4}}$}
    \psfrag{a}[Bc]{$\la{.01}$}
    \psfrag{b}[Bc]{$\la{1}$}
    \psfrag{c}[Bc]{$\la{100}$}
    \psfrag{d}[Bc]{$\la{10^{4}}$}
    \psfrag{e}[Bc]{$\la{10^{6}}$}
    \psfrag{f}[Bc]{$\la{10^{8}}$}
    \includegraphics{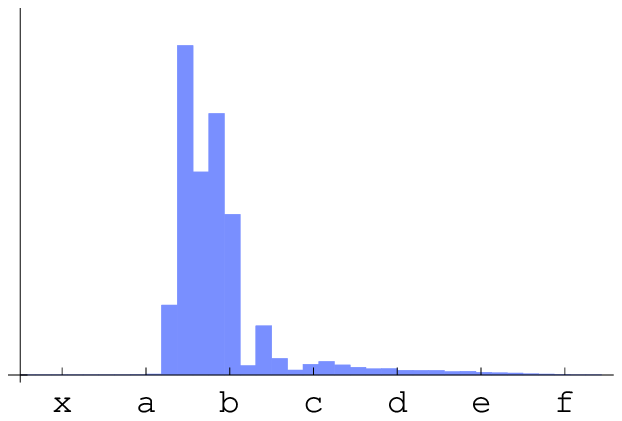}
    \caption{\label{fig:h11-15-metric-bulk}Complete spectrum}
  \end{subfigure}
  \caption{\label{fig:h11-15-metric-maxes-bulk}The largest eigenvalues
    and the entire spectrum of the metric on the K\"ahler moduli space
    of a threefold with $h^{1,1}=15$.  In (a), the tail of the
    distribution was truncated to demonstrate the bulk of the
    distribution.  The entire spectrum has support from $2.6\times
    10^{-5}$ to $1.2\times 10^{9}$.}
\end{center}
\end{figure}

To study the spectrum of metric eigenvalues we fix the basis of
K\"{a}hler moduli to a basis of independent toric divisors, as
discussed in Appendix~\ref{app:toric}.  We then compute the
eigenvalues of the moduli space metric in this basis, projected to the
$\cV = 1$ hypersurface of the K\"ahler cone.  The metric scales
homogeneously with the volume, so restricting to any other fixed value
of the volume will only change the overall scale of the metric
eigenvalues. In Figures~\ref{fig:h11-15-metric-mins}
and~\ref{fig:h11-15-metric-maxes-bulk}, we present the eigenvalue
spectra for an ensemble of moduli space metrics for a Calabi-Yau
threefold with $h^{1,1} = 15$. There are two notable features that
appear in all of the examples we examined: the spectra manifest large
hierarchies, with the largest and smallest eigenvalues being very
different from the median, and the distributions of the smallest few
eigenvalues are highly peaked.

\begin{figure}
\begin{center}
  \begin{subfigure}{0.45\textwidth}
    \psfrag{a}[Bc]{$\la{200}$}
    \psfrag{b}[Bc]{$\la{400}$}
    \psfrag{c}[Bc]{$\la{600}$}
    \psfrag{d}[Bc]{$\la{800}$}
    \psfrag{e}[Br]{$\la{50}$}
    \psfrag{f}[Br]{$\la{100}$}
    \psfrag{g}[Br]{$\la{150}$}
    \psfrag{h}[Br]{$\la{200}$}
    \psfrag{V}[Bc]{$\cV$}
    \psfrag{T}[Bl]{$\tau_L^{3/2}$}
    \includegraphics{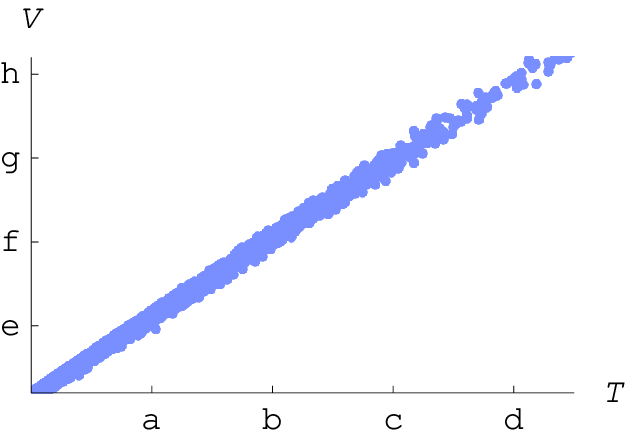}
    \caption{\label{fig:VvsT}Volume scaling}
  \end{subfigure}
  \quad
  \begin{subfigure}{0.45\textwidth}
    \psfrag{a}[Bc]{$\la{0.05}$}
    \psfrag{b}[Bc]{$\la{0.1}$}
    \psfrag{c}[Bc]{$\la{0.15}$}
    \psfrag{d}[Br]{$\la{0.01}$}
    \psfrag{e}[Br]{$\la{0.02}$}
    \psfrag{f}[Br]{$\la{0.03}$}
    \psfrag{L}[Bc]{$\lambda^g_{\text{min}}$}
    \psfrag{T}[Bl]{$\tau_L^{-2}$}
    \includegraphics{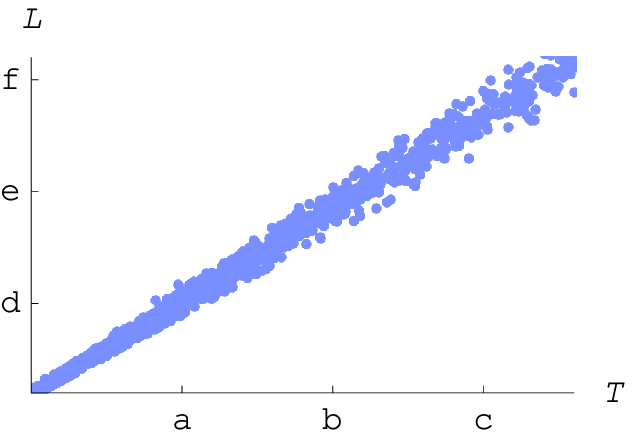}
    \caption{\label{fig:LvsT}Scaling of the smallest metric eigenvalue}
  \end{subfigure}
  \caption{\label{fig:xvsT}Scalings of the volume and of the smallest metric
    eigenvalue with the largest 4-cycle volume, $\tau_L$.  Both
    relations are approximately linear in this regime.}
\end{center}
\end{figure}

The sharply-localized distributions of the smallest few eigenvalues
can be attributed to the particular dependence of ${\cal V}$ on the
4-cycle volumes.  It is convenient to work in a basis that
diagonalizes the matrix $A_{a{b}} = -\partial^2 \cV/\partial
\tau^a \partial \tau^{{b}}$. We refer to the diagonalized A-matrix as
$\hat{A}$.  In Figure~\ref{fig:VvsT} we show the relationship between
the largest 4-cycle volume $\tau_L$ and the compactification volume
${\cal V}$, where $\tau_L$ is measured in a basis where $A_{ab} =
\hat{A}_{ab}$.  Because $\cV$ is approximately linear in
$\tau_L^{3/2}$, it appears that the single parameter $\tau_L$ controls
the overall volume of the Calabi-Yau in this region of moduli space.
The smallest eigenvalue of the metric, $\lambda^g$, is strongly
correlated with ${\cal V}$ --- see Figure~\ref{fig:LvsT} --- and so
the rescaling to ${\cal V}=1$ makes the distribution of $\lambda^g$
highly localized.  The smaller 4-cycles, on the other hand, do not
change the volume very much, and the corresponding metric eigenvalues
have a larger spread.  As $\tau_{L}$ increases, changes to the other
4-cycle volumes result in a smaller relative change to the overall
volume, and so these tendencies become more pronounced.

The hierarchies in the metric eigenvalues are an intriguing and important feature that we will discuss further in the following sections.

\subsection{The curvature Hessian in an example}

We next turn to examining the curvature contribution
$\cH^{\Othree}_{R}$~\eqref{eq:curvature_Hessian} to the Hessian
matrices in our ensemble.  We will begin with the example of a
particular Calabi-Yau hypersurface with $h^{1,1}=25$, which we call
$\ex$.

\begin{figure}
\begin{center}
  \psfrag{a}[Bc]{$\la{-10}$}
  \psfrag{b}[Bc]{$\la{-8}$}
  \psfrag{c}[Bc]{$\la{-6}$}
  \psfrag{d}[Bc]{$\la{-4}$}
  \psfrag{e}[Bc]{$\la{-2}$}
  \psfrag{f}[Bc]{$\la{2}$}
  \psfrag{g}[Bc]{$\la{4}$}
  \includegraphics{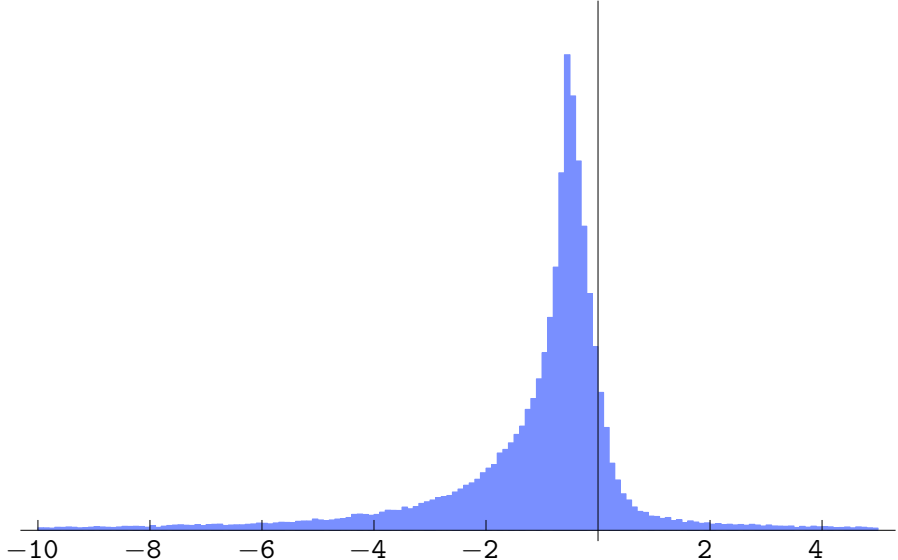}
  \caption{\label{fig:h11_25-HR}Distribution of eigenvalues of the
    curvature contribution to the Hessian $\cH^{\Othree}_{R}$ in an
    ensemble of 5000 curvature Hessians derived from $\ex$, a
    Calabi-Yau with $h^{1,1}=25$.  The tails of the distribution
    extend for a few decades in either direction (see
    Figure~\ref{fig:h11_25_HR_minmax}) and so have been truncated to
    make the bulk of the spectrum more visible.  This distribution is
    to be contrasted with Figure~\ref{fig:WW_spectra}, which is the
    distribution of eigenvalues for the Wigner+Wishart ensemble
    suggested by the null hypothesis.}
\end{center}
\end{figure}

\begin{figure}
\begin{center}
  \begin{subfigure}{0.45\textwidth}
    \psfrag{a}[Bc]{$\la{-2000}$}
    \psfrag{b}[Bc]{$\la{-1500}$}
    \psfrag{c}[Bc]{$\la{-1000}$}
    \psfrag{d}[Bc]{$\la{-500}$}
    {\includegraphics{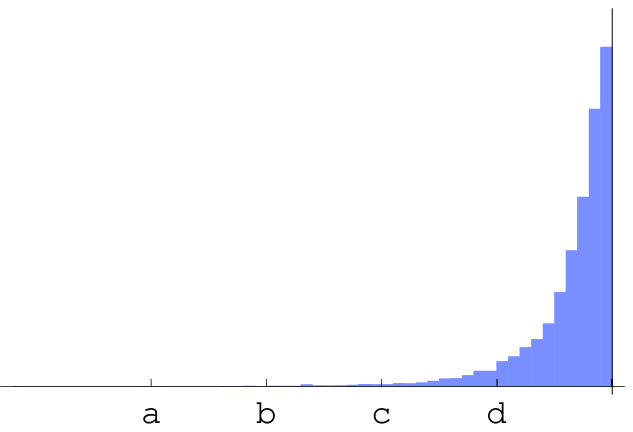}}
    \caption{Minimum eigenvalues }
  \end{subfigure}
  \quad
  \begin{subfigure}{0.45\textwidth}
    \psfrag{a}[Bc]{$\la{50}$}
    \psfrag{b}[Bc]{$\la{100}$}
    \psfrag{c}[Bc]{$\la{150}$}
    {\includegraphics{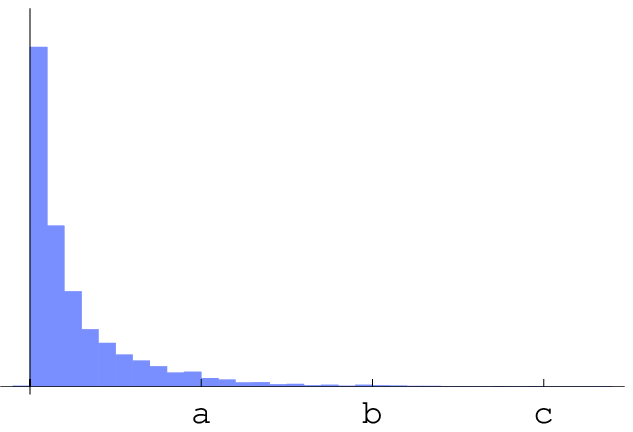}}
    \caption{Maximum eigenvalues}
  \end{subfigure}
  \caption{\label{fig:h11_25_HR_minmax}The spectra of minimum and
    maximum eigenvalues of $\cH^{\Othree}_{R}$ for $\ex$.  For our sample, the
    minimum eigenvalues lie in the range $\left[-2573,-3.2\right]$ and
    the maximum eigenvalues lie in the range $\left[-.03,167\right]$.}
\end{center}
\end{figure}

\begin{figure}
\begin{center}
\begin{subfigure}{0.45\textwidth}
  \psfrag{x}[Bc]{$\la{-1000}$}
  \psfrag{a}[Bc]{$\la{-10}$}
  \psfrag{b}[Bc]{$\la{-.1}$}
  \psfrag{d}[Bc]{$\la{.1}$}
  \psfrag{e}[Bc]{$\la{10}$}
  \psfrag{f}[Bc]{$\la{1000}$}
  \psfrag{g}[Bc]{$\la{4}$}
  \includegraphics{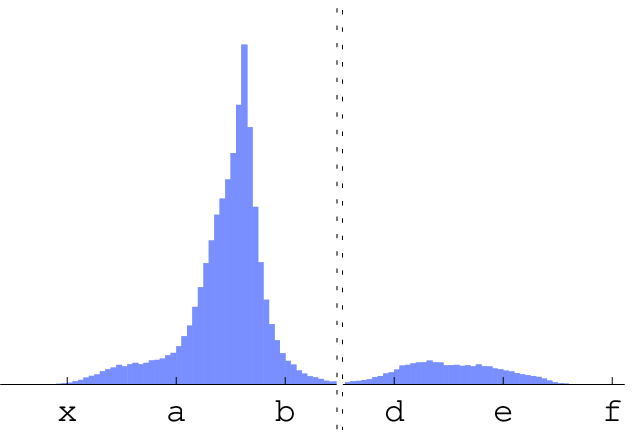}
  \caption{\label{fig:h11_25-HR_log_CY}$\cH^{\Othree}_{R}$}
  \end{subfigure}
\quad
\begin{subfigure}{0.45\textwidth}
  \psfrag{x}[Bc]{$\la{-1000}$}
  \psfrag{a}[Bc]{$\la{-10}$}
  \psfrag{b}[Bc]{$\la{-.1}$}
  \psfrag{d}[Bc]{$\la{.1}$}
  \psfrag{e}[Bc]{$\la{10}$}
  \psfrag{f}[Bc]{$\la{1000}$}
  \psfrag{g}[Bc]{$\la{4}$}
  \includegraphics{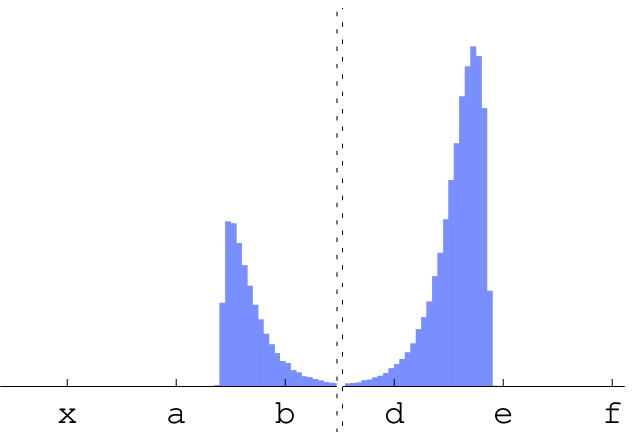}
  \caption{\label{fig:h11_25-HR_log_WW}$\cH^{\WW}_{R}$}
  \end{subfigure}
  \caption{\label{fig:h11_25-HR_log}Distribution of eigenvalues of the
    curvature contribution to the Hessian $\cH^{\Othree}_{R}$ and
    $\cH^{\WW}_{R}$, presented on a logarithmic scale.  Note the
    lengthy tail toward negativity in the spectrum of
    $\cH^{\Othree}_{R}$.}
\end{center}
\end{figure}

The eigenvalue spectrum of $\cH^{\Othree}_{R}$, which we denote by
$\rho_{R}$ for brevity, is shown for the example of $\ex$ in
Figures~\ref{fig:h11_25-HR}, \ref{fig:h11_25_HR_minmax}, and
Figure~\ref{fig:h11_25-HR_log}.  The spectrum exhibits dramatic tails,
as illustrated in Figure~\ref{fig:h11_25_HR_minmax}.  The bulk of the
eigenvalue spectrum, with the tails removed to emphasize the shape
near the peak of the spectrum, appears in Figure~\ref{fig:h11_25-HR},
while a logarithmic plot of the entire spectrum is shown in
Figure~\ref{fig:h11_25-HR_log}.

\begin{figure}
\begin{center}
  \begin{subfigure}{0.45\textwidth}
  \psfrag{a}[Bc]{$\la{-1000}$}
  \psfrag{b}[Bc]{$\la{-100}$}
  \psfrag{c}[Bc]{$\la{-10}$}
  \psfrag{d}[Bc]{$\la{-1}$}
  \psfrag{e}[Bc]{$\la{-.1}$}
    {\includegraphics{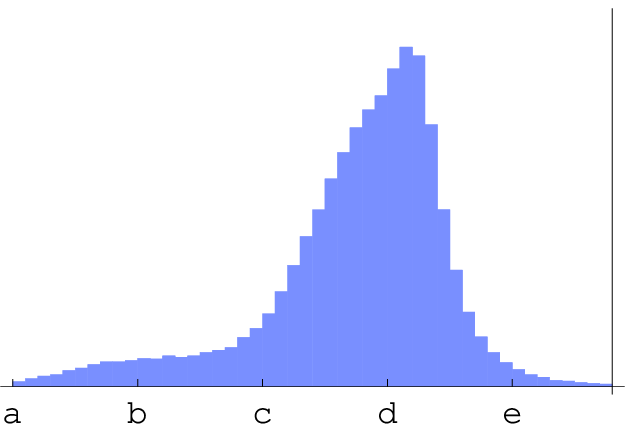}}
    \caption{$\cH^{\Othree}_{\cK^{\left(4\right)}}$}
  \end{subfigure}
  \quad
  \begin{subfigure}{0.45\textwidth}
  \psfrag{a}{$\la{.1}$}
  \psfrag{b}{$\la{1}$}
  \psfrag{c}{$\la{10}$}
  \psfrag{d}{$\la{100}$}
  \psfrag{e}{$\la{1000}$}
    {\includegraphics{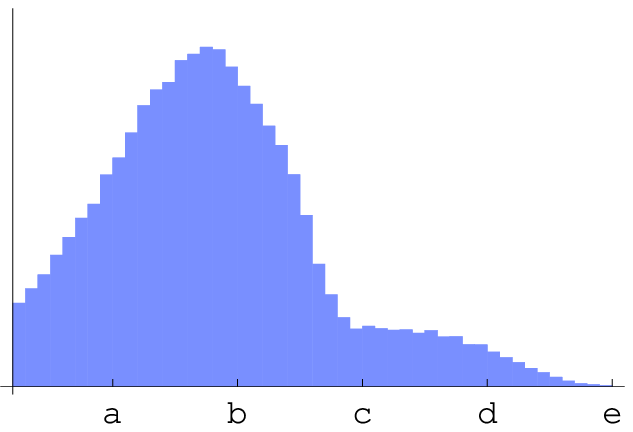}}
    \caption{$\cH^{\Othree}_{\cK^{\left(3\right)}}$}
  \end{subfigure}
  \caption{\label{fig:h11-25-k3k4}The spectra of the
    $\cK^{\left(3\right)}$ and $\cK^{\left(4\right)}$ contributions to
    $\cH_{R}^{\Othree}$ for $\ex$. The eigenvalues of
    $\cH^{\Othree}_{\cK^{\left(4\right)}}$ can be of either sign, but
    only the negative eigenvalues have been
    plotted. $\cH^{\Othree}_{\cK^{\left(3\right)}}$ is non-negative.}
\end{center}
\end{figure}

Although $\rho_{R}$ displays significant tails in both directions, the
tail towards negativity is substantially longer than the tail towards
positivity.  This may come as a surprise, given the
positive-semidefinite contribution to $\cH^{\Othree}_{R}$ coming from
$\cH^{\Othree}_{\cK^{\left(3\right)}}$~\eqref{eq:curvature_Hessian}.
Indeed, the eigenvalue distribution for $\cH^{\Othree}_{\cK^{\left(3\right)}}$
for $\ex$ is positive and exhibits large tails towards positivity (see
Figure~\ref{fig:h11-25-k3k4}).  However, because of the correlations
between $\cH^{\Othree}_{\cK^{\left(3\right)}}$ and
$\cH^{\Othree}_{\cK^{\left(4\right)}}$, this positive tail is almost
entirely removed in $\cH^{\Othree}_{R}$, and the longest tail that
remains is the tail towards negativity.  These correlations can be
seen explicitly in the similar structures in the derivatives of the
K\"ahler potential~\eqref{eq:K3_squared} and~\eqref{eq:K4}.  Indeed,
the form of the combined Riemann tensor~\eqref{eq:Riemann2}
demonstrates that significant cancellations have occurred.

\begin{figure}
\begin{center}
  \begin{subfigure}{0.20\textwidth}
    \psfrag{a}[Bc]{$\la{-5}$}
    \psfrag{b}[Bc]{$\la{5}$}
    {\includegraphics{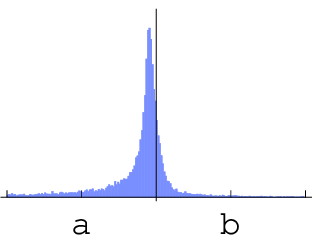}}
  \end{subfigure}
  \quad
  \begin{subfigure}{0.20\textwidth}
    \psfrag{a}[Bc]{$\la{-5}$}
    \psfrag{b}[Bc]{$\la{5}$}
    {\includegraphics{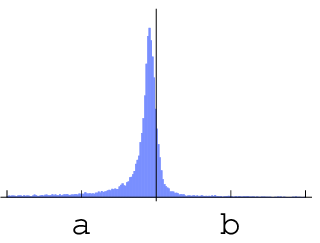}}
  \end{subfigure}
  \quad
  \begin{subfigure}{0.20\textwidth}
    \psfrag{a}[Bc]{$\la{-5}$}
    \psfrag{b}[Bc]{$\la{5}$}
    {\includegraphics{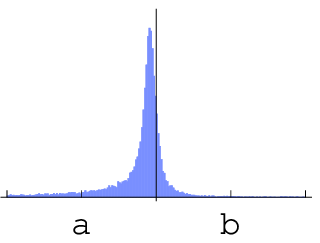}}
  \end{subfigure}
  \quad
  \begin{subfigure}{0.20\textwidth}
    \psfrag{a}[Bc]{$\la{-5}$}
    \psfrag{b}[Bc]{$\la{5}$}
    {\includegraphics{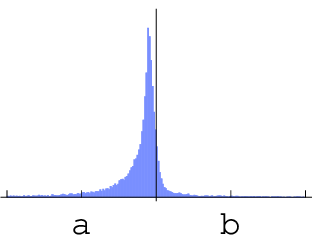}}
  \end{subfigure}
\\
Top row: $h^{1,1}=25$
\\
  \begin{subfigure}{0.20\textwidth}
    \psfrag{a}[Bc]{$\la{-5}$}
    \psfrag{b}[Bc]{$\la{5}$}
    {\includegraphics{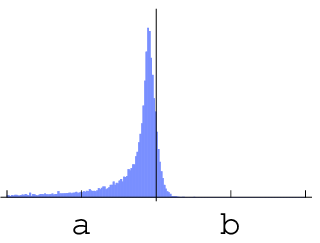}}
  \end{subfigure}
  \quad
  \begin{subfigure}{0.20\textwidth}
    \psfrag{a}[Bc]{$\la{-5}$}
    \psfrag{b}[Bc]{$\la{5}$}
    {\includegraphics{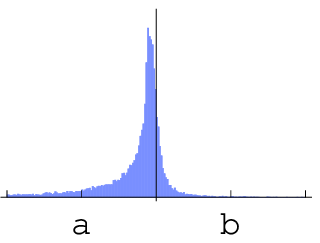}}
  \end{subfigure}
  \quad
  \begin{subfigure}{0.20\textwidth}
    \psfrag{a}[Bc]{$\la{-5}$}
    \psfrag{b}[Bc]{$\la{5}$}
    {\includegraphics{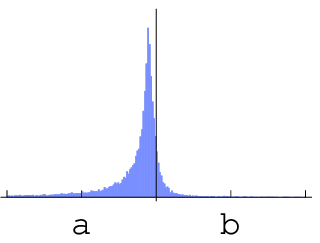}}
  \end{subfigure}
  \quad
  \begin{subfigure}{0.20\textwidth}
    \psfrag{a}[Bc]{$\la{-5}$}
    \psfrag{b}[Bc]{$\la{5}$}
    {\includegraphics{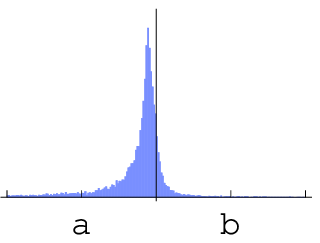}}
  \end{subfigure}
\\
Second row: $h^{1,1}=20$
\\
  \begin{subfigure}{0.20\textwidth}
    \psfrag{a}[Bc]{$\la{-5}$}
    \psfrag{b}[Bc]{$\la{5}$}
    {\includegraphics{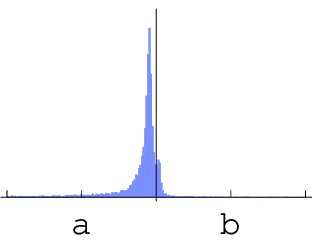}}
  \end{subfigure}
  \quad
  \begin{subfigure}{0.20\textwidth}
    \psfrag{a}[Bc]{$\la{-5}$}
    \psfrag{b}[Bc]{$\la{5}$}
    {\includegraphics{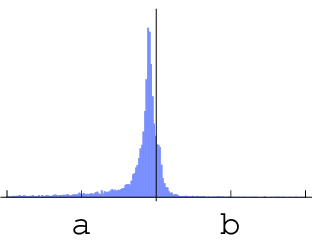}}
  \end{subfigure}
  \quad
  \begin{subfigure}{0.20\textwidth}
    \psfrag{a}[Bc]{$\la{-5}$}
    \psfrag{b}[Bc]{$\la{5}$}
    {\includegraphics{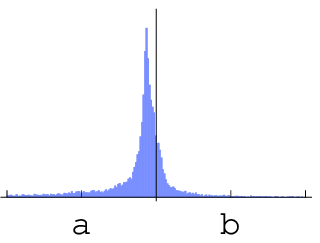}}
  \end{subfigure}
  \quad
  \begin{subfigure}{0.20\textwidth}
    \psfrag{a}[Bc]{$\la{-5}$}
    \psfrag{b}[Bc]{$\la{5}$}
    {\includegraphics{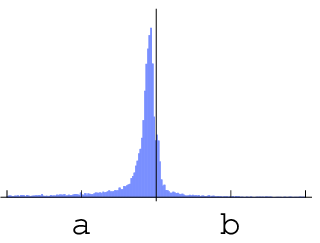}}
  \end{subfigure}
\\
Third row: $h^{1,1}=15$
\\
  \begin{subfigure}{0.20\textwidth}
    \psfrag{a}[Bc]{$\la{-5}$}
    \psfrag{b}[Bc]{$\la{5}$}
    {\includegraphics{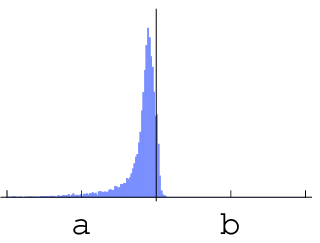}}
  \end{subfigure}
  \quad
  \begin{subfigure}{0.20\textwidth}
    \psfrag{a}[Bc]{$\la{-5}$}
    \psfrag{b}[Bc]{$\la{5}$}
    {\includegraphics{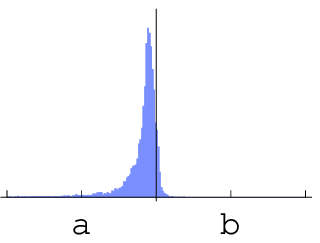}}
  \end{subfigure}
  \quad
  \begin{subfigure}{0.20\textwidth}
    \psfrag{a}[Bc]{$\la{-5}$}
    \psfrag{b}[Bc]{$\la{5}$}
    {\includegraphics{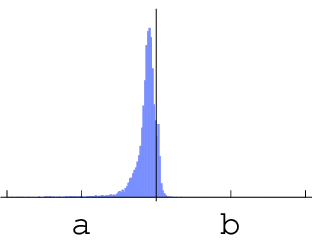}}
  \end{subfigure}
  \quad
  \begin{subfigure}{0.20\textwidth}
    \psfrag{a}[Bc]{$\la{-5}$}
    \psfrag{b}[Bc]{$\la{5}$}
    {\includegraphics{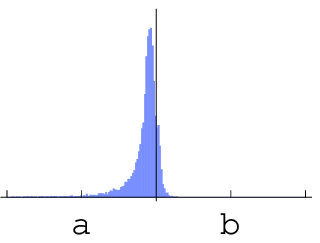}}
  \end{subfigure}
\\
Fourth row: $h^{1,1}=10$
\\
  \begin{subfigure}{0.20\textwidth}
    \psfrag{a}[Bc]{$\la{-5}$}
    \psfrag{b}[Bc]{$\la{5}$}
    {\includegraphics{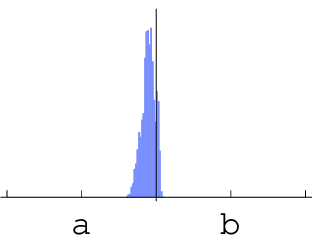}}
  \end{subfigure}
  \quad
  \begin{subfigure}{0.20\textwidth}
    \psfrag{a}[Bc]{$\la{-5}$}
    \psfrag{b}[Bc]{$\la{5}$}
    {\includegraphics{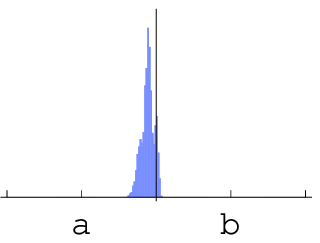}}
  \end{subfigure}
  \quad
  \begin{subfigure}{0.20\textwidth}
    \psfrag{a}[Bc]{$\la{-5}$}
    \psfrag{b}[Bc]{$\la{5}$}
    {\includegraphics{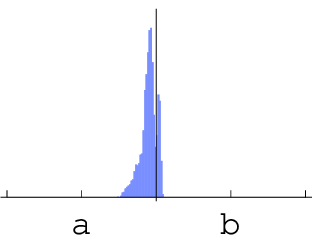}}
  \end{subfigure}
  \quad
  \begin{subfigure}{0.20\textwidth}
    \psfrag{a}[Bc]{$\la{-5}$}
    \psfrag{b}[Bc]{$\la{5}$}
    {\includegraphics{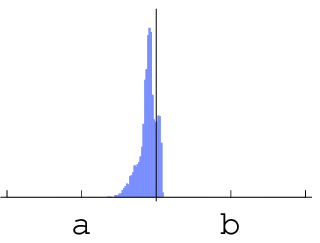}}
  \end{subfigure}
\\
Bottom row: $h^{1,1}=5$
\\
  \caption{\label{fig:universality}Distributions of eigenvalues
    $\cH^{\Othree}_{R}$ for a collection of Calabi-Yau manifolds. The
    tails of all of the distributions have been truncated to focus on
    the bulks of the distributions, which display common features consistent with universality.}
\end{center}
\end{figure}

Although we have focused on a particular
example, $\ex$, the bulk of the spectra of $\cH_{R}^{\Othree}$
calculated on other Calabi-Yau manifolds are qualitatively very
similar, especially for larger $h^{1,1}$ (see
Figure~\ref{fig:universality}).  However, the tails of the
distributions behave in quantitatively different ways: as we
demonstrate in \S\ref{sec:explaining}, the greater the curvature of
moduli space, the longer the corresponding tail of
$\cH_{R}^{\Othree}$.

\subsection{\label{sec:heavytails}Evidence for heavy-tailed spectra}

The tails in $\rho_{R}$ are clearly extensive --- see
Figure~\ref{fig:h11_25-HR}.  Indeed, the kurtosis of our sample of
$\cH_{R}^{\Othree}$ on $\ex$ is $230$. In this section we quantify the
extent of these tails in comparison to those in a number of reference
distributions.

\begin{figure}
\begin{center}
    \psfrag{a}[Bc]{$\la{50}$}
    \psfrag{b}[Bc]{$\la{100}$}
    \psfrag{c}[Bc]{$\la{150}$}
    \psfrag{d}[Bc]{$\la{200}$}
    \psfrag{e}[Br]{$\la{50}$}
    \psfrag{f}[Br]{$\la{100}$}
    \psfrag{g}[Br]{$\la{150}$}
    \psfrag{h}[Br]{$\la{200}$}
    \psfrag{N}[Bl]{$\cN_{\sig}$}
    \psfrag{Y}[Bc]{$\ex$}
    \includegraphics{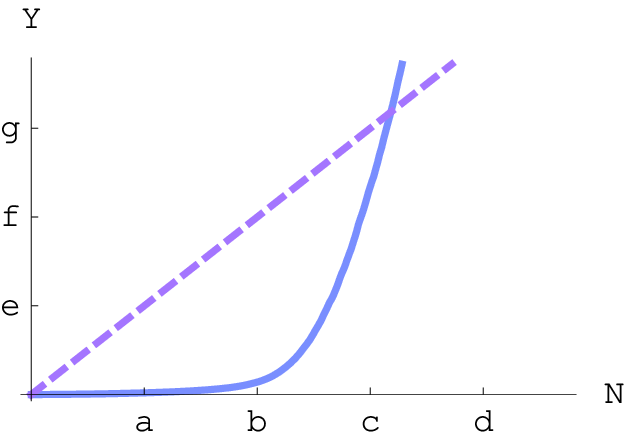}
    \caption{\label{fig:qq}Quantiles of the left half of the spectrum
      $\rho_R$ of eigenvalues of $\cH^{\Othree}_{R}$ computed in the
      example of $\ex$, against the same quantiles of a half-normal
      distribution $\cN_{\sig}$.  The variance of $H$ is chosen to give the
      best fit to the left tail of $\rho_R$.  The dashed line is the
      curve that would result if $\rho_{R}$ were half-normal.}
\end{center}
\end{figure}

Comparison of the quantiles of the left tail of $\rho_{R}$ to the
quantiles of a half-normal distribution (Figure~\ref{fig:qq})
establishes that the tails of $\rho_{R}$ contain more of the
population than do the tails of a half-normal distribution.  The
kurtosis of the left tail ($K\approx 149$) is much larger than that of
a half-normal distribution ($K\approx 3.87$), and the kurtosis of the
right tail is also significant ($K\approx 77.4$). In the
remainder of this section we will focus on the tail towards negativity
and so for graphical purposes, and only within this section, we will
reverse the sign of $\rho_{R}$ and shift the origin, so that the tail
toward negativity in $\rho_{R}$ is displayed extending to the right
and peaking at zero.

By definition, a distribution is {\it{heavy-tailed}} when the tail is
not exponentially bounded.  Specifically, the survival (or tail)
function of a distribution with probability distribution function
(pdf) $f$ is defined as
\begin{equation}
  S\bigl(x\bigr)=\int_{x}^{\infty}\ud t\,f\left(t\right)\,.
\end{equation}
The distribution is heavy-tailed if and only if for all $\ga>0$,
$\ue^{\ga x}S\left(x\right)$ diverges as $x\to\infty$.
A canonical example of a heavy-tailed distribution is the Pareto
distribution, for which the pdf $f_{\beta;k}\left(x\right)=\beta\,
k^{\beta}x^{-\beta-1}$ has support for $x>k$.  For a normal
distribution, $\ue^{\gamma x}S\left(x\right)\to 0$ for any $\ga>0$,
so the tails are not heavy.  Exponential distributions
$f_{\ka}=\ka\ue^{-\ka x}$ ($\ka>0$) are a marginal case in which the
limit vanishes, diverges, or converges to a finite value depending on
whether $\ga$ is less than, greater than, or equal to $\ka$.

\begin{figure}
\begin{center}
  \begin{subfigure}{0.45\textwidth}
    \psfrag{a}[Bc]{$\la{500}$}
    \psfrag{b}[Bc]{$\la{1000}$}
    \psfrag{c}[Bc]{$\la{1500}$}
    \psfrag{d}[Bc]{$\la{2000}$}
    \psfrag{e}[Bc]{$\la{2500}$}
    \psfrag{f}[Br]{$\la{2}$}
    \psfrag{g}[Br]{$\la{4}$}
    \psfrag{h}[Br]{$\la{6}$}
    \psfrag{i}[Br]{$\la{8}$}
    \psfrag{j}[Br]{$\la{10}$}
    \psfrag{N}[Bl]{}
    \psfrag{Y}[Bc]{}
    \psfrag{q}[Bl]{$\ex$}
    \psfrag{t}[Bl]{Exp}
    \psfrag{r}[Bl]{HN}
    \psfrag{s}[Bl]{Pow}
    \includegraphics{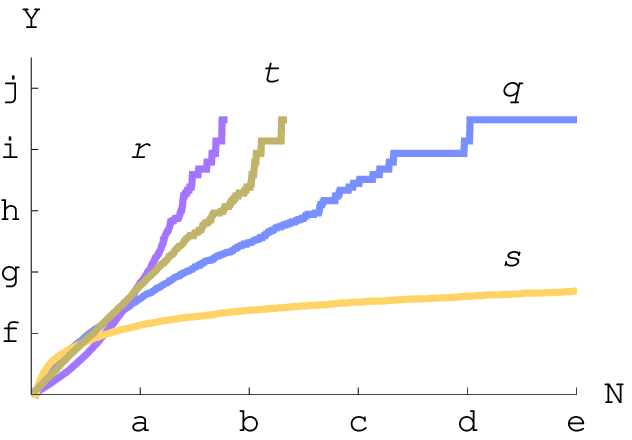}
    \caption{\label{fig:hazard}Hazard functions}
  \end{subfigure}
  \begin{subfigure}{0.45\textwidth}
  \psfrag{a}[Bc]{$\la{-.5}$}
  \psfrag{b}[Bc]{$\la{.5}$}
  \psfrag{c}[Bc]{$\la{1.0}$}
  \psfrag{d}[br]{$\la{100}$}
  \psfrag{e}[br]{$\la{200}$}
  \psfrag{f}[br]{$\la{300}$}
  \psfrag{g}[br]{$\la{400}$}
  \psfrag{h}[br]{$\la{600}$}
  \psfrag{j}[br]{$\la{800}$}
  \psfrag{A}[bl]{$\ w$}
  \psfrag{X}{$\bar{\chi}^{2}$}
  \psfrag{Y}{$\ex$}
  \psfrag{E}[bc]{\,Exp}
    \includegraphics{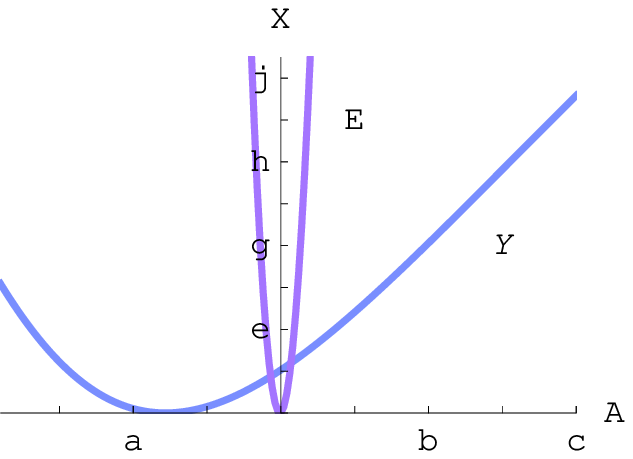}
    \caption{\label{fig:chisq}$\chi^{2}$ per degree of freedom}
  \end{subfigure}
  \caption{\label{fig:concavity}(a): Hazard functions for the last
    $10\%$ of the distribution of $\cH_{R}^{\Othree}$ eigenvalues of
    $\ex$ compared to best-fit half-normal, exponential, and power
    (Pareto) distributions. (b): Goodness of fits ($\chi^{2}$ per
    degree of freedom) of the hazard functions of $\rho_{R}$ and of an
    exponential distribution to the form $a+b\,x^{1+w}$. Nonnegative
    concavity ($w\ge 0$) in the case of $\rho_{R}$ is rejected with
    overwhelming significance.}
\end{center}
\end{figure}

A sufficient condition for a distribution to be heavy-tailed is that
the hazard function $h\!\left(x\right):=-\log S\!\left(x\right)$ is
strictly concave downward. In Figure~\ref{fig:hazard}, we compare
simulated hazard functions for the Pareto, half-normal, and
exponential distributions to that of the last $10\%$ of the left
(negative) tail of the eigenvalue spectrum $\rho_R$ of
$\cH_{R}^{\Othree}$ computed for $\ex$.  The hazard curves for the
half-normal and power law distributions are strictly concave up and
down, indicating light and heavy tails, respectively.  The simulation
of the hazard function for the exponential distribution does not
clearly possess either concavity, and indeed the analytic result is
$h(x)=\ka\, x$.

The hazard function for $\rho_R$ in Figure~\ref{fig:concavity}
appears to be very slightly concave
downward.  A  straightforward way to determine the concavity is to fit the
hazard function to the form $\hat{h}=a+b\,x^{1+w}$.  When $w>0$, the
model curve is concave upward, while when $w<0$, the model curve is
concave downward. In Figure~\ref{fig:chisq}, we show the $\chi^{2}$
calculated from the best fit for a variety of values of $w$,
indicating that the data are well-fitted by a curve that is concave
down:  by this measure, $w\ge 0$ is excluded with extremely high confidence.  A rough
estimate of errors was obtained by subdividing the populations into
five equal groups and combining the resulting hazard functions using
the standard error.

However, using a parametric test for concavity is potentially
problematic, as a concave-upward curve may in principle give a better
fit to the data.  A non-parametric test for concavity was described
in~\cite{Abrevaya}.  Given a function $y=f\!\left(x\right)$ and three
points $\left\{x_{1},x_{2},x_{3}\right\}$, the concavity
$\cC\left(x_{1},x_{2},x_{3}\right)$ is equal to $+1$ ($-1$) if the
quadratic interpolation between the three points
$\left(x_{1},y_{1}\right)$, $\left(x_{2},y_{2}\right)$, and
$\left(x_{3},y_{3}\right)$ is concave up (down), and equals 0 if the
points are collinear.  The method described in~\cite{Abrevaya} yields
an approximation to the expectation value of the concavity of a
function.  For the hazard function of $\rho_R$, sampled with $100$
points, we find $\left\langle\cC\right\rangle=-0.62\pm .06$, giving
strong evidence for downward concavity. Removing the points in the
beginning of the distribution (which constitute the bulk of the
distribution rather than the tail) or the end (which are statistically
limited) tends to increase the significance of concavity. For
comparison, an analysis for the hazard function of a corresponding
amount of simulated data from an exponential distribution gives
$\left\langle\cC\right\rangle=0.02 \pm 0.06$, and so an absence of
curvature cannot be statistically rejected.

In summary, the eigenvalue spectrum $\rho_R$ of $\cH_{R}^{\Othree}$ in the
example of $\ex$ has large positive excess kurtosis (i.e., is
leptokurtic), with much longer tails than those of a normal
distribution.  Moreover, from the negative concavity of the hazard
function we conclude that $\rho_R$ is in fact a heavy-tailed
distribution, with a survival function that falls off more slowly than
any exponential.

While we will not present here the corresponding analyses for other
examples of Calabi-Yau threefolds, we find similarly pronounced tails
in the eigenvalue spectra $\rho_R$ of $\cH_{R}^{\Othree}$ in all of
the threefolds.  Indeed, in an ensemble of threefolds with
$h^{1,1}=10,\ldots,25$ and five threefolds at each $h^{1,1}$, we find
that the average best fit value to $\hat{h}=a+b\,x^{1+w}$ is
$w=-0.52\pm 0.15$.

We find comparable results for the metric eigenvalue spectrum as well:
not only are the tails of the metric distributions quite extensive
(see Figure~\ref{fig:h11-15-metric-bulk}), they are formally heavy.

\subsection{\label{sec:explaining}Explaining the tails in the spectra}

Having established the existence of heavy tails in the spectra $\rho_R$
of $\cH_{R}^{\Othree}$ in Calabi-Yau hypersurface examples, we turn to
explaining the causes of these tails.  From the form
$\left(\cH_{R}\right)_{a\bar{b}}=-R_{a\bar{b}c\bar{d}}\bar{F}^{c}F^{\bar{d}}$,
it is easy to see that the length of the tails in $\cH_{R}$ is
intimately related to the curvature of moduli space as measured by
\begin{equation}
\Ri^2 := R_{a\overline{b}c\overline{d}}R^{a\overline{b}c\overline{d}}\,,
\end{equation}  which,
like the Ricci scalar
$R=-\cK^{a\bar{d}}\cK^{\bar{b}c}R_{a\bar{b}c\bar{d}}$, is invariant
under the homogeneous rescaling $\vec{t}\to\lambda\vec{t}$.
Let us explore this connection in detail. At any given point in moduli space, we can
adopt K\"ahler normal coordinates in which
$\cK_{a\bar{b}}=\delta_{a\bar{b}}$, in which case the curvature
contribution to the Hessian becomes
\begin{equation}
  (\cH_{R})_{a\bar{b}} = -R_{a\bar{b}c\bar{d}}\bar{F}^{c}F^{\bar{d}} =
  -R_{a\bar{b}c\bar{d}}\delta^{c\bar{e}}\delta^{f\bar{d}}F_f
  \bar{F}_{\bar{e}}\,.
\end{equation}
For fixed $F^{2}$, parametrically large eigenvalues occur when
$F_{a}$ points in directions of large curvature.  In K\"{a}hler normal
coordinates, $\Ri^{2}$ is just the sum of the squares of the entries
in the Riemann curvature tensor, and so the minimum eigenvalues satisfy
\begin{equation}
\lambda_{R,\text{min}} \geq -\Ri \left\lvert F\right\rvert^2\,.  \label{curvaturebound}
\end{equation}
The precise alignment of the F-terms with the Riemann tensor required
to saturate the bound (\ref{curvaturebound}) will not occur in generic
cases: instead there will be a misalignment by a random
rotation. Therefore, generically the eigenvalues of $\cH_{R}$ will
respect the approximate bound
\begin{equation}
\label{eq:eigenvalue_bound}
  \lambda_{R}\gtrsim-\Ri\left\lvert F\right\rvert^{2}/h^{1,1}\,.
\end{equation}

\begin{figure}
\begin{center}
    \psfrag{a}[Bc]{$\la{34000}$}
    \psfrag{b}[Bc]{$\la{36000}$}
    \psfrag{c}[Bc]{$\la{38000}$}
    \psfrag{d}[Bc]{$\la{40000}$}
    \psfrag{e}[Bc]{$\la{42000}$}
    \psfrag{f}[Bc]{$\la{44000}$}
    \psfrag{g}[br]{$\la{-52000}$}
    \psfrag{h}[br]{$\la{-57000}$}
    \psfrag{i}[br]{$\la{-47000}$}
    \psfrag{R}[Bc]{$\Ri$}
    \psfrag{S}[Bc]{$R$}
  {\includegraphics{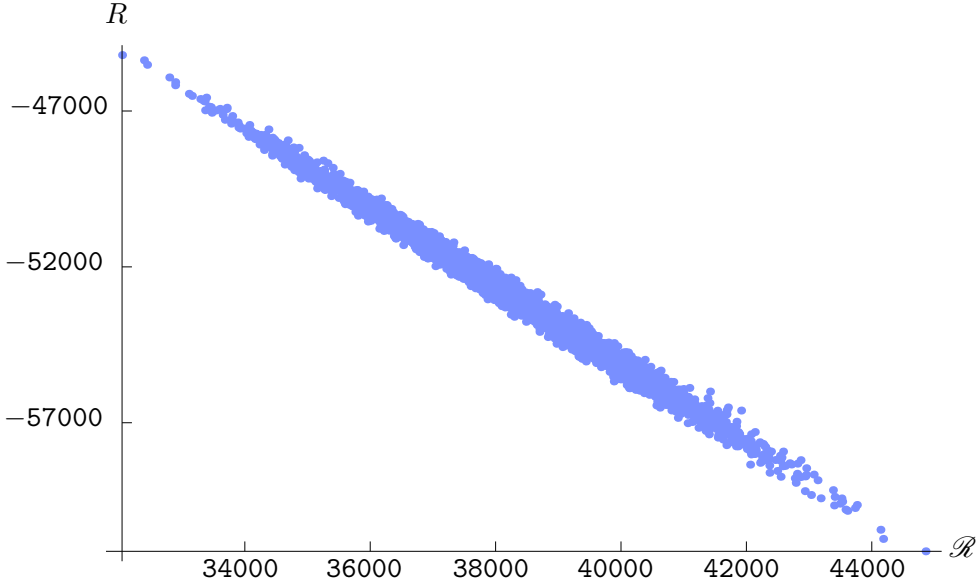}}
  \caption{\label{fig:rieR}Values of
    $\Ri=\left(R_{a\bar{b}c\bar{d}}R^{a\bar{b}c\bar{d}}\right)^{1/2}$
    and the Ricci scalar $R$ taken from an ensemble of points in $\ex$.
    Contrast Figure~\ref{fig:randomRiemann}.}
\end{center}
\end{figure}

\begin{figure}
\begin{center}
  \begin{subfigure}{0.45\textwidth}
  \psfrag{a}[Bc]{$\la{1955}$}
  \psfrag{b}[Bc]{$\la{1960}$}
  \psfrag{c}[Bc]{$\la{1965}$}
  \psfrag{d}[br]{$\la{-150}$}
  \psfrag{e}[br]{$\la{-75}$}
  \psfrag{f}[br]{$\la{0}$}
  \psfrag{g}[br]{$\la{75}$}
  \psfrag{h}[br]{$\la{150}$}
  \psfrag{R}{$\Ri$}
  \psfrag{S}{$R$}
    \includegraphics{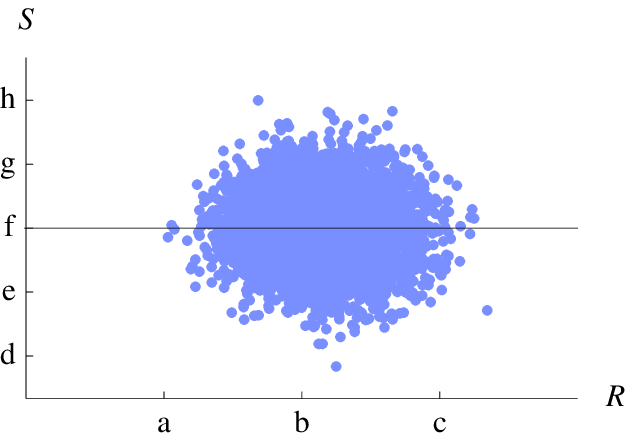}
    \caption{\label{fig:randomRie}
      Riemann tensors with i.i.d.~entries,  in spaces of real dimension 50}
  \end{subfigure}
  \quad
  \begin{subfigure}{0.45\textwidth}
  \psfrag{a}[Bc]{$\la{40}$}
  \psfrag{b}[Bc]{$\la{42}$}
  \psfrag{c}[Bc]{$\la{44}$}
  \psfrag{d}[Bc]{$\la{46}$}
  \psfrag{e}[Bc]{$\la{48}$}
  \psfrag{f}[br]{$\la{-15\!}$}
  \psfrag{g}[br]{$\la{-10\!}$}
  \psfrag{h}[br]{$\la{-5\!}$}
  \psfrag{i}[br]{$\la{0}$}
  \psfrag{R}{$\Ri$}
  \psfrag{S}{$R$}
  \includegraphics{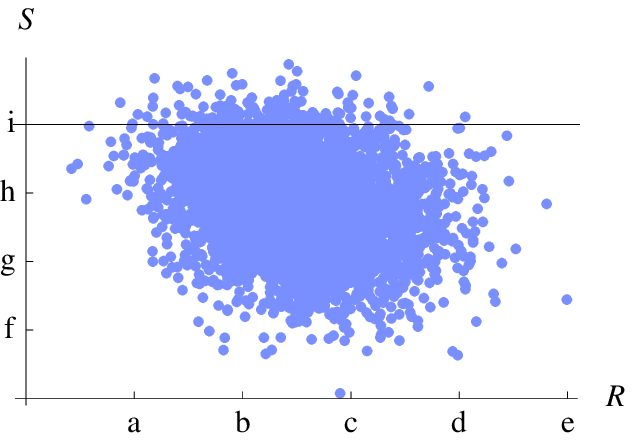}
  \caption{\label{fig:randomKahRie} Riemann tensors with
    i.i.d.~entries, in K\"ahler manifolds of complex dimension 15}
  \end{subfigure}
  \begin{center}
  \begin{subfigure}{0.45\textwidth}
  \psfrag{a}[Bc]{$\la{40}$}
  \psfrag{b}[Bc]{$\la{42}$}
  \psfrag{c}[Bc]{$\la{44}$}
  \psfrag{d}[Bc]{$\la{46}$}
  \psfrag{e}[Bc]{$\la{48}$}
  \psfrag{x}[br]{$\la{-40}$}
  \psfrag{y}[br]{$\la{-35}$}
  \psfrag{f}[br]{$\la{-30}$}
  \psfrag{g}[br]{$\la{-25}$}
  \psfrag{h}[br]{$\la{-20}$}
  \psfrag{R}{$\Ri$}
  \psfrag{S}{$R$}
  \includegraphics{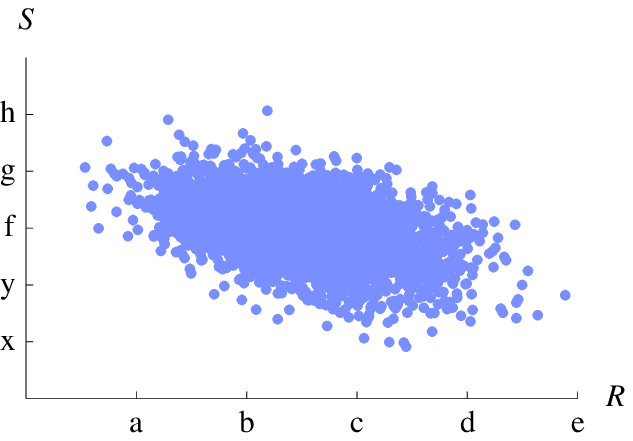}
  \caption{\label{fig:randomSignedKahRie} Riemann tensors with
    i.i.d.~positive entries for $R_{a\bar{b}c\bar{d}}$, in K\"ahler
    manifolds of complex dimension 15}
  \end{subfigure}
  \end{center}
  \caption{\label{fig:randomRiemann}Curvature invariants of random
    Riemann tensors with increasing amounts of structure.  These
    simple reference models are to be contrasted with the strong
    correlation exhibited in a Calabi-Yau (Figure~\ref{fig:rieR}).}
\end{center}
\end{figure}

One might think that an even stronger bound may result due to
cancellations between opposite-sign terms in the curvature tensor.
However, we find empirically that in the deep interior of the
K\"{a}hler cone, the components $R_{a\bar{b}c\bar{d}}$ of the Riemann
tensor (in the toric basis used throughout this work) are mostly
positive.\footnote{Note that this does not violate the antisymmetry of
  the Riemann tensor, which requires for example that
  $R_{a\bar{b}c\bar{d}}=-R_{\bar{b}ac\bar{d}}.$} Indeed, if the
components of the Riemann tensor had no statistical preference for
either sign then an ensemble of Ricci scalars taken from a Calabi-Yau
example would be clustered around zero. As shown in
Figure~\ref{fig:rieR}, $R$ is negative and linearly correlated with
$\Ri$, which is a manifestation of the preference of the components of
Riemann $R_{a\bar{b}c\bar{d}}$ to have a positive sign.\footnote{That
  the curvatures are large compared to unity might cause concern since the
  curvature of moduli space often diverges near the boundaries of the
  K\"ahler cone, where cycles collapse and the supergravity
  approximation is no longer trustable.  However, the points that we
  consider here are in the bulk of the cone, where $\alpha'$ corrections
  are under control.}  To understand this finding, we compare to a few
simple reference cases.  First, the Ricci scalar of a Riemann tensor
that has i.i.d.~components with zero mean has no preference for either
sign, and there is no significant correlation between the invariants:
see Figure~\ref{fig:randomRie}.  In the special case of a K\"ahler
manifold for which the metric depends only on the real part of the
coordinates, the Ricci scalar has a slight preference to be negative,
but there is little correlation between $\Ri$ and $R$
(Figure~\ref{fig:randomKahRie}).  Finally, in an ensemble of Riemann
tensors on a K\"ahler manifold with i.i.d.~{\it{positive}}
coefficients, the Ricci scalar indeed turns out to be negative, but
with only a very slight correlation
(Figure~\ref{fig:randomSignedKahRie}).  Clearly, the structure of the
Riemann tensor on a Calabi-Yau K\"ahler moduli space is more intricate
than that of these reference models.  The empirical results just
described are related to the conjectured bounds on the sectional
curvature proposed in~\cite{Wilsonsign} (see also~\cite{Ooguri:2006in}
for a related proposal and~\cite{Wilsonsign2} for counterexamples).
However, not all examples that we examined have a negative Ricci
scalar.

\begin{figure}
\begin{center}
  \begin{subfigure}{0.45\textwidth}
    \psfrag{a}[Bc]{$\la{-200}$}
    \psfrag{b}[Bc]{$\la{-150}$}
    \psfrag{c}[Bc]{$\la{-100}$}
    \psfrag{d}[Bc]{$\la{-50}$}
    {\includegraphics{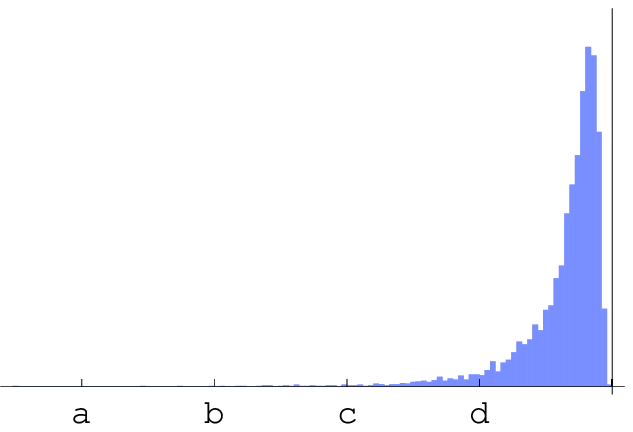}}
    \caption{Minimum eigenvalues}
  \end{subfigure}
  \quad\quad
  \begin{subfigure}{0.45\textwidth}
    \psfrag{a}[Bc]{$\la{700}$}
    \psfrag{b}[Bc]{$\la{900}$}
    \psfrag{c}[Bc]{$\la{1100}$}
    \psfrag{g}[br]{$\la{-1350}\!$}
    \psfrag{h}[br]{$\la{-1550}\!$}
    \psfrag{i}[br]{$\la{-1750}$}
    \psfrag{R}[Bc]{$\Ri$}
    \psfrag{S}[Bc]{$R$}
    {\includegraphics{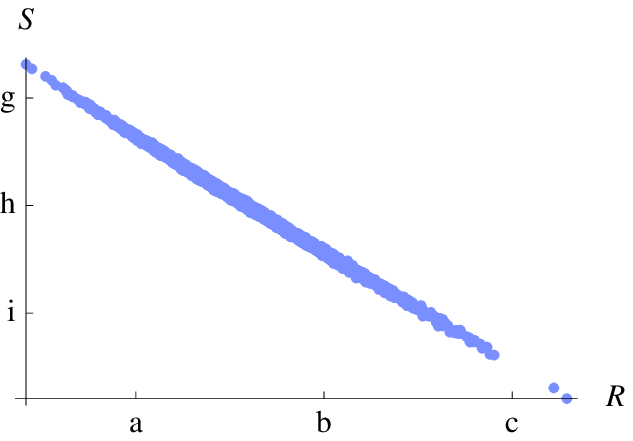}}
    \caption{Curvature invariants}
  \end{subfigure}
  \caption{\label{fig:h11-20}The spectrum of minimum eigenvalues of
    $\cH^{\Othree}_{R}$ and the correlation between the Ricci scalar and $\Ri$
    for a Calabi-Yau with $h^{1,1}=20$. This ensemble consists of 5000
    matrices.}
\end{center}
\end{figure}

Similar behavior occurs for other threefolds, though the
characteristic sizes of the curvatures can be quite different for
different threefolds, with no simple scaling with $h^{1,1}$.  An
example Calabi-Yau with $h^{1,1}=20$ exhibits smaller curvatures and
correspondingly less negative eigenvalues of $\cH^{\Othree}_{R}$: see
Figure~\ref{fig:h11-20}.

\begin{figure}
\begin{center}
    \psfrag{a}[Bc]{$\la{1}$}
    \psfrag{b}[Bc]{$\la{10}$}
    \psfrag{c}[Bc]{$\la{100}$}
    \psfrag{d}[Bc]{$\la{10^{3}}$}
    \psfrag{e}[Bc]{$\la{10^{4}}$}
    \psfrag{f}[Bc]{$\la{10^{5}}$}
    \psfrag{g}[Bc]{$\la{10^{6}}$}
    \psfrag{i}[br]{$\la{1}$}
    \psfrag{j}[br]{$\la{10}$}
    \psfrag{k}[br]{$\la{100}$}
    \psfrag{l}[br]{$\la{10^{3}}$}
    \psfrag{m}[br]{$\la{10^{4}}$}
    \psfrag{n}[br]{$\la{10^{5}}$}
    \psfrag{o}[br]{$\la{10^{6}}$}
    \psfrag{R}{$\cR/h^{1,1}$}
    \psfrag{L}{$\left\lvert\lambda^{\Othree}_{R,\text{min}}\right\rvert$}
  \includegraphics{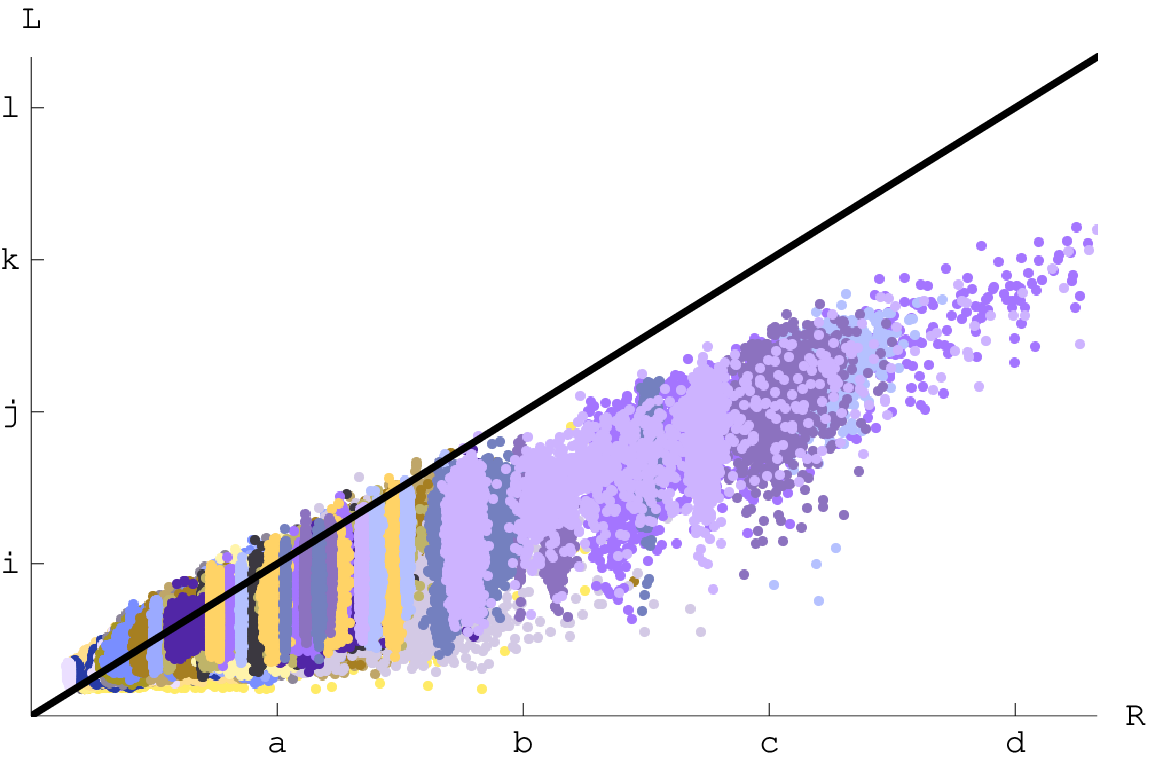}
  \caption{\label{fig:fracs}Comparison of the minimum eigenvalue of
    $\cH_{R}^{\Othree}$ (in units of $\left\lvert F\right\rvert^{2}$)
    to the rescaled curvature $\Ri/h^{1,1}$.  The data were gathered
    from a collection of Calabi-Yau threefolds between $h^{1,1}=2$ and
    $h^{1,1}=25$, with five distinct threefolds for each value of
    $h^{1,1}$ and about $1000$ points in moduli space from each
    threefold. Each color corresponds to a different value of
    $h^{1,1}$.   Well inside the K\"ahler cone there is typically little variation in curvature,
    and hence curvature invariants for each Calabi-Yau
    tend to cluster in vertical stripes. Those points below the line
    satisfy the bound~\eqref{eq:eigenvalue_bound}.}
    \end{center}
\end{figure}

The bound (\ref{eq:eigenvalue_bound}) is simply a scaling result, with
the factor of $h^{1,1}$ determined by modeling a random rotation in a space
of dimension $h^{1,1}$, and we can only expect it to hold up to an
$\cO\left(1\right)$ multiplicative factor.  Figure~\ref{fig:fracs}
compares the minimum eigenvalues of an ensemble of spectra of
$\cH^{\Othree}_{R}$, for a number of threefolds, to the curvature
invariants $\Ri$.  We observe that $76\%$ of the data satisfy the
bound (\ref{eq:eigenvalue_bound}), while $96\%$ satisfy the weaker
bound $\lambda^{\Othree}_{R,\mathrm{min}}<2\Ri/h^{1,1}$.

\begin{figure}
\begin{center}
    \psfrag{a}[Bc]{$\la{0.2}$}
    \psfrag{b}[Bc]{$\la{0.4}$}
    \psfrag{c}[Bc]{$\la{0.6}$}
    \psfrag{d}[Bc]{$\la{0.8}$}
    \psfrag{e}[Bc]{$\la{1.0}$}
    \psfrag{f}[Br]{$\la{-0.8}$}
    \psfrag{g}[Br]{$\la{-0.6}$}
    \psfrag{h}[Br]{$\la{-0.4}$}
    \psfrag{i}[Br]{$\la{-0.2}$}
    \psfrag{j}[Br]{$\la{-1.0}$}
    \psfrag{l}[Bc]{$\lambda^{\Othree}_{R,\text{min}}\bigl(\theta\bigr)/
      \lambda_{R,\mathrm{min}}\bigl(\theta=0\bigr)$}
    \psfrag{s}[Bl]{$\sin\theta$}
    \includegraphics[scale=1]{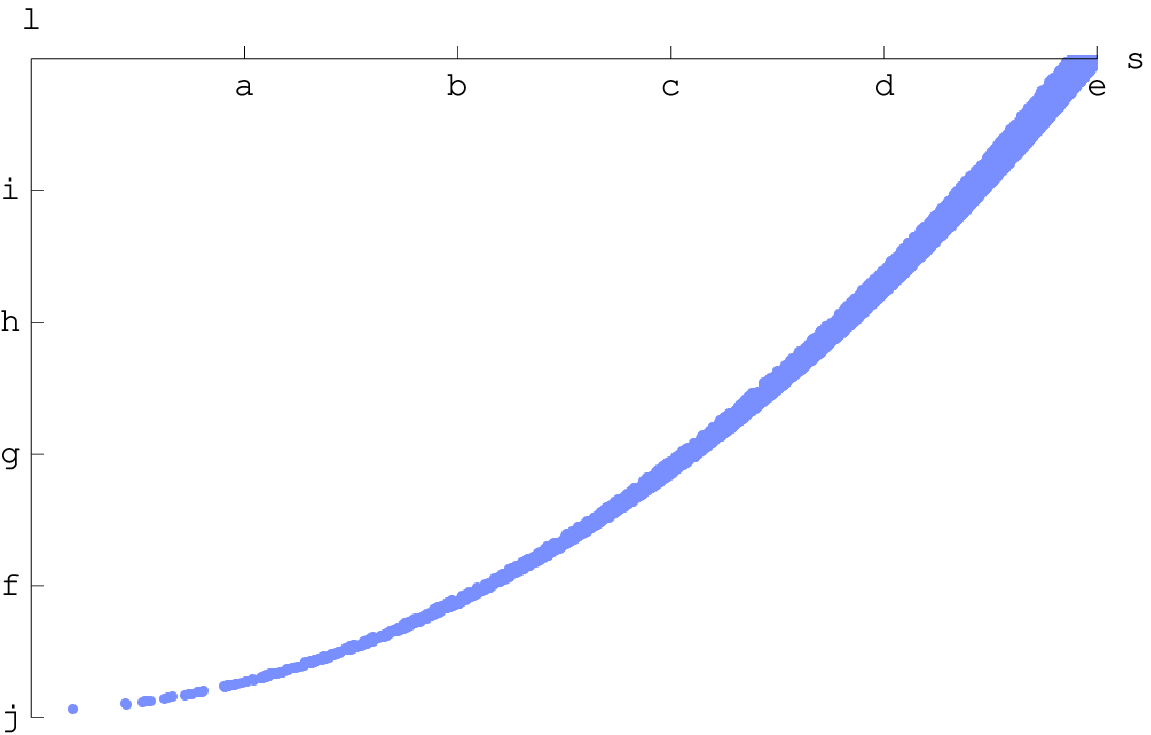}
    \caption{\label{fig:angle}Magnitude of the most negative
      eigenvalue of $\cH^{\Othree}_{R}$ as a function of the angle of
      $\hat{F}$ with respect to the direction that produces the most
      negative eigenvalue. The eigenvalues are normalized with respect
      to the most negative eigenvalue obtained. This sample results
      from a Calabi-Yau with $h^{1,1}=4$.}
  \end{center}
\end{figure}

The scaling of the Hessian eigenvalues with curvature is modulated by
the fact that alignment of the F-terms with the directions of maximum
curvature is non-generic.  It is then interesting to consider the
effect that changing the direction of
$\hat{F}_{a}=F_{a}/\left\lvert{F}\right\rvert$ has on the tail
population.  Finding the direction of $\hat{F}$ that minimizes the
smallest eigenvalue is computationally challenging at even modest
$h^{1,1}$, and so we consider an example with $h^{1,1}=4$.  In a
particular basis and at a fixed point in moduli space, the maximal
sectional curvature is $R_{1\bar{1}1\bar{1}}\approx 103$, where we
have rotated coordinates so that the direction of largest sectional
curvature is the $1$ direction.  The corresponding minimum eigenvalue
of $\cH_{R}^{\Othree}$ is $\lambda^{\Othree}_{R,\mathrm{min}}\approx
-200$.  The direction that results in the most negative
$\cH^{\Othree}_{R}$, $\hat{F}^{\mathrm{min}}$, has a large overlap
with the $1$ direction, $\hat{F}^{\mathrm{min}}_{1}\approx 0.8$.
Remaining at the same point in moduli space, we generate an ensemble
of Hessians by choosing randomly oriented F-terms.  In
Figure~\ref{fig:angle} we plot the smallest eigenvalue of
$\cH^{\Othree}_{R}$ against the angle between $\hat{F}$ and
$\hat{F}^{\mathrm{min}}$.  One can see that the orientation of
$\hat{F}$ plays an important role in the population of the tail. For
larger $h^{1,1}$, finding the direction of $\hat{F}$ that results in the
largest tails towards negativity is computationally challenging, but
we observe empirically that when the tails are particularly extensive,
$\hat{F}$ has large overlap with the directions in which the curvature is
large.

The maximum value of $\Ri$ sets an approximate bound for the minimum
eigenvalue of $\cH^{\Othree}_{R}$.  Although the curvature is well
behaved, albeit often large, in the interior of the cone, the metric
may degenerate and the curvature diverge near the boundary (typically
near boundaries that have codimension greater than $1$).  The
boundaries of the K\"ahler cone are marked by the collapse of one or
more 2-cycles where corrections to the leading-order K\"ahler
potential become important, and so we have remained away from the
boundaries in all of our analysis thus far.  However, to expose the
relationships between curvature and eigenvalue hierarchies derived
from the leading-order K\"ahler potential, it is useful to approach
the boundaries, with the understanding that in actual string
compactifications there would be large corrections to these results.

\begin{figure}
\begin{center}
  \begin{subfigure}{0.45\textwidth}
    \psfrag{a}[Bc]{$\la{.005}$}
    \psfrag{b}[Bc]{$\la{.05}$}
    \psfrag{c}[Bc]{$\la{.5}$}
    \psfrag{d}[Br]{$\la{10^{7}}$}
    \psfrag{e}[Br]{$\la{10^{10}}$}
    \psfrag{f}[Br]{$\la{10^{13}}$}
    \psfrag{t}[Bc]{$\Ri$}
    \psfrag{R}[Bc]{$t$}
    {\includegraphics{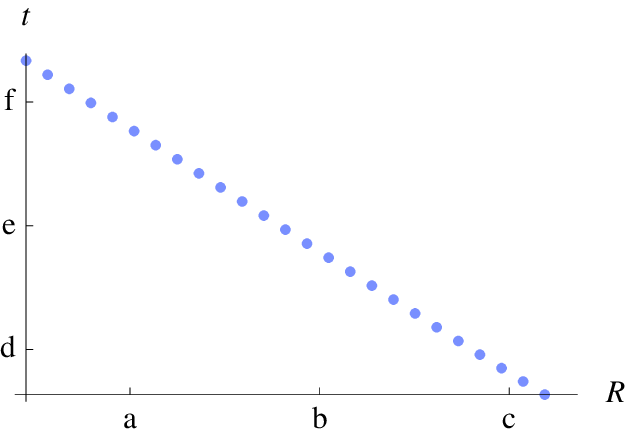}}
    \caption{\label{fig:walls}Divergence of the curvature}
  \end{subfigure}
  \quad\quad
  \begin{subfigure}{0.45\textwidth}
    \psfrag{a}[Bc]{$\la{10^{7}}$}
    \psfrag{b}[Bc]{$\la{10^{10}}$}
    \psfrag{c}[Bc]{$\la{10^{13}}$}
    \psfrag{d}[Br]{$\la{400}$}
      \psfrag{e}[Br]{$\la{4000}$}
    \psfrag{f}[Br]{$\la{4 \times 10^{4}}$}
    \psfrag{R}[Bc]{$\Ri$}
    \psfrag{L}[Bc]{$\lambda^g_{\text{max}}/\lambda^g_{\text{median}}$}
    \includegraphics{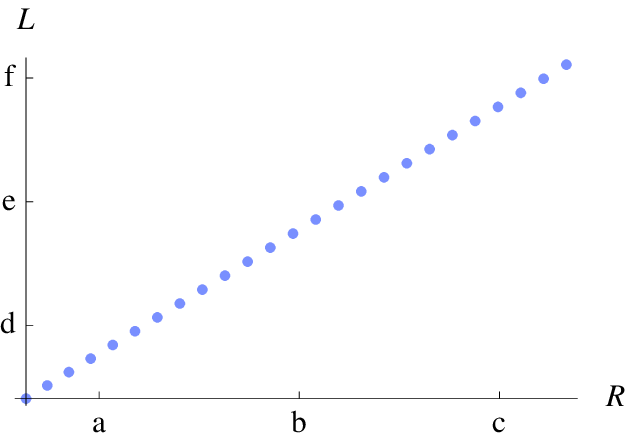}
    \caption{\label{eq:metric_explosion} Correlation between the
      metric hierarchies and curvature $\Ri$}
  \end{subfigure}
  \begin{center}
  \begin{subfigure}{0.45\textwidth}
    \psfrag{a}[Bc]{$\la{10^{7}}$}
    \psfrag{b}[Bc]{$\la{10^{10}}$}
    \psfrag{c}[Bc]{$\la{10^{13}}$}
    \psfrag{d}[Br]{$\la{10^{3}}$}
      \psfrag{e}[Br]{$\la{10^{8}}$}
    \psfrag{f}[Br]{$\la{10^{13}}$}
     \psfrag{L}[Bc]{$|\lambda^{\Othree}_{R,\mathrm{min}}|$}
    \psfrag{R}[Bc]{$\Ri$}
    \includegraphics[scale=1]{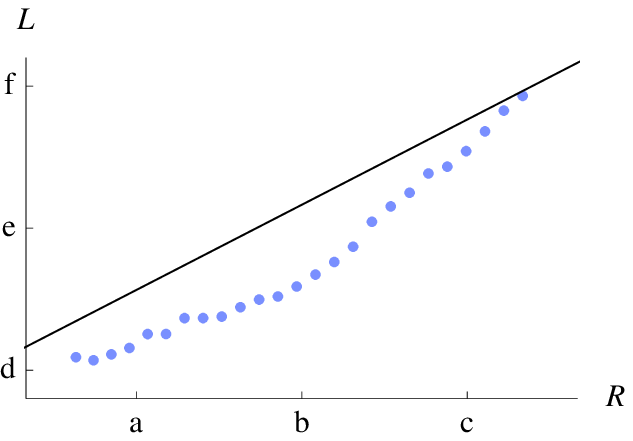}
    \caption{\label{fig:min-wall}Correlation of the smallest
      eigenvalues of $\rho_{R}$ with the curvature, with the line
      indicating the bound~\eqref{eq:eigenvalue_bound}}
\end{subfigure}
\end{center}
    \caption{\label{fig:wall_approach}Behavior of the curvature,
      metric eigenvalues, and curvature Hessian eigenvalues as one
      approaches a particular boundary of the K\"ahler cone by taking
      $t\to 0$.  The hierarchies in the metric eigenvalues are measured
      by comparing the largest eigenvalue to the median
      eigenvalue. Corrections to the leading-order K\"ahler potential
      are large in this regime, as explained in the main text.}
\end{center}
\end{figure}

We illustrate this in an example with $h^{1,1} = 15$, approaching
a codimension two boundary of the K\"{a}hler cone where
$A_{ab}=\partial t_{a}/\partial\tau^{b}$ degenerates.  The distance to
the wall is parameterized by $t$, which measures the volume of the two
2-cycles that are shrinking. In Figure~\ref{fig:walls} we demonstrate
the relationship between $\Ri$ and $t$: the curvature diverges as we
approach this boundary.  A related effect is that $A^{-1}$ develops a
zero eigenvalue and consequently, via~\eqref{eq:metric}, one of the
eigenvalues of the metric $\cK_{a\bar{b}}$ diverges
(Figure~\ref{eq:metric_explosion}). To explore the behavior of the
curvature Hessian, at each point in moduli space we generate $10^5$
random F-terms, and calculate the eigenvalues of
$\cH^{\Othree}_{R}$. In figure~\ref{fig:min-wall}, we demonstrate the
correlation between the minimum eigenvalue and $\Ri$: as $t \rightarrow 0$ and $\Ri$ grows, the length of the tail in
$\cH^{\Othree}_{R}$ also grows.

\subsection{\label{sec:stability}Implications for stability}

The presence of heavy tails in the distribution of negative eigenvalues
of the curvature contribution $\cH^{\Othree}_{R}$ to the total Hessian
${\cal H}$ may have considerable impact on the probability $P_+$ that
a Hessian drawn from the ensemble has only positive eigenvalues.  In
this section we will comment briefly on the effect of heavy tails on
$P_+$.

Direct study of $P_+$ requires a fully-specified stochastic model for
the superpotential and its derivatives.  While in a characterization of
$\cH_{R}$ the details of the superpotential model have little impact,
in the full Hessian ${\cal H}$ they can be determinative.  To apply
the i.i.d.~model of $W$ used in~\cite{Denef:2004cf,Marsh:2011aa}, it
is necessary to fix a physically meaningful ambiguity, namely the
relative r.m.s.~sizes of $F_a$, $Z_{ab}$, and $U_{abc}$.  As explained
in this context in \cite{Marsh:2011aa}, the relative magnitudes of the
F-terms and of the supersymmetric masses encoded in $Z_{ab}$ reflect
the strength of supersymmetry breaking, and the prevalence of
corresponding instabilities.  Modeling supersymmetry breaking in this
manner necessarily introduces a large degree of model-dependence, and
we defer a detailed analysis to future work.

Even without modeling the non-curvature Hessian ${\cal H}-\cH_{R}$, we
can draw a few qualitative conclusions about the impact of the heavy
tails toward negativity in $\cH_{R}$.  We consider the fraction of
eigenvalues that are negative in our ensemble of curvature Hessians
$\cH_{R}$, and compare this to the fraction of negative eigenvalues in
the i.i.d.~Wigner+Wishart model discussed in \S\ref{sec:RMT_sugra},
i.e.~for the null hypothesis (\ref{eq:WW_curvature_Hessian_NH}).

\begin{figure}
\begin{center}
    \psfrag{a}[Bc]{$\la{0.2}$}
    \psfrag{b}[Bc]{$\la{0.4}$}
    \psfrag{c}[Bc]{$\la{0.6}$}
    \psfrag{d}[Bc]{$\la{0.8}$}
    \psfrag{e}[Bc]{$\la{1.0}$}
    \psfrag{I}{$\ex$}
    \psfrag{C}{WW}
  \includegraphics{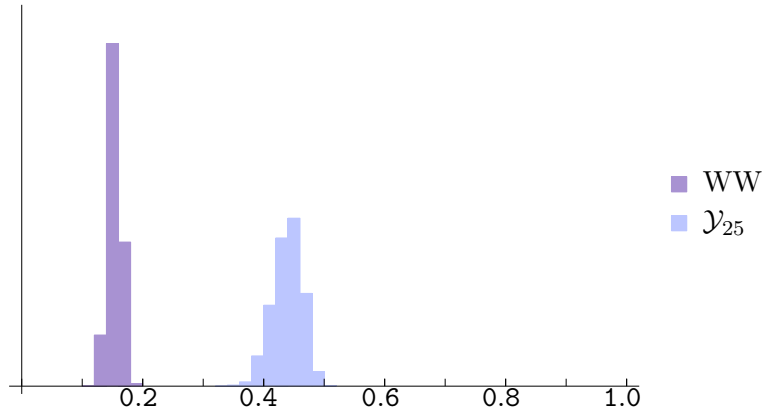}
  \caption{\label{fig:frac_negs}Distribution of the fraction of
    negative eigenvalues in the curvature Hessian.  The Hessians are
    taken from ensembles of $\cH_{R}^{\Othree}$ for $\ex$ and of
    $\cH_{R}^{\WW}$ for $25$ complex scalar
    fields~\eqref{eq:WW_curvature_Hessian_NH}.  For the Calabi-Yau data,
    the mean and standard deviation of this distribution are $43\%$
    and $1.2\%$, while for the WW distribution, the mean and standard
    deviation are $2.5\%$ and $1.2\%$.}
    \end{center}
\end{figure}

\begin{figure}
\begin{center}
  \psfrag{a}{$\la{5}$}
  \psfrag{b}{$\la{10}$}
  \psfrag{c}{$\la{15}$}
  \psfrag{d}{$\la{20}$}
  \psfrag{e}{$\la{25}$}
  \psfrag{f}[bc]{$\la{10}$}
  \psfrag{g}[bc]{$\la{20}$}
  \psfrag{h}[bc]{$\la{30}$}
  \psfrag{i}[bc]{$\la{40}$}
  \psfrag{k}[bc]{$\la{50}$}
  \psfrag{t}{$\tau$}
  \psfrag{n}{$h^{1,1}$}
  \psfrag{x}{$\left\langle f\right\rangle/\sig_{f}$}
  {\includegraphics{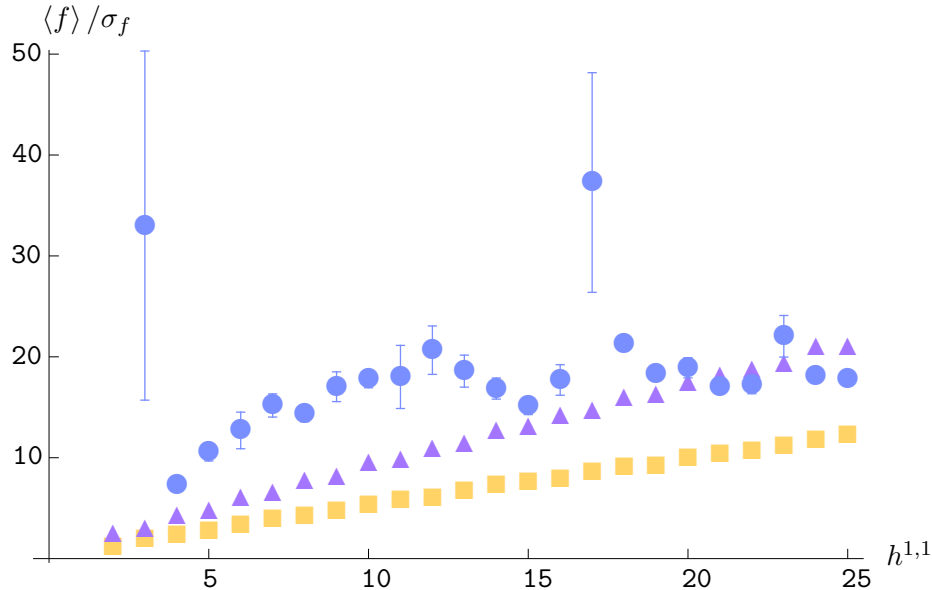}}
  \caption{\label{fig:positivityproxyplot}A proposed proxy for the
    probability of large fluctuations of $\cH_{R}$ in the direction of
    positivity: the expectation value of the fraction of negative
    eigenvalues divided by the standard deviation of the distribution
    of this fraction.  The growth of this quantity with $h^{1,1}$
    should trace the growth of instabilities with $h^{1,1}$.  Shown
    are results for Wigner matrices
    (\textcolor[RGB]{163,117,255}{$\blacktriangle$}), the
    Wigner+Wishart model
    (\textcolor[RGB]{255,211,102}{$\blacksquare$}) and Calabi-Yau
    examples (\textcolor[RGB]{121,142,255}{$\bullet$}). The Wigner and
    Wigner+Wishart data were gathered from ensembles of $5000$
    matrices at each point, while the Calabi-Yau data were gathered
    from ensembles of $5$ threefolds at each $h^{1,1}$ with $1000$
    Hessians from each threefold.}
\end{center}
\end{figure}

Focusing first on $\ex$, the mean fraction of negative eigenvalues is
$43\%$, which is significantly higher than the $14\%$ fraction we find
in Wigner+Wishart matrices of the same size.  However, the
distribution in the case of $\ex$ has a wider spread, with a standard
deviation of $2.5\%$ for the Calabi-Yau data and $1.2\%$ for the
Wigner+Wishart matrices (see Figure~\ref{fig:frac_negs}).  For a more
meaningful comparison we therefore rescale the mean fraction of
negative eigenvalues by the standard deviation of the corresponding
distribution.  Repeating this analysis for several different
threefolds at different values of $h^{1,1}$ yields
Figure~\ref{fig:positivityproxyplot}.  We find that the rescaled
fraction of negative eigenvalues generally increases more quickly with
$h^{1,1}$ in the threefold examples than it does in Wigner+Wishart or
Wigner reference models.
Although this proxy is suggestive of decreased metastability, a more
precise result requires detailed study of the full Hessian that
incorporates superpotential data.

\subsection{\label{sec:special_geometry}Other string theories}

\begin{figure}
\begin{center}
  \begin{subfigure}{0.45\textwidth}
    \psfrag{a}[Bc]{$\la{-1.5}$}
    \psfrag{b}[Bc]{$\la{-1.3}$}
    \psfrag{c}[Bc]{$\la{-1.1}$}
    {\includegraphics{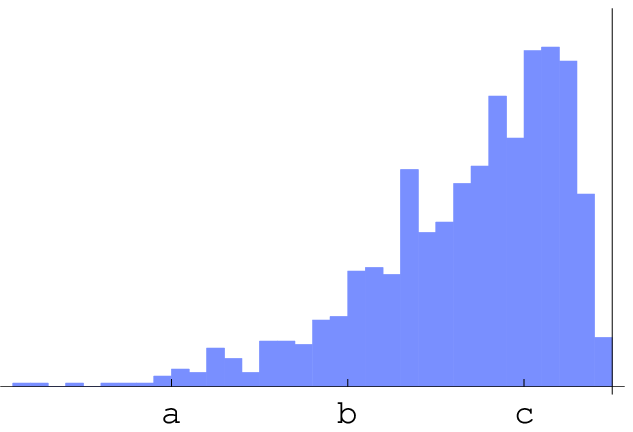}}
    \caption{Minimum eigenvalues, which in this sample fall in
      $\left[-1.7,-1.0\right]$}
  \end{subfigure}
  \quad\quad
   \begin{subfigure}{0.45\textwidth}
    \psfrag{a}[Bc]{$\la{2000}$}
    \psfrag{b}[Bc]{$\la{4000}$}
    \psfrag{c}[Bc]{$\la{6000}$}
    \psfrag{d}[Bc]{$\la{8000}$}
    {\includegraphics{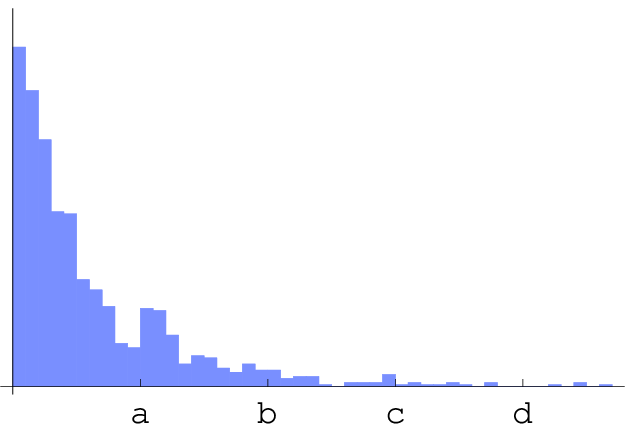}}
    \caption{Maximum eigenvalues, which in this sample fall in
      $\left[29,9300\right]$}
  \end{subfigure}
   \begin{center}
     \begin{subfigure}{0.45\textwidth}
       \psfrag{a}[Bc]{$\la{-1}$}
       \psfrag{b}[Bc]{$\la{1}$}
       \psfrag{c}[Bc]{$\la{2}$}
       \includegraphics{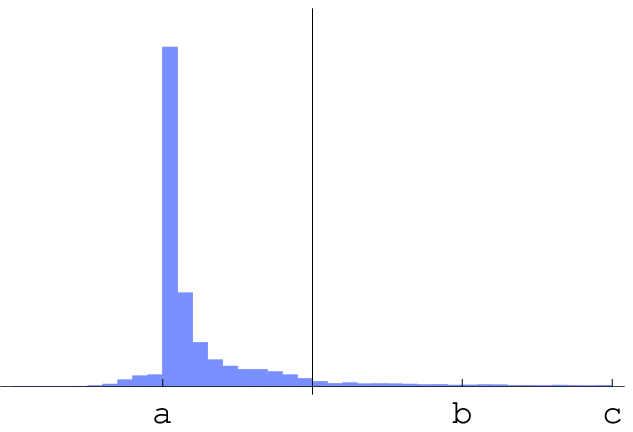}
       \caption{Bulk of the distribution, with the tail towards
         positivity truncated}
     \end{subfigure}
   \end{center}
  \caption{\label{fig:h11_25_special_HR_minmax}The spectrum of an
    ensemble of $5000$ curvature Hessians $\cH^{\Othree'}_{R}$ for
    $\ex$, with  the K\"ahler  coordinates  that arise in compactification of
    the heterotic  string, O5/O9 projections of type~IIB, or
    O6 projections of type IIA.}
\end{center}
\end{figure}

In all of the above, we focused on the properties of the Hessian on
the K\"ahler moduli space of O3/O7 orientifolds of a Calabi-Yau
threefold, in which the K\"ahler coordinates are complexifications of
4-cycle volumes.  In other corners of the landscape, including the
heterotic string, O5/O9 projections of type IIB, and O6 projections of
type IIA, the complexified K\"ahler moduli are complexifications of
$2$-cycle volumes, $J+i\,B_{2}=\varrho^{i}\omega_{i}$ (in O5/O9
projections, $C_{2}$ rather than $B_{2}$ is used).  Our methods
readily apply to these constructions, and it is interesting to compare
the spectra shown in~\S\ref{sec:results} to these cases.  We denote
the resulting Hessians by $\cH_{R}^{\Othree'}$.  The eigenvalue
spectrum of an ensemble of matrices $\cH_{R}^{\Othree'}$ for $\ex$ is
shown in Figure~\ref{fig:h11_25_special_HR_minmax}.  Just as in the
O3/O7 projection of type IIB, we find extended tails.  One noteworthy
difference from the O3/O7 projection is that the Ricci scalar in the
case of $\cH_{R}^{\Othree'}$ is typically positive, and
$\cH_{R}^{\Othree'}$ exhibits corresponding positive tails.  We defer
a more complete consideration of these classes of compactifications to
future work.

\section{\label{sec:alt}Reference Models}

In the previous sections, we identified key statistical properties of
the K\"ahler metrics and curvature Hessians $\cH_{R}$ arising in
compactifications on (orientifolds of) explicit Calabi-Yau
hypersurfaces in toric varieties.  The presence of extensive tails ---
toward positivity for $\cK_{a\bar{b}}$, and in both directions for
$\cH^{\CY}_{R}$ --- is the most striking feature of these ensembles.

Our findings establish that the i.i.d.~Wigner+Wishart model
\eqref{eq:WW_curvature_Hessian_NH} for $\cH_{R}$ proposed
in~\cite{Marsh:2011aa} must be modified in order to reflect the
properties of actual Calabi-Yau compactifications. The i.i.d.~model,
however, does have the important virtues of simplicity and extremely
low computational cost in comparison to the analysis described in this
paper.  It would be instructive to identify a stochastic model that
shares some of the simplicity of the i.i.d.~Wigner+Wishart model, but
more accurately models the geometry of Calabi-Yau moduli spaces.  In
this section we will describe several candidate models for $\cH_{R}$,
among which the Bergman metric discussed in~\S\ref{sec:bergman} has
the most qualitative success, though in its simplest form it lacks the
heavy tails of the Calabi-Yau spectrum.

\subsection{Random intersection numbers}

A central result of this paper is that correlations between the
various contributions to the Hessian will impact the features of the
eigenvalue spectrum.  Although we focused on the particular example of
O3/O7 orientifolds of Calabi-Yau manifolds, correlations should be a
universal feature, and further elucidation of their effect could be
obtained by a study of i.i.d.~K\"ahler potentials.  One could hope to
model the K\"ahler potential as
\begin{equation}
  \label{eq:random_prepot}
  \cK=-2\log\, \biggl(\frac{1}{3!}\,c^{abc}t_{a}t_{b}t_{c}\biggr)\,,
\end{equation}
where $c^{abc}$ is a totally symmetric i.i.d.~tensor. This approach is
immediately problematic: for almost all choices of $c^{abc}$, and for
almost all points in the space spanned by $t_{a}$, the metric
following from~\eqref{eq:random_prepot} has negative eigenvalues.
Finding the interior of the K\"ahler cone numerically is then very
difficult.  In fact there is no guarantee that the K\"ahler cone will
not be empty for a generic set of random intersection
numbers.\footnote{Further refinement could be obtained by
  incorporating the consistency requirements for the intersection
  numbers on a Calabi-Yau discussed in~\cite{Wilsonintersect}.}

\subsection{Random positive Riemann curvature}

\begin{figure}
\begin{center}
  \begin{subfigure}{0.45\textwidth}
  \psfrag{a}[Bc]{$\la{-8}$}
  \psfrag{b}[Bc]{$\la{-6}$}
  \psfrag{c}[Bc]{$\la{-4}$}
  \psfrag{d}[Bc]{$\la{-2}$}
  \includegraphics{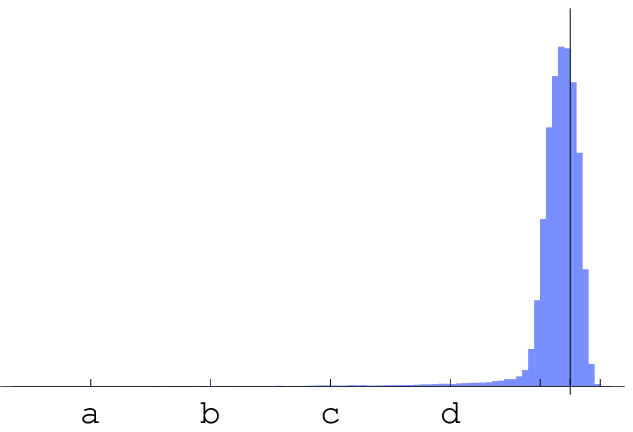}
  \caption{\label{fig:posRiemann}Eigenvalue spectrum}
  \end{subfigure}
  \quad
  \begin{subfigure}{0.45\textwidth}
  \psfrag{a}[Bc]{$\la{-8}$}
  \psfrag{b}[Bc]{$\la{-6}$}
  \psfrag{c}[Bc]{$\la{-4}$}
  \psfrag{d}[Bc]{$\la{-2}$}
    \includegraphics{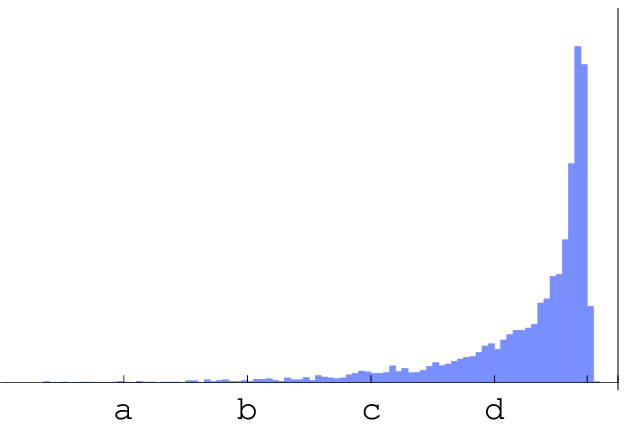}
    \caption{\label{fig:posRiemannMins}Spectrum of minimum eigenvalues}
  \end{subfigure}
  \caption{\label{fig:singlesign}The spectrum of
    $\cH_{R}^{\mathrm{i.i.d.}+}$, in which $R_{a\bar{b}c\bar{d}}$ are
    taken to be i.i.d. and positive. Here $h^{1,1} = 15$, and the
    ensemble consists of 5000 matrices.}
\end{center}
\end{figure}

The components $R_{a\bar{b}c\bar{d}}$ of the Riemann tensors on the
K\"ahler moduli spaces that we study are nearly always positive.  An
ad hoc modification of the i.i.d.~approach would be to impose
positivity on the i.i.d.~components.  We accomplish this in a very
crude way by generating an ensemble of Riemann tensors by adding
i.i.d.~$\cK^{\left(3\right)}$ and $\cK^{\left(4\right)}$ contributions
and then taking the absolute values of the resulting tensor
components.  This prescription yields an ensemble of matrices
$\cH_{R}^{\iid +}$ which manifest an increased correlation between the
curvature tensors (Figure~\ref{fig:randomSignedKahRie}).
Correspondingly, the curvature Hessian spectrum displays a tail
towards negativity as shown in Figure~\ref{fig:singlesign}.  However,
these features are not as pronounced as in the Calabi-Yau case.

\subsection{\label{sec:bergman}Random Bergman metrics}

A construction that allows for a definition of a random K\"ahler
potential is the Bergman metric, which comes from the embedding of
compact K\"ahler manifolds into projective space along the lines of
Kodaira.\footnote{See~\cite{Phong} for a mathematical review.}  In
practice, the Bergman metric on a K\"ahler manifold with coordinates
$z^{i}$ is specified by the K\"{a}hler potential:
\begin{equation}
  \cK(z,\overline{z}) = \frac{1}{k}\log f\bigl(z,\bar{z}\bigr),\quad
  f\bigl(z,\bar{z}\bigr):=\bar{s}_\alpha(\bar{z})P_{\alpha\beta}s_\beta(z)\,,
  \label{eq:Bergman}
\end{equation}
in which $\left\{s_{\alpha}\left(z\right)\right\}$ is a basis of
polynomials of degree $\le k$ and $P$ is a positive-definite Hermitian
matrix.  $f$ is called the Bergman kernel.  A particular Bergman
metric is defined by a choice of positive-definite matrix
$P_{\alpha\beta}$, and by taking the entries of this matrix to be
i.i.d.,~one can construct a {\it{random Bergman metric}}, in the
spirit of~\cite{Ferrari:2011we,*Ferrari:2011is,*Ferrari:2011sf}.

One might be tempted to use~\eqref{eq:Bergman} to construct an
ensemble of metrics that scale homogeneously under $z^{i}\to \lambda
z^{i}$, and thereby model the homogeneous metrics on the moduli space
of a Calabi-Yau.  However, positive-definiteness of the Bergman metric
is only guaranteed when $P$ is positive-definite, and in particular is
not degenerate.  Indeed, if $P$ is such that $f$ is a homogeneous
function under $z^{i}\to \lambda z^{i}$, then the resulting metric is
degenerate.\footnote{The argument is very similar to the argument used
  in~\S\ref{sec:rhess} to argue for the negativity of $\cH_{R}$ ---
  for a homogeneous kernel, contraction of the metric with $z^{i}$
  shows that the metric is degenerate.}  More generally, if $P$ were not
positive-definite, then the metric would only be positive-definite in
certain regions in moduli space, and finding such regions is
computationally difficult when the dimension of the space is large.

One way to study a Bergman metric is to fix the matrix $P$, and
then scan over a neighborhood in the coordinates $z^i$.
 However, when $z^{i}\bar{z}^{\bar{i}}$ is large and
$k$ is finite, the kernel is dominated by the largest-power monomials
in
the basis,
and so $f$ is
well-approximated by a homogeneous function of degree $2k$.  As
discussed above, homogeneous kernels are degenerate, and so the metric
is guaranteed to have singularities for sufficiently large $z$.

An alternative to exploring a small region with a fixed kernel is to
generate a new $P_{\alpha\beta}$ at each point in moduli space, and
then redefine the coordinates such that the point under consideration
is $z^{i}=0$.  The philosophy is quite similar to the i.i.d.~approach
in that we assume that we work with an ensemble of Hessians generated
at points in moduli space that have a separation that is large in
comparison to the correlation length of $P$.  Therefore, as in the
i.i.d.~approach, we lose information about correlations between
Hessians at different points but retain correlations between the
various derivatives of the K\"ahler potential at each
point.\footnote{Correlations between different points could be
  incorporated, at least in a statistical sense, by implementing the
  techniques of~\cite{Bachlechner:2014rqa}.}  Since calculation of the
Riemann tensor of a K\"ahler metric requires no more than two
holomorphic or anti-holomorphic derivatives (see~\eqref{eq:Riemann})
and we are evaluating all tensors at $z=0$, we can take the
$s_{\alpha}$ to be quadratic polynomials.

We now consider an ensemble of Bergman metrics as described above,
taking $P$ to be a random Wishart matrix.  The Riemann tensor can be
calculated entirely in terms of $P_{\alpha\beta}$.  In order to
construct the associated contribution to the Hessian ${\cal
  H}_R^{\uB}$ we need to include F-terms. As in the
O3/O7 case, we take the F-terms to be
\begin{equation}
  F_{a}\in F\,
  \Omega\bigl(0,\sig\bigr),\quad \sig=1/\sqrt{N}\,,
\end{equation}
and then report on the distribution of eigenvalues of the
canonically-normalized and rescaled curvature Hessian
(cf.~\eqref{eq:defHRothree}).

\begin{figure}
\begin{center}
  \begin{subfigure}{0.45\textwidth}
    \psfrag{a}[Bc]{$\la{.02}$}
    \psfrag{b}[Bc]{$\la{.04}$}
    \psfrag{c}[Bc]{$\la{.06}$}
    \psfrag{d}[Bc]{$\la{.08}$}
    {\includegraphics{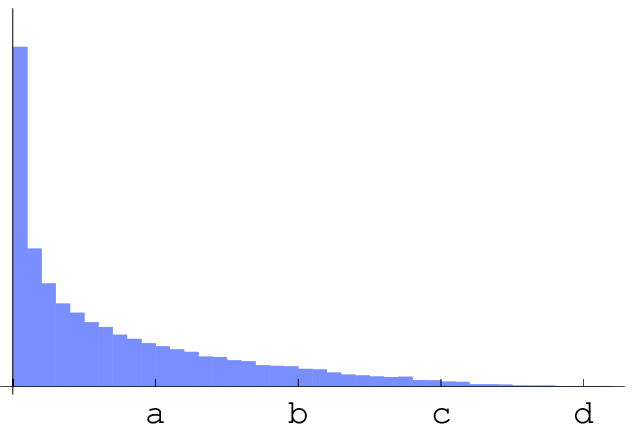}}
    \caption{$\cH_{\cK^{\left(3\right)}}^{\uB}$}
      \end{subfigure}
  \quad\quad
  \begin{subfigure}{0.45\textwidth}
    \psfrag{a}[Bc]{$\la{-1.75}$}
    \psfrag{b}[Bc]{$\la{-1.0}$}
    \psfrag{c}[Bc]{$\la{-0.25}$}
    \psfrag{d}[Bc]{$\la{0.5}$}
    {\includegraphics{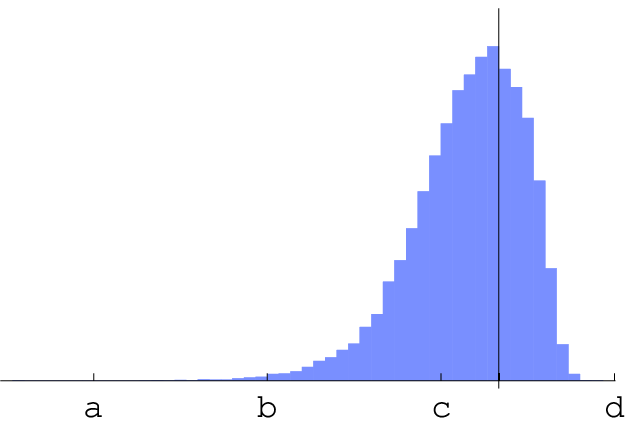}}
  \caption{$\cH_{\cK^{\left(4\right)}}^{\uB}$}
      \end{subfigure}
  \caption{\label{fig:berg_pieces}Eigenvalue spectra for
    the $\cK^{(3)}$ and $\cK^{(4)}$ contributions to the Hessian in the
    case of Bergman metrics.  Each ensemble consists of 1000 $20\times
    20$ matrices.}
\end{center}
\end{figure}

\begin{figure}
\begin{center}
  \begin{subfigure}{0.45\textwidth}
    \psfrag{a}[Bc]{$\la{-1.5}$}
    \psfrag{b}[Bc]{$\la{-0.5}$}
    \psfrag{c}[Bc]{$\la{.0.5}$}
    {\includegraphics{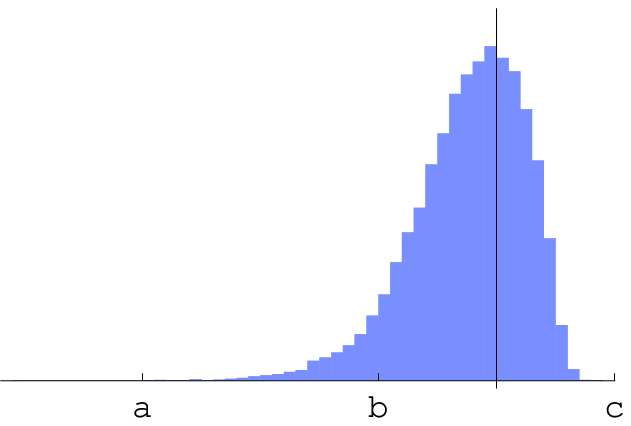}}
    \caption{Eigenvalues of $\cH^{\uB}_R$}
      \end{subfigure}
  \quad\quad
  \begin{subfigure}{0.45\textwidth}
    \psfrag{a}[Bc]{$\la{9}$}
    \psfrag{b}[Bc]{$\la{11}$}
    \psfrag{c}[Bc]{$\la{13}$}
    \psfrag{d}[Br]{$\la{-20}$}
    \psfrag{e}[Br]{$\la{-60}$}
    \psfrag{f}[Br]{$\la{-100}$}
    \psfrag{R}[Bc]{$\Ri$}
    \psfrag{S}[Bc]{$R$}
    {\includegraphics{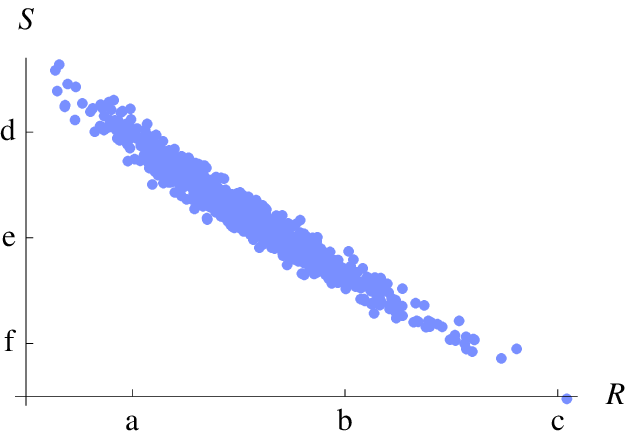}}
    \caption{Curvature invariants}
          \end{subfigure}
 \caption{\label{fig:bergman_spectra}Distribution of eigenvalues for
   ${\cal H}_R^\uB$ at $h^{1,1} = 20$, and the  correlation between the curvature
   invariants.}
    \end{center}
\end{figure}

The spectrum of eigenvalues of $\cH_{R}^{\uB}$ has qualitative
features that resemble those of the Calabi-Yau hypersurface
examples. In particular, the log-type K\"{a}hler potential leads to
correlations between $\cH_{\cK^{\left(3\right)}}^\uB$ and
$\cH_{\cK^{\left(4\right)}}^\uB$, as shown in
Figure~\ref{fig:berg_pieces}, that are very similar to the
correlations shown in Figure~\ref{fig:h11-25-k3k4}.  In addition, both
the spectrum of ${\cal H}_R^\uB$ and the correlation between curvature
invariants shown in Figure~\ref{fig:bergman_spectra} mirror the
Calabi-Yau case. Since the curvature invariants are small, the tails
are not as extensive as in the Calabi-Yau case. Indeed, repeating the
analysis in \S\ref{sec:heavytails}, we find no significant evidence
for the tails to be heavy.\footnote{If one instead takes the entries
  of the $P_{\alpha\beta}$ appearing in the Bergman K\"ahler
  potential~\eqref{eq:Bergman} to be drawn from a heavy-tailed
  distribution, the resulting metric spectra can have heavy tails.}

The Bergman metric has features that make it appealing as a
model for moduli space metrics: it incorporates the correlations
between $\cK_{(3)}$ and $\cK_{(4)}$, and appears to have similar
relationships between $R$ and $\Ri$ as in the Calabi-Yau
case. However, although fluctuations to positivity in $\cH_{R}^{\uB}$
are rare, they are not prevented from occurring.  Furthermore, at
large ``volume'' $f$, where the argument of the logarithmic K\"{a}hler
potential is well-modeled by a degree-three homogeneous function, one
should expect a non-positive spectrum of ${\cal H}_R$.  Said
differently, it is not clear how to connect the preferred coordinates
in terms of which the volume is homogeneous to the coordinates used in
the Bergman metric.

\section{\label{sec:axions}Spectra of Axion Decay Constants}

A significant application of our results is to the problem of
determining axion decay constants.  The axion decay constant spectrum
is relevant for a wide range of questions, including analysis of the
``axiverse'' proposed in~\cite{Arvanitaki:2009fg}, the study of axion
dark matter, late time decays, etc.  For the particular purpose of
developing models of inflation in string theory, it is important to
understand whether axion decay constants can reach or exceed the
Planck scale, $f \gtrsim M_{\rm pl}$.  While axion decay constants in
string theory have been the subject of intensive study
(see~\cite{Banks:2003sx,*Svrcek:2006yi} and the overview in
\cite{Baumann:2014nda}), most results to date have involved either
single axions; pairs of axions; or, in systems with $N\gg 1$ axions,
``democratic'' estimates in which the $N$ decay constants are taken to
be essentially comparable and
uncorrelated~\cite{Dimopoulos:2005ac,Easther:2005zr}.  Here we
initiate a more systematic study of the {\it{statistics}} of axion
decay constants, incorporating the critical phenomenon of eigenvalue
repulsion.

In a compactification of type~IIB string theory on a Calabi-Yau
orientifold with O3-plane and O7-plane fixed loci, the intersection
numbers determine the classical effective action for the K\"ahler
moduli $\rho^{a} = \tau^{a} + i\, c^{a}$, including in particular the
axions\footnote{The same data determine the classical action for
  axionic fields descending from $C_2$ and $B_2$, in compactifications
  with $h^{1,1}_{-}>0$, but for simplicity we will describe only the
  case $h^{1,1}_{-}=0$.}
\begin{equation}
  c^{a} = \int_{D^{a}} C_4\,.
\end{equation}
The kinetic term for these axions is given by the K\"ahler metric
$\cK_{a\bar{b}}$ (\ref{eq:metric}): taking the $\tau^a$ to be constants,
the low-energy Lagrangian includes
\begin{equation}
  {\cal L} \supset \cK_{a\bar{b}}(\tau^c) \, \partial c^{a}\partial c^{b}
  - \sum_a \Lambda_a^4(\tau^c)\,
  {\rm{cos}}(c_a/\zeta_a)\,.  \label{equ:polyaxion}
\end{equation}
where the potential term represents instanton contributions that break
the perturbative shift symmetries $c_a \to c_a + const.$ to discrete
shifts $c_a \to c_a + 2\pi \zeta_a$.  In the absence of a preferred
basis for the K\"ahler moduli space, one could rescale the $\rho_a$ to
arrange that each of the $c_a$ has period $2\pi$, i.e.~setting
$\zeta_a \to 1$, but we will not do so here: instead, the metric
$\cK_{a\bar{b}}(\tau^c)$ is taken to be expressed in the basis used
throughout this work.

Determining the decay
constant $f$ in a model with $h^{1,1}_+=1$ and a
corresponding single axion $c$ is straightforward: defining
$\cK_{c\bar{c}}=f^2$, the Lagrangian takes the form
\begin{equation}
  {\cal L} = \frac{1}{2} f^2 (\partial c)^2 -
  \Lambda^{4}_c \, {\rm{cos}}(c/\zeta)\,.
\end{equation}
so that by defining $\phi=f c$ we arrive at
\begin{equation}
  {\cal L} = \frac{1}{2} (\partial \phi)^2 -
  \Lambda^{4}_c \, {\rm{cos}}\left(\frac{\phi}{f\zeta}\right)\,.
\end{equation}
The periodicity of the canonically-normalized field $\phi$ is then
$2\pi f\zeta$.  As explained in \cite{Bachlechner:2014hsa}, in
generalizing this result to multiple axions one must carefully account
for kinetic mixing, i.e. off-diagonal terms $\cK_{a\bar{b}}$, $b\neq
a$: the orthogonal rotation that diagonalizes $\cK_{a\bar{b}}$ does
not preserve the form of the potential in (\ref{equ:polyaxion}), in
which each field appears in only a single cosine term in the
potential.  Nevertheless, we will define the decay constants in an
$N$-axion system to be the (positive square roots of the) eigenvalues
$\lambda_a$ of the K\"ahler metric: $f_a = \sqrt{\lambda_a}$.  We will
order the eigenvalues for convenience, $f_1 \le f_2 \le \ldots \le
f_N$, with $N=h^{1,1}$.

A natural goal would be to derive a matrix model for the K\"ahler
metric analogous to the Wigner+Wishart+Wishart matrix model for the
Hessian obtained in~\cite{Marsh:2011aa}, and to obtain analytic
results for the eigenvalue spectrum, and hence for the spectrum of
decay constants.  This problem is beyond the scope of the present
work, and we limit ourselves to conclusions that can be extracted from
the ensemble of K\"ahler metrics presented in \S\ref{sec:results}.

For the purpose of characterizing the spectra of axion decay
constants, the principal qualitative lesson of our analysis is that
the eigenvalue spectrum of the K\"ahler metric displays a pronounced
tail to the right of the median.  Because the
extreme eigenvalues are very different from typical eigenvalues, a
computation (or an estimate of the moduli dependence of) an
{\it{individual}} decay constant $f_{\star}$, along the lines
of~\cite{Banks:2003sx,*Svrcek:2006yi}, does not directly constrain the
full spectrum.  In particular, depending on whether $f_{\star}$ is a
typical member of the spectrum, or belongs to the left or right tail,
the remaining decay constants may extend over several decades in
either direction.

Large outliers, i.e.~decay constants belonging to the right tail of the spectrum,  provide the tantalizing prospect of achieving super-Planckian decay constants in a compactification in which all parametric expansions are under good control, and {\it{typical}}  decay constants are very small in Planck units.  However, although super-Planckian effective decay constants  appear possible  via collective excitations
in systems with $N \gg 1$ axions \cite{Dimopoulos:2005ac} (and can be enlarged via kinetic alignment  as in \cite{Bachlechner:2014hsa})  or via decay constant  alignment \cite{Kim:2004rp}  of two axions,   in neither  of these cases  is a super-Planckian eigenvalue of the K\"ahler metric  responsible for the large decay constant.  In generic  examples in the interior of the K\"ahler  cone,
the parametric bound $f\lesssim 1$ of~\cite{Banks:2003sx} appears to apply to the maximal eigenvalues $f_N$ in our ensemble, not to the typical eigenvalues.   Identifying non-generic exceptions  with super-Planckian decay constants,  while ensuring the validity of the $\alpha^{\prime}$  expansion, is an  intriguing problem for future work.

\newpage

\section{\label{sec:conc}Conclusions}

The primary results of this work are statistical models for the metric
on K\"ahler moduli space in string compactifications on Calabi-Yau
hypersurfaces in toric varieties, and for the corresponding Riemann
curvature contribution ${\cal H}_R$ to the Hessian
matrix.\footnote{Our most significant results were obtained for O3/O7
  orientifolds of type~IIB, but other string theories were considered
  in~\S\ref{sec:special_geometry}.}  Our starting point was the
assumption made in \cite{Marsh:2011aa} (and implicitly in much earlier
work) that for the purpose of studying vacuum statistics, the
derivatives of $\cK$ may be taken to be independent and identically
distributed random variables.  At the level of random matrix models,
assuming i.i.d.~entries allows one to model ${\cal H}_R$ as the sum of
a Wishart matrix and a Wigner matrix \cite{Marsh:2011aa},
\begin{equation}
  {\cal H}^{\iid}_R \approx {\cal H}^{\WW}_{R} \, ,
\end{equation}
for which the eigenvalue spectrum can be computed analytically, and
the fluctuation probabilities can be obtained.  In this work we did
not take the derivatives of $\cK$ to be independent,
but instead accounted for the correlations that follow from the
well-known form of the classical K\"ahler potential,
\begin{equation}
  \cK = - 2  \log {\cal{V}} =
  -2\log\, \biggl(\frac{1}{3!}\,\ka^{abc}t_{a}t_{b}t_{c}\biggr)\,.
  \label{classicalK}
\end{equation}
One key finding at this level is that although ${\cal H}^{\WW}_{R}$
has a nonvanishing (albeit small) probability to fluctuate to
positivity, ${\cal H}_R^{\Othree}$ is {\it{necessarily non-positive}},
with at least one negative eigenvalue.  Thus, incorporating the
correlations in $\cK$ reduces the probability of
metastability.\footnote{See \cite{Farquet:2012cs} for a study of the
  incidence of instabilities in a related class of geometries.}

The non-positivity of ${\cal H}_R^{\Othree}$ follows directly from the
form (\ref{classicalK}), combined with the fact that ${\cal V}$ is a
homogeneous function of the K\"ahler moduli. No knowledge of the
intersection numbers $c^{ijk}$ is required.  On the other hand,
bounding a single eigenvalue is less informative than a statistical
description of the entire eigenvalue spectrum, which does require
additional geometric input.

To go further, we examined the actual form of ${\cal H}^{\Othree}_R$
in an ensemble of compactifications on Calabi-Yau hypersurfaces in
toric varieties.  We found that ${\cal H}_R^{\Othree}$ is strikingly
different from ${\cal H}^{\WW}_{R}$: in particular, the eigenvalue
spectrum of ${\cal H}_R^{\Othree}$ displays {\it heavy tails} toward
negative eigenvalues (see Figures~\ref{fig:h11_25-HR},
\ref{fig:h11_25_HR_minmax}, and~\ref{fig:h11_25-HR_log_CY}), in
contrast to the well-localized spectrum of ${\cal H}^{\WW}_{R}$
(Figures~\ref{fig:WW_spectra} and~\ref{fig:h11_25-HR_log_WW}).  We
showed that the tails in ${\cal H}_R^{\Othree}$ are closely correlated
to heavy tails in the eigenvalues of the K\"ahler
metric\footnote{Recall that we worked in a fixed coordinate system
  defined by a specified basis of cycles, so the eigenvalues of the
  metric are meaningful.} (Figure~\ref{fig:h11-15-metric-maxes-bulk}),
which in turn are tied to hierarchies of 2-cycle volumes
(Figure~\ref{fig:wall_approach}).  Correspondingly, the tails in
${\cal H}_R^{\Othree}$ are largest at points in moduli space that are
close to a wall of the K\"ahler cone.  However, we argued that the
tails are not caused by failure of the $\alpha^{\prime}$ expansion:
heavy tails are present when there are modest hierarchies in cycle
volumes, even if all cycles are large in string units.

Our results shed light on the characteristic properties of mass
matrices and metrics on moduli space in Calabi-Yau compactifications
of string theory.  We discussed two significant implications: first,
the heavy tail toward negativity in ${\cal H}_R^{\Othree}$ very
plausibly leads to an increased incidence of instabilities in
comparison to the reference model ${\cal H}^{\WW}_{R}$.  Second, the
heavy tails in the eigenvalues of the K\"ahler metric imply that in
Calabi-Yau compactifications with many axions, the axion decay
constants are hierarchically separated, with the largest eigenvalue
orders of magnitude larger than the smallest eigenvalue.

Exploring the consequences of these findings, as well as extending our
results to broader classes of compactifications, are important
problems for the future.

\acknowledgments

We are grateful to Michael Douglas, Juan Maldacena, and Itamar Yaakov
for comments that provided inspiration for a portion of this work.  We
thank Lara Anderson, Thomas Bachlechner, James Gray, Jim Halverson,
Ben Heidenreich, Jim Sethna, Gary Shiu, John Stout, and Timm Wrase for
related discussions.  We are especially indebted to Julia Goodrich for
technical assistance.  This work was supported by NSF grant
PHY-0757868.

\appendix

\section{\label{app:toric}Toric Geometry}

In this appendix we provide a brief review of the aspects of toric
varieties and Calabi-Yau hypersurfaces that are used in our
work.  For more complete expositions of toric varieties
see~\cite{Witten:1993yc,Greene:1996cy,Cox:2000vi,Hori:2003ic,Denef:2008wq,coxnotes,*coxbook}.

A physical definition of a toric variety is as the moduli space of a
2d $\cN=\left(2,2\right)$ field theory.
In general, the moduli space of such a
theory is a K\"ahler manifold.  To specialize to a toric variety, we
consider a $\U{1}^{s}$ gauge group and $r$ chiral superfields $X^{i}$,
where the charge of the $i$th chiral superfield under the $a$th
$\U{1}$ factor is $Q_{i}^{a}$.  Each $\U{1}$ factor has a
Fayet-Iliopoulos term $\xi^{a}$, so that the D-flatness condition is
\begin{equation}
  \sum_{i}Q_{i}^{a}\bigl\lvert X^{i}\bigr\rvert=\xi^{a}\,.
\end{equation}
The resulting moduli space is a toric variety.

A useful collection of divisors are the toric divisors $\hat{T}^{i}$
defined by $X^{i}=0$.  Because of the $\U{1}^{s}$ gauge symmetry, the
coordinates $X^{i}$ on $\CC^{r}$, known as homogeneous coordinates,
are not functions on the toric variety.  On the other hand,
gauge-invariant combinations of the $X^{i}$ are well-defined
functions, and the corresponding vanishing loci are homologically
trivial.  Thus, not all of the toric divisors $\hat{T}^{i}$ are
homologically independent, and indeed the number of independent
divisors is equal to the rank of the gauge group.  We will use
$\{\hat{D}^{i}\}$ to refer to a set of independent toric divisors.  To
our knowledge, there is no natural or canonical basis choice for the
$\hat{D}^{i}$, and we work in the bases adopted in the software
packages that we use.  The intersection numbers among the toric
divisors $\hat{T}^{i}$ (and therefore the independent divisors
$\hat{D}^{i}$) can be calculated by simultaneously solving $X^{i}=0$
subject to the D-term constraint.

The Chern class of a toric variety $V$ is
\begin{equation}
  c\bigl(V\bigr)=\prod_{i=1}^{r}\bigl(1+\bigl[\hat{T}^{i}\bigr]\bigr)\,,
\end{equation}
where $\bigl[\hat{T}\bigr]$ denotes the Poincar\'e dual of $\hat{T}$.
From the adjunction formula, it follows that the hypersurface
\begin{equation}
  S=\sum_{i}\hat{T}^{i}  \label{defhypersurface}
\end{equation}
has vanishing first Chern class, and so is Calabi-Yau.  Calabi-Yau
hypersurfaces of the form~\eqref{defhypersurface} are the only
Calabi-Yau manifolds considered in our analysis.

The intersections of toric divisors $\hat{D}^{a}$ with the Calabi-Yau
hypersurface furnish divisors of the hypersurface.  When all of the
divisors of the Calabi-Yau are inherited from the ambient toric
variety in this way, the polytope is said to be {\it{favorable}}, and
we limit our analysis to such cases.  One can  easily
check whether a
  polytope is favorable by
 calculating $h^{1,1}$ from the
  toric data~\cite{Batyrev}.

The classes of holomorphic curves in an $n$-dimensional toric variety
form a cone called the Mori cone,  which can be determined by considering
the intersections of $\left(n-1\right)$ toric divisors.  The  Mori cone is
generated by a set of curves $\hat{C}_{i}$, and the intersections
between the divisors of the toric variety and these curves
are given by the charge matrix
\begin{equation}
  Q_{i}^{a}=\hat{C}_{i}\cdot \hat{D}^{a}\,.
\end{equation}
Note that the Mori cone of the Calabi-Yau is not in general the same
as the Mori cone of the ambient toric variety.  For example, if a
curve of the ambient variety does not intersect the Calabi-Yau
hypersurface, then the curve may be safely taken to zero volume and
even be allowed to flop.\footnote{It has been conjectured that after
  taking into account this additional freedom, the K\"ahler cone of a
  Calabi-Yau will always be simplicial (see, for example,~\cite{Cox:2000vi}).} However, positivity of curves
of the ambient space implies positivity of the curves of the
hypersurface.\footnote{See~\cite{Braun} for more work on determining
  the Mori cone of the hypersurface itself.}

The data of a compact toric variety can be encoded by a polytope,
which is a higher-dimensional generalization of a polyhedron.
We will consider $n$-dimensional polytopes
$\Delta$ that are {\it{reflexive}},
meaning that the only point of $\ZZ^{n}$ that is in the
interior of $\Delta$ is the origin, and that $\Delta$ can be expressed as an
intersection of half-spaces
\begin{equation}  \label{reflexivepolydef}
  \Delta=\bigcap_{F}\bigl\{\mathbf{m}\in\RR^{n}\,\mid\,
  \mathbf{m}\cdot \mathbf{n}_{F}\ge -1\bigr\}\,,
\end{equation}
where $F$ denotes a codimension-$1$ face of $\Delta$, and
$\mathbf{n}_{F}$ is the shortest vector in $\ZZ^{n}$ that is an inward-facing normal to $F$.
Given a polytope $\Delta$, its dual polytope $\Delta^{\circ}$ is
defined by taking the points $\mathbf{n}_{F}$ as the vertices of
$\Delta^{\circ}$.  When $\Delta$ is reflexive, $\Delta^{\circ}$ is
reflexive.

The next step is to triangulate $\Delta^{\circ}$: that is, we divide
$\Delta^{\circ}$ into simplices.  In particular, we consider
triangulations of $\Delta^{\circ}$ that are {\it{star}}, which means
that all simplices share one common point as a vertex. For our
purposes this point will be the origin, as each simplex defines a cone
with the tip being the origin, and one can form a fan by pasting
together such cones.  To each line segment $\mathbf{v}^{i}$ that
connects to the origin, we associate one of the variables $X^{i}$.
The $\mathbf{v}^{i}$ satisfy a set of linear relations,
\begin{equation}
  Q^{a}_{i}\mathbf{v}^{i}=0\,,
\end{equation}
which give the charge matrices.  Note that there are multiple ways of
triangulating a polytope, and thus a polytope does not define a unique
toric variety.  However, the various varieties are related to each
other via blowups and blow-downs.

A necessary condition for a non-empty K\"{a}hler cone is that the triangulation is {\it{regular}}, meaning that the
triangulation is a projection of the bottom part of an
$\left(n+1\right)$-dimensional polytope.  More precisely, let
$\left\{\mathbf{m}_{k}\right\}$ be the points of the triangulation of
$\Delta$.  Then the triangulation is regular if there exists an $\left(n+1\right)$-dimensional polytope
$\Xi$ containing points $\left(\mathbf{m}_{k},\lambda_{k}\right)$ whose inward-facing normals
have positive $\left(n+1\right)$-dimensional components.

Toric varieties are generally singular, but when the triangulation
is {\it{maximal}}, meaning that every point in $\ZZ^{n}$ that is in $\Delta$ is a vertex
of a simplex of the triangulation, then the variety becomes smooth.  In
such a situation, the Calabi-Yau hypersurface will also be smooth.
In fact, total smoothness of the variety is not necessary to ensure
 smoothness of a generic hypersurface: at generic points in the complex
 structure moduli space of the Calabi-Yau, the hypersurface will miss pointlike
singularities of the ambient variety.  Thus, our triangulations need not include
the points in the interior of codimension-1 faces (facets) of the polytope.
We are therefore led to consider complete regular star
triangulations of favorable reflexive polytopes,
 ignoring points interior to facets.

\section{\label{app:algorithm}Triangulation Algorithm}

A common way to obtain complete, regular, star triangulations is to
calculate all complete, regular triangulations and then select those
that are star.  This is extremely inefficient if one is interested
only in star triangulations. In order to directly obtain
triangulations that are star, we recall that by definition, each point
in a star triangulation is connected to the origin by a line.  One can
therefore obtain a complete star triangulation by triangulating each
of the facets, drawing a line from the origin to each point on the
facets, and enforcing that the triangulations of each facet agree on
the overlaps of facets.

The  resulting triangulation  is automatically
maximal and star, but  it is still  necessary to check for regularity. We present our algorithm below,  illustrated for concreteness in the case of a polytope $P$ with four facets, $F_a, F_b, F_c$, and $F_d$:
\begin{enumerate}
\item Find all regular, complete triangulations of each facet.  We
  obtained these triangulations from
  \texttt{TOPCOM}~\cite{Rambau:TOPCOM-ICMS:2002}, using \texttt{Sage}
  as an interface.

\item Organize the facets into pairs, such that each facet in a pair
  has nonzero intersection with its partner.  If facets $F_a$ and
  $F_b$ intersect, and facets $F_c$ and $F_d$ intersect, create the
  pairs $(F_a,F_b)$ and $(F_c,F_d)$.
\item For each pair of facets, compare the triangulations of each
  member of the pair on the intersection, retaining only those
  triangulations that are regular and agree on the overlap.  That is,
  let $S$ be the overlap of the facets $F_a$ and $F_b$, and let
  $\Delta_a$ and $\Delta_b$ be triangulations of facets $F_a$ and
  $F_b$, respectively. Determine whether $\Delta_a$ and $\Delta_b$
  agree on $S$, and whether $\Delta_a \cup \Delta_b :=
  \Delta_{ab}$ is regular (including the lines drawn to the origin).
  Retain $\Delta_{ab}$ if and only if it is regular.  Repeat this
  process for all other pairs of facets.
\item Combine $F_a$ and $F_b$ into a single \emph{intermediate
  polytope} $F_{ab}$ with regular triangulations $\{\Delta^i_{ab}\}$,
  and combine $F_c$ and $F_d$ into an intermediate polytope $F_{cd}$
  with regular triangulations $\{\Delta^{i^{\prime}}_{cd}\}$. Do the
  same for all other pairs of facets.
\item Repeat steps 2, 3, and 4 for the new pairs, such as $(F_{ab},F_{cd})$, until the full polytope is triangulated.
\end{enumerate}

\begin{figure}
\begin{center}
 \psfrag{a}[Bc]{$a$}
    \psfrag{b}[Bc]{$b$}
    \psfrag{c}[Bc]{$c$}
    \psfrag{d}[Bc]{$d$}
    \psfrag{F1}[Bc]{Facet 1}
    \psfrag{F2}[Bc]{Facet 2}
  {\includegraphics{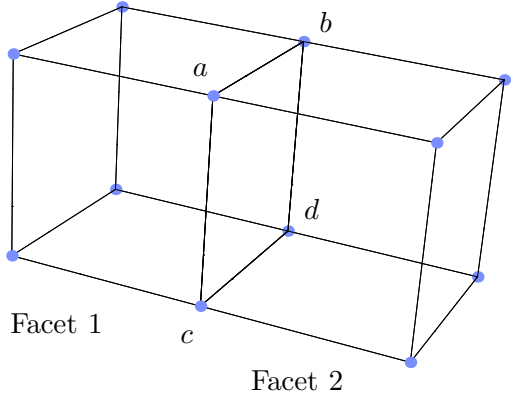}}
  \caption{\label{fig:boxes}Facet 1 and Facet 2 are three-dimensional
    facets of a four-dimensional polytope.  The facets overlap along
    the surface $ \langle abcd \rangle$.}
\end{center}
\end{figure}

The above method yields all complete, regular, star triangulations of
the polytope $P$. We will illustrate step $2$ in toy example. Consider
a four-dimensional polytope, where two of the facets are cubes that
overlap on a square surface. In Figure~\ref{fig:boxes} we show the two
facets that overlap along the surface $\langle abcd \rangle$.

\begin{figure}
\begin{center}
 \psfrag{a}[Bc]{$a$}
    \psfrag{b}[Bc]{$b$}
    \psfrag{c}[Bc]{$c$}
    \psfrag{d}[Bc]{$d$}
    \psfrag{x}[Bc]{Facet 1}
    \psfrag{y}[Bc]{Facet 2}
     \psfrag{g}[Bc]{$\checkmark$}
  {\includegraphics{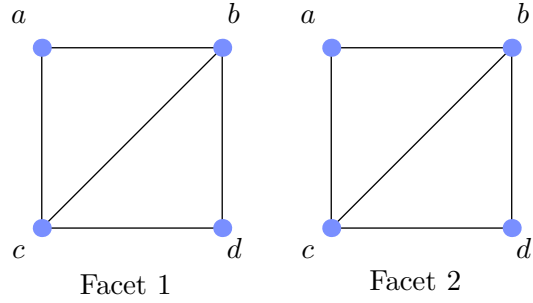}}
  \caption{\label{fig:good-tri}The triangulations $\Delta_1$ and $\Delta_2$
  agree on the overlap $\langle abcd \rangle$.}
\end{center}
\end{figure}

\begin{figure}
\begin{center}
    \psfrag{a}[Bc]{$a$}
    \psfrag{b}[Bc]{$b$}
    \psfrag{c}[Bc]{$c$}
    \psfrag{d}[Bc]{$d$}
    \psfrag{x}[Bc]{Facet 1}
    \psfrag{y}[Bc]{Facet 2}
    \psfrag{g}[Bc]{$\checkmark$}
  {\includegraphics{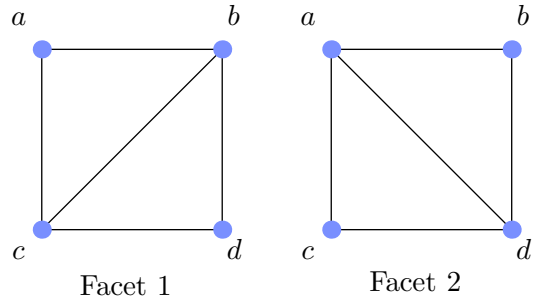}}
  \caption{\label{fig:bad-tri}The triangulations $\Delta_1$ and $\Delta_2$
  do not agree on the overlap $\langle abcd \rangle$.}
\end{center}
\end{figure}

After triangulating each facet individually, we need to ensure that
the triangulations agree on their overlap. Consider triangulations
of facet 1 and facet 2, labeled $\Delta_1$ and $\Delta_2$,
respectively. If $\Delta_1$ and $\Delta_2$ induce the same
triangulation on $\langle abcd \rangle$, as shown in
Figure~\ref{fig:good-tri}, then the pair of triangulations is good,
and we store the pair.  If instead the triangulations do not agree on
the overlap, as in Figure~\ref{fig:bad-tri}, we discard the pair.

A challenge in the above algorithm is that storing all
possible pairs of triangulations at each step becomes costly for
$h^{1,1} \gtrsim 15$.  To access a {\it{subset}} of complete, regular, star
triangulations in examples with larger $h^{1,1}$, one can keep only a
subset of the possible pairs at each step, in the following way.
\begin{enumerate}
\item Start with two overlapping facets $F_a$ and $F_b$, and determine all regular  triangulations of each.
Of the regular triangulations of each facet that agree on the overlap, retain the first $j_1$ pairs, for some number $j_1$.
\item Form the intermediate polytope $F_{ab}$.
\item Select another overlapping facet $F_c$,  and compute all regular  triangulations of  $F_c$.  Compare each of these to the $j$  triangulations of $F_{ab}$, retaining the first $j_2$
triangulations of $F_c$ that agree with a  triangulation of $F_{ab}$ on the overlap of $F_{ab}$ with $F_c$, and that are regular.
\item Repeat steps 2 and 3, adding in facets, until one has obtained  one or more triangulations of the full polytope.
\end{enumerate}
Since one only keeps $j_k$ pairs at step $k$, this second method is
not guaranteed to yield a triangulation of the full polytope: one can
obtain pairs early on for which there is no regular completion of the
triangulation. Even so, this method is much faster than the available
alternatives, particularly if the $j_k$ are chosen to be small.  We
have used this approach to obtain complete, regular, star
triangulations of polytopes with $h^{1,1}$ as large as 25 on a desktop
computer.  For $h^{1,1} \gtrsim 25$ even triangulating the individual
facets becomes expensive, and more sophisticated tools are required to
obtain triangulations.

We note that our method of triangulation is most successful with
polytopes that have many facets. In typical cases, the more facets there
are, the fewer triangulations each facet admits, so that the time to triangulate each individual
facet is decreased. This gives the algorithm more opportunities
to fail at each step, but since the method is fast once all of the facet
triangulations are found, we nevertheless find the greatest success for many-faceted polytopes.

After a triangulation is found the resulting fan can be provided as input
to \texttt{Sage}, which then constructs the corresponding toric
variety data, including the intersection ring and Mori cone
generators  of the ambient toric space. The intersection ring  can then be
pulled back to the Calabi-Yau hypersurface.

\bibliography{lmm}

\providecommand{\href}[2]{#2}\begingroup\raggedright\begin{thebibliography}{10}

\bibitem{Douglas:2003um}
M.~R. Douglas, ``{The Statistics of String/M Theory Vacua}.''  {\em JHEP} {\bf
  0305} (2003) 046, [\href{http://arxiv.org/abs/hep-th/0303194}{{\tt
  hep-th/0303194}}].

\bibitem{Grana:2005jc}
M.~Gra{\~n}a, ``{Flux Compactifications in String Theory: A Comprehensive
  Review}.''  {\em Phys.Rept.} {\bf 423} (2006) 91--158,
  [\href{http://arxiv.org/abs/hep-th/0509003}{{\tt hep-th/0509003}}].

\bibitem{Douglas:2006es}
M.~R. Douglas and S.~Kachru, ``{Flux Compactification}.''  {\em Rev.Mod.Phys.}
  {\bf 79} (2007) 733--796, [\href{http://arxiv.org/abs/hep-th/0610102}{{\tt
  hep-th/0610102}}].

\bibitem{Denef:2008wq}
F.~Denef, ``{Les Houches Lectures on Constructing String Vacua}.''
  \href{http://arxiv.org/abs/0803.1194}{{\tt arXiv:0803.1194}}.

\bibitem{Ashok:2003gk}
S.~Ashok and M.~R. Douglas, ``{Counting Flux Vacua}.''  {\em JHEP} {\bf 0401}
  (2004) 060, [\href{http://arxiv.org/abs/hep-th/0307049}{{\tt
  hep-th/0307049}}].

\bibitem{Denef:2004ze}
F.~Denef and M.~R. Douglas, ``{Distributions of Flux Vacua}.''  {\em JHEP} {\bf
  0405} (2004) 072, [\href{http://arxiv.org/abs/hep-th/0404116}{{\tt
  hep-th/0404116}}].

\bibitem{Denef:2004cf}
F.~Denef and M.~R. Douglas, ``{Distributions of Nonsupersymmetric Flux
  Vacua}.''  {\em JHEP} {\bf 0503} (2005) 061,
  [\href{http://arxiv.org/abs/hep-th/0411183}{{\tt hep-th/0411183}}].

\bibitem{Douglas:2004zu}
M.~R. Douglas, B.~Shiffman, and S.~Zelditch, ``{Critical Points and
  Supersymmetric Vacua}.''  {\em Commun. Math. Phys.} {\bf 252} (2004)
  325--358, [\href{http://arxiv.org/abs/math/0402326}{{\tt math/0402326}}].

\bibitem{Ferrari:2011we}
F.~Ferrari, S.~Klevtsov, and S.~Zelditch, ``{Random Geometry, Quantum Gravity
  and the K\"ahler Potential}.''  {\em Phys.Lett.} {\bf B705} (2011) 375--378,
  [\href{http://arxiv.org/abs/1107.4022}{{\tt arXiv:1107.4022}}].

\bibitem{Ferrari:2011is}
F.~Ferrari, S.~Klevtsov, and S.~Zelditch, ``{Random K{\"a}hler Metrics}.''
  {\em Nucl.Phys.} {\bf B869} (2013) 89--110,
  [\href{http://arxiv.org/abs/1107.4575}{{\tt arXiv:1107.4575}}].

\bibitem{Ferrari:2011sf}
F.~Ferrari, S.~Klevtsov, and S.~Zelditch, ``{Simple Matrix Models for Random
  Bergman Metrics}.''  {\em J.Stat.Mech.} {\bf 2012} (2012) P04012,
  [\href{http://arxiv.org/abs/1112.4382}{{\tt arXiv:1112.4382}}].

\bibitem{Marsh:2011aa}
D.~Marsh, L.~McAllister, and T.~Wrase, ``{The Wasteland of Random
  Supergravities}.''  {\em JHEP} {\bf 1203} (2012) 102,
  [\href{http://arxiv.org/abs/1112.3034}{{\tt arXiv:1112.3034}}].

\bibitem{Bachlechner:2014rqa}
T.~C. Bachlechner, ``{On Gaussian Random Supergravity}.''  {\em JHEP} {\bf
  1404} (2014) 054, [\href{http://arxiv.org/abs/1401.6187}{{\tt
  arXiv:1401.6187}}].

\bibitem{KS_database}
M.~Kreuzer and H.~Skarke, ``{Calabi-Yau Data}.''
  {\url{http://hep.itp.tuwien.ac.at/~kreuzer/CY/}}.

\bibitem{Gray:2012jy}
J.~Gray, Y.-H. He, V.~Jejjala, B.~Jurke, B.~D. Nelson, and J.~Sim{\'o}n,
  ``{Calabi-Yau Manifolds with Large Volume Vacua}.''  {\em Phys.Rev.} {\bf
  D86} (2012) 101901, [\href{http://arxiv.org/abs/1207.5801}{{\tt
  arXiv:1207.5801}}].

\bibitem{Balasubramanian:2005zx}
V.~Balasubramanian, P.~Berglund, J.~P. Conlon, and F.~Quevedo, ``{Systematics
  of Moduli Stabilisation in Calabi-Yau Flux Compactifications}.''  {\em JHEP}
  {\bf 0503} (2005) 007, [\href{http://arxiv.org/abs/hep-th/0502058}{{\tt
  hep-th/0502058}}].

\bibitem{Banks:2003sx}
T.~Banks, M.~Dine, P.~J. Fox, and E.~Gorbatov, ``{On the Possibility of Large
  Axion Decay Constants}.''  {\em JCAP} {\bf 0306} (2003) 001,
  [\href{http://arxiv.org/abs/hep-th/0303252}{{\tt hep-th/0303252}}].

\bibitem{Svrcek:2006yi}
P.~Svr{\u c}ek and E.~Witten, ``{Axions in String Theory}.''  {\em JHEP} {\bf
  0606} (2006) 051, [\href{http://arxiv.org/abs/hep-th/0605206}{{\tt
  hep-th/0605206}}].

\bibitem{Wigner}
E.~P. Wigner, ``{On the Statistical Distribution of the Widths and Spacings of
  Nuclear Resonance Levels}.''  {\em {Math. Proc. Cambridge}} {\bf 47} (1951)
  790--798. {DOI:
  \href{http://dx.doi.org/10.1017/S0305004100027237}{10.1017/S0305004100027237}}.

\bibitem{Wigner2}
E.~P. Wigner, ``{Characteristic Vectors of Bordered Matrices with Infinite
  Dimensions}.''  {\em {Ann. Math.}} {\bf 62} (1955) 548--564. {JSTOR:
  \url{http://www.jstor.org/stable/1970079}}.

\bibitem{Wigner3}
E.~P. Wigner, ``{Results and Theory of Resonance Absorption}.''  in {\em
  {Gatlinberg Conference on Neutron Physics by Time-of-Flight}}, pp.~59--70,
  Oak Ridge National Laboratory, 1957.

\bibitem{Wigner4}
E.~P. Wigner, ``{Statistical Properties of Real Symmetric Matrices with Many
  Dimensions}.''  in {\em {Proceedings of the Fourth Canadian Mathematical
  Congress}}, pp.~174--184.
\newblock {University of Toronto Press}, 1957.

\bibitem{Wishart}
J.~Wishart, ``{The Generalised Product Moment Distribution in Samples from a
  Normal Multivariate Population}.''  {\em Biometrika} {\bf 20A} (1928) 32--52.

\bibitem{Kachru:2003aw}
S.~Kachru, R.~Kallosh, A.~D. Linde, and S.~P. Trivedi, ``{de Sitter Vacua in
  String Theory}.''  {\em Phys.Rev.} {\bf D68} (2003) 046005,
  [\href{http://arxiv.org/abs/hep-th/0301240}{{\tt hep-th/0301240}}].

\bibitem{Grimm:2004uq}
T.~W. Grimm and J.~Louis, ``{The Effective Action of $N=1$ Calabi-Yau
  Orientifolds}.''  {\em Nucl.Phys.} {\bf B699} (2004) 387--426,
  [\href{http://arxiv.org/abs/hep-th/0403067}{{\tt hep-th/0403067}}].

\bibitem{Blumenhagen:2006ci}
R.~Blumenhagen, B.~K{\"o}rs, D.~L{\"u}st, and S.~Stieberger,
  ``{Four-Dimensional String Compactifications with D-Branes, Orientifolds and
  Fluxes}.''  {\em Phys.Rept.} {\bf 445} (2007) 1--193,
  [\href{http://arxiv.org/abs/hep-th/0610327}{{\tt hep-th/0610327}}].

\bibitem{Conlon:2006gv}
J.~P. Conlon, ``{Moduli Stabilisation and Applications in IIB String Theory}.''
   {\em Fortsch.Phys.} {\bf 55} (2007) 287--422,
  [\href{http://arxiv.org/abs/hep-th/0611039}{{\tt hep-th/0611039}}].

\bibitem{PALP}
M.~Kreuzer and H.~Skarke, ``{PALP: A Package for Analyzing Lattice Polytopes
  with Applications to Toric Geometry}.''  {\em Comput.Phys.Commun.} {\bf 157}
  (2004) 87--106, [\href{http://arxiv.org/abs/math/0204356}{{\tt
  math/0204356}}].

\bibitem{PALP2}
A.~P. Braun, J.~Knapp, E.~Scheidegger, H.~Skarke, and N.-O. Walliser,
  ``{PALP---A User Manual}.''  \href{http://arxiv.org/abs/1205.4147}{{\tt
  arXiv:1205.4147}}.

\bibitem{Dine:1985he}
M.~Dine and N.~Seiberg, ``{Is the Superstring Weakly Coupled?}''  {\em
  Phys.Lett.} {\bf B162} (1985) 299. {DOI:
  \href{http://dx.doi.org/10.1016/0370-2693(85)90927-X}{10.1016/0370-2693(85)90927-X}}.

\bibitem{Rummel:2013yta}
M.~Rummel and Y.~Sumitomo, ``{Probability of Vacuum Stability in Type IIB
  Multi-K{\"a}hler Moduli Models}.''  {\em JHEP} {\bf 1312} (2013) 003,
  [\href{http://arxiv.org/abs/1310.4202}{{\tt arXiv:1310.4202}}].

\bibitem{Abrevaya}
J.~Abrevaya and W.~Jiang, ``{A Nonparametric Approach to Measuring and Testing
  Curvature}.''  {\em Journal of Business and Economic Statistics} {\bf 23}
  (205) 1--19. {JSTOR: \url{http://www.jstor.org/stable/27638790}.}

\bibitem{Wilsonsign}
P.~Wilson, ``{Sectional Curvatures of K\"ahler Moduli}.''  {\em {Mathematical
  Annalen}} {\bf 330} (2004) 631--664,
  [\href{http://arxiv.org/abs/math/0307260}{{\tt math/0307260}}].

\bibitem{Ooguri:2006in}
H.~Ooguri and C.~Vafa, ``{On the Geometry of the String Landscape and the
  Swampland}.''  {\em Nucl.Phys.} {\bf B766} (2007) 21--33,
  [\href{http://arxiv.org/abs/hep-th/0605264}{{\tt hep-th/0605264}}].

\bibitem{Wilsonsign2}
T.~Trenner and P.~Wilson, ``{Asymptotic Curvature of Moduli Spaces for
  Calabi-Yau Threefolds}.''  {\em {Journal of Geometric Analysis}} {\bf 21}
  (2011) 409--428, [\href{http://arxiv.org/abs/0902.4611}{{\tt
  arXiv:0902.4611}}].

\bibitem{Wilsonintersect}
A.~Kanazawa and P.~Wilson, ``{Trilinear Forms and Chern Classes of Calabi-Yau
  Threefolds}.''  {\em Osaka J. Math.} {\bf 51} (2014) 203--215,
  [\href{http://arxiv.org/abs/1201.3266}{{\tt arXiv:1201.3266}}].

\bibitem{Phong}
D.~Phong and J.~Sturm, ``{Lectures on Stability and Constant Scalar
  Curvature}.''  {\em Current Developments in Mathematics} {\bf 2007} (2009)
  101--176, [\href{http://arxiv.org/abs/0801.4179}{{\tt arXiv:0801.4179}}].

\bibitem{Arvanitaki:2009fg}
A.~Arvanitaki, S.~Dimopoulos, S.~Dubovsky, N.~Kaloper, and J.~March-Russell,
  ``{String Axiverse}.''  {\em Phys.Rev.} {\bf D81} (2010) 123530,
  [\href{http://arxiv.org/abs/0905.4720}{{\tt arXiv:0905.4720}}].

\bibitem{Baumann:2014nda}
D.~Baumann and L.~McAllister, ``{Inflation and String Theory}.''
  \href{http://arxiv.org/abs/1404.2601}{{\tt arXiv:1404.2601}}.

\bibitem{Dimopoulos:2005ac}
S.~Dimopoulos, S.~Kachru, J.~McGreevy, and J.~G. Wacker, ``{N-flation}.''  {\em
  JCAP} {\bf 0808} (2008) 003, [\href{http://arxiv.org/abs/hep-th/0507205}{{\tt
  hep-th/0507205}}].

\bibitem{Easther:2005zr}
R.~Easther and L.~McAllister, ``{Random Matrices and the Spectrum of
  N-flation}.''  {\em JCAP} {\bf 0605} (2006) 018,
  [\href{http://arxiv.org/abs/hep-th/0512102}{{\tt hep-th/0512102}}].

\bibitem{Bachlechner:2014hsa}
T.~C. Bachlechner, M.~Dias, J.~Frazer, and L.~McAllister, ``{A New Angle on
  Chaotic Inflation}.''  \href{http://arxiv.org/abs/1404.7496}{{\tt
  arXiv:1404.7496}}.

\bibitem{Kim:2004rp}
J.~E. Kim, H.~P. Nilles, and M.~Peloso, ``{Completing Natural Inflation}.''
  {\em JCAP} {\bf 0501} (2005) 005,
  [\href{http://arxiv.org/abs/hep-ph/0409138}{{\tt hep-ph/0409138}}].

\bibitem{Farquet:2012cs}
D.~Farquet and C.~A. Scrucca, ``{Scalar Geometry and Masses in Calabi-Yau
  String Models}.''  {\em JHEP} {\bf 1209} (2012) 025,
  [\href{http://arxiv.org/abs/1205.5728}{{\tt arXiv:1205.5728}}].

\bibitem{Witten:1993yc}
E.~Witten, ``{Phases of {$N=2$} Theories in Two Dimensions}.''  {\em
  Nucl.Phys.} {\bf B403} (1993) 159--222,
  [\href{http://arxiv.org/abs/hep-th/9301042}{{\tt hep-th/9301042}}].

\bibitem{Greene:1996cy}
B.~R. Greene, ``{String Theory on {C}alabi-{Y}au Manifolds}.''
  \href{http://arxiv.org/abs/hep-th/9702155}{{\tt hep-th/9702155}}.

\bibitem{Cox:2000vi}
D.~Cox and S.~Katz, {\em {Mirror Symmetry and Algebraic Geometry}}.
\newblock {American Mathematical Society}, 2000.

\bibitem{Hori:2003ic}
K.~Hori, S.~Katz, A.~Klemm, R.~Pandharipande, C.~Vafa, R.~Vakil, and E.~Zaslow,
  {\em {Mirror Symmetry}}.
\newblock American Mathematical Society and Clay Mathematics Institute, 2003.

\bibitem{coxnotes}
D.~Cox, ``{What is a Toric Variety?}'' Unpublished lecture notes available at
  {\url{http://www3.amherst.edu/\textasciitilde dacox/} }.

\bibitem{coxbook}
D.~A. Cox, J.~B. Little, and H.~K. Schenck, {\em {Toric Varieties}}.
\newblock American Mathematical Society, 2011.

\bibitem{Batyrev}
V.~V. Batryev, ``{Dual Polyhedra and Mirror Symmetry for Calabi-Yau
  Hypersurfaces in Toric Varieties}.''  {\em {J. Alg. Geom}} (1996) 493--535,
  [\href{http://arxiv.org/abs/alg-geom/9310003}{{\tt alg-geom/9310003}}].

\bibitem{Braun}
V.~Braun, ``{The Moricone of a Calabi-Yau Space from Toric Geometry}.''
  Master's thesis, The University of Texas at Austin, 1998.

\bibitem{Rambau:TOPCOM-ICMS:2002}
J.~Rambau, ``Topcom: Triangulations of point configurations and oriented
  matroids.''  in {\em Mathematical Software---ICMS 2002} (A.~M. Cohen, X.-S.
  Gao, and N.~Takayama, eds.), pp.~330--340, World Scientific, 2002.

\end{thebibliography}\endgroup

\end{document}